\documentclass[suppldata]{interact}
\PassOptionsToPackage{hyphens}{url}\usepackage{hyperref}
\newcommand{\fnref}[1]{\textsuperscript{\hyperref[#1]{\ref*{#1}}}}
\usepackage{anyfontsize}
\usepackage{epstopdf}
\usepackage{subfigure}
\usepackage{xcolor}
\usepackage{soul}

\usepackage{natbib}
\bibpunct[, ]{(}{)}{,}{a}{}{,}
\renewcommand\bibfont{\fontsize{10}{12}\selectfont}
\usepackage{endnotes}

\theoremstyle{plain}

\theoremstyle{definition}

\theoremstyle{remark}

\usepackage{lineno,hyperref}
\usepackage{amsmath}
\usepackage{mathtools}
\usepackage{floatrow}  
\usepackage[utf8]{inputenc}
\usepackage{csquotes}
\usepackage{color,soul}
\usepackage{titlesec}
\usepackage[shortlabels]{enumitem}
\usepackage{multirow}
\usepackage{amsfonts}
\usepackage{amssymb}
\usepackage{tikz}
\usepackage{tabularx}
\usepackage{adjustbox}
\usepackage{url}
\usepackage{makecell}
\usepackage{bm}
\usepackage[normalem]{ulem}
\usepackage{floatrow}
\usepackage{caption}
\usepackage{todonotes}
\usepackage{CJKutf8}
\usepackage{pdflscape}
\usepackage{longtable}
\usepackage{booktabs}  
\usepackage{efbox,graphicx}
\usepackage{array} 
\usepackage{pdflscape} 
\usepackage{url}
\urldef\myurl\url{foo%.com}

\definecolor{beige}{RGB}{245,245,220}
\sethlcolor{beige}

\begin{document}

\articletype{REVIEW ARTICLE}

\title{
A Deep Dive into OpenStreetMap Research Since its Inception (2008–2024): Contributors, Topics, and Future Trends
}

\author{
\name{Yao Sun\textsuperscript{a,c}\thanks{CONTACT Yao Sun Email: yao.sun@tum.de},
Liqiu Meng\textsuperscript{b},
Andrés Camero\textsuperscript{c},
Stefan Auer\textsuperscript{c},
and Xiao Xiang Zhu\textsuperscript{a}}
\affil{\textsuperscript{a}Data Science in Earth Observation, Technical University of Munich, Arcisstraße 21, 80333 Munich, Germany; 
\textsuperscript{b}Cartography and Visual Analytics, 
Technical University of Munich, Arcisstraße 21, 80333 Munich, Germany;
\textsuperscript{c}Remote Sensing Technology Institute, German Aerospace Center, Münchener Straße 20, 82234 Weßling, Germany
}
}

\maketitle

\begin{abstract}
\textcolor{blue}{This paper has been accepted for publication in the \textit{International Journal of Geographical Information Science (IJGIS)}.} 
OpenStreetMap (OSM) has transitioned from a pioneering volunteered geographic information (VGI) project into a global, multi-disciplinary research nexus. This study presents a bibliometric and systematic analysis of the OSM research landscape, examining its development trajectory and key driving forces. By evaluating 1,926 publications from the Web of Science (WoS) Core Collection and 782 State of the Map (SotM) presentations 
up to June 2024, we quantify publication growth, collaboration patterns, and thematic evolution.
Results demonstrate simultaneous consolidation and diversification within the field. While a stable core of contributors continues to anchor OSM research, themes have shifted from initial concerns over data production and quality toward advanced analytical and applied uses. Comparative analysis of OSM-related research in WoS and SotM reveals distinct but complementary agendas between scholars and the OSM community. 
Building on these findings, we identify six emerging research directions and discuss how evolving partnerships among academia, the OSM community, and industry are poised to shape the future of OSM research. 
This study establishes a structured reference for understanding the state of OSM studies and offers strategic pathways for navigating its future trajectory.
\end{abstract}




\section{Introduction}\label{sec:intro}

OpenStreetMap (OSM), founded in 2004 by Steve Coast with the vision of democratizing geographic data and fostering an open data environment~\citep{OpenStreetMapWiki2024}, {is a free, open map project created and maintained by a global community of volunteers}~\citep{haklay2008openstreetmap}. 
Over the past 20 years, OSM has evolved into an exemplary project within Volunteered Geographic Information (VGI), significantly impacting the mapping landscape~\citep{girres2010quality,neis2014recent,schott2023analyzing}. Its influence is particularly notable in underdeveloped regions where authoritative mapping resources are scarce or deprioritized~\citep{hagen2019}. By encouraging open data sharing, OSM has amassed a vast dataset and a growing user base, solidifying its role as a pioneering VGI project~\citep{mooney2017review}.

{As OSM reaches its 20-year milestone,} its significance extends beyond practical applications into the field of research. 
Research on OpenStreetMap, {which began with the first academic publication in 2008}~\citep{haklay2008openstreetmap}, encompasses a broad spectrum of studies, addressing technical, motivational, and community-related aspects. 
On one hand, OSM data has become a valuable resource for scientific studies across various fields. 
As a large, openly available geographic dataset, OSM presents an attractive data source for various studies. 
For instance, generation of 3D city models using OSM data~\citep{over2010generating}, building geometry~\citep{sun2017Building, bagheri2019fusion, zhuo2018optimization}, building information from street-view images~\citep{kang2018building, biljecki2021street, hoffmann2023using, sun2023flickrstr}. 
On the other hand, research has increasingly focused on OSM itself, examining aspects such as data quality~\citep{fan2014quality}, 
the motivations of OSM contributors~\citep{budhathoki2013motivation}, the corporate involvement in OSM~\citep{anderson2019corporate}, 
and vandalism such as popular game Pokemon related vandalism~\citep{juhasz2020cartographic}. 
These research areas are essential for understanding the efficacy and development of OSM as both a tool and a community-driven project.

There have been several articles introducing specific aspects of OSM research. 
For instance, the review by Neis and Zielstra \citep{neis2014recent} and the review by Mooney and Minghini \citep{mooney2017review} provided in-depth analyses of OSM's growth, data quality, and the diverse tools and applications developed within its ecosystem.
A systematic review by Kaur and Antony \citep{kaur2017systematic} emphasized the need for intrinsic methods to assess OSM data quality. 
{Yan et al.~\citep{Yan01092020} provided a comprehensive narrative review of VGI research from 2007 to 2017 to reveal trends, categorize research topics, and identify gaps and future research directions}. 
Vargas-Munoz et al.~\citep{vargas2020openstreetmap} conducted an extensive review exploring the integration of OSM with machine learning and remote sensing.
In 2022, Grinberger et al. \citep{grinberger2022bridges} examined the engagement between academic researchers and the OSM community.
Also in 2022, an editorial~\citep{grinberger2022osm} introduced the concept of ``OSM Science," proposing a unified approach to studying OSM as a multidisciplinary nexus, based on insights from academic conferences, and emphasizing the interconnectedness of research on OSM applications, data quality, and community dynamics.
These reviews of specific aspects of OSM research along with their valuable insights have set the cornerstones for a holistic review of OSM research in this paper. 
Despite the extensive research using OSM, studying OSM, and reviewing OSM, as we reflect on OSM research at this juncture, 
there are questions remain to be answered: 
\textit{Who has been contributing to and driving OSM research? 
What are the primary research themes and how are they evolving? 
How have the research and the use of OSM influenced each other?
Moreover, what are the future directions for {OSM research}?}

On the other hand, although most of the research on OSM is conducted within academia, a significant portion of its contributors come from outside the academic sphere. The OSM community is vast, and its activities often reflect real, applicable needs for OSM, which can help define research topics. However, for academia, much of the community's discussions remain unknown. This is partly due to the use of different platforms and the decentralized nature of the community's discussions and activities, making thorough analysis challenging.
As Mooney et al. discussed~\citep{mooney2018coordinating}, there is a recognized gap and a need for communication between academia and the OSM community, and they suggest establishing meetings, discussions, and collaborations between these groups. In \citep{grinberger2022bridges}, the efforts in research and the community were labeled as OSM-R and OSM-C, respectively, and a preliminary analysis of their relationship was conducted, emphasizing the importance of establishing and strengthening their interaction.
However, academic researchers often still lack a clear understanding of the community’s dynamics. For example, questions remain about \textit{who the key players in the community are and what topics are being discussed within the community}.

More importantly, we are in an era of fast-growing, widely applied artificial intelligence, which is also influencing OSM: 
in data production, AI-assisted mapping tools are transforming traditional workflows~\citep{housel2022rapidx}; 
in terms of contributors, the structure of participation is shifting, with increasing involvement from institutions such as tech companies~\citep{microsoft2025globalml,sirko2021continental}. 
These changes raise important questions: 
\textit{What impact will they have on OSM research and the OSM community? Is OSM at a turning point? And how might OSM evolve in the future?}

This paper aims to explore these questions through a statistical analysis of OSM publications. 
For OSM-related academic research, we conduct a systematic analysis within the Web of Science (WoS) Core Collection. 
Additionally, for discussions in OSM community, we statistically analyse the talks given in the State of the Map conferences (SotM). 

We aim to address the research landscape in OSM by depicting the who and the what, 
while also identifying emerging areas of relevance for the future. 
Specifically: 
\begin{itemize}
    \item 
    {Contributors - the Who}: 
    
    Who is conducting research on OSM? Which countries, institutions, and individuals are involved, and how does the movement of researchers reflect institutional and national priorities?
    \item 
    {Topics - the What}: 
    
    What are the core themes and trends in OSM research? How do these topics evolve, and what are their interconnections?
    \item 
    {Community Priorities}: 

    Additional questions, especially for academic researchers who are less involved in OSM community,  include: 
    What topics are emerging from sources other than academic journals and proceedings, such
    as SotM? And who are the key players in the community?

    \item 
    {Future trends}:  
    
    Which research topics in OSM are emerging? 
    What topics discussed in SotM are likely to become hot topics in research? 
    {Is OSM research entering a new phase, and where is it headed?}

\end{itemize}

Contributions of this work are fourfolds:

\begin{enumerate}
    \item First,  
    we employ quantitative methods to analyze OSM-related research, providing a data-driven study that serves as a foundational reference for the field. Our approach is designed to be repeatable, allowing for updates and comparisons every few years as the field continues to progress and evolve. 

    \item Second, 
    we examine the key contributors and the key research topics, focusing on their impact and collaborations. We highlights the evolution of the field and emerging trends. 

    \item Third, 
    we systematically analyze talks at the SotM conference since 2007, comparing it with academic research. This comparison helps researchers understand the key players, topics, and trends within the broader OSM user base, bridging the gap between academic research and community-driven initiatives.

    \item 
    Fourth, we predict future trends in OSM research by identifying and analyzing evolving research topics within the WoS collection and recent discussions at SotM. 
    {We also reflect on whether OSM is entering a turning point in the age of AI and growing institutional involvement, and what this means for future research and community efforts. This foresight aims to guide both scholars and contributors in navigating a rapidly evolving landscape.}
    
\end{enumerate}

 The remainder of this paper proceeds as follows. 
 Section~\ref{sec:method} to~\ref{sec:topics} focus on academic research: 
 Section~\ref{sec:method} introduces the methodology and data we used in this study, and the general facts are presented in Section~\ref{sec:3}; 
 Section~\ref{sec:who} is concerned with the contributors by analysing the academic contributions from countries, institutions, and authors;   Section~\ref{sec:topics} discusses the research topics and how they are evolved over time. 
 In Section~\ref{sec:sotm}, we change to research beyond academia by analysing the contributors and topics in the annual OSM conference SotM. 
 Based on the study, in Section~\ref{sec:future} we predict future trends. 
 Finally, Section~\ref{sec:conclude} concludes this paper.

\section{Methodology}\label{sec:method}

\begin{figure}[H]
    \centering
    \includegraphics[width=.8\linewidth]{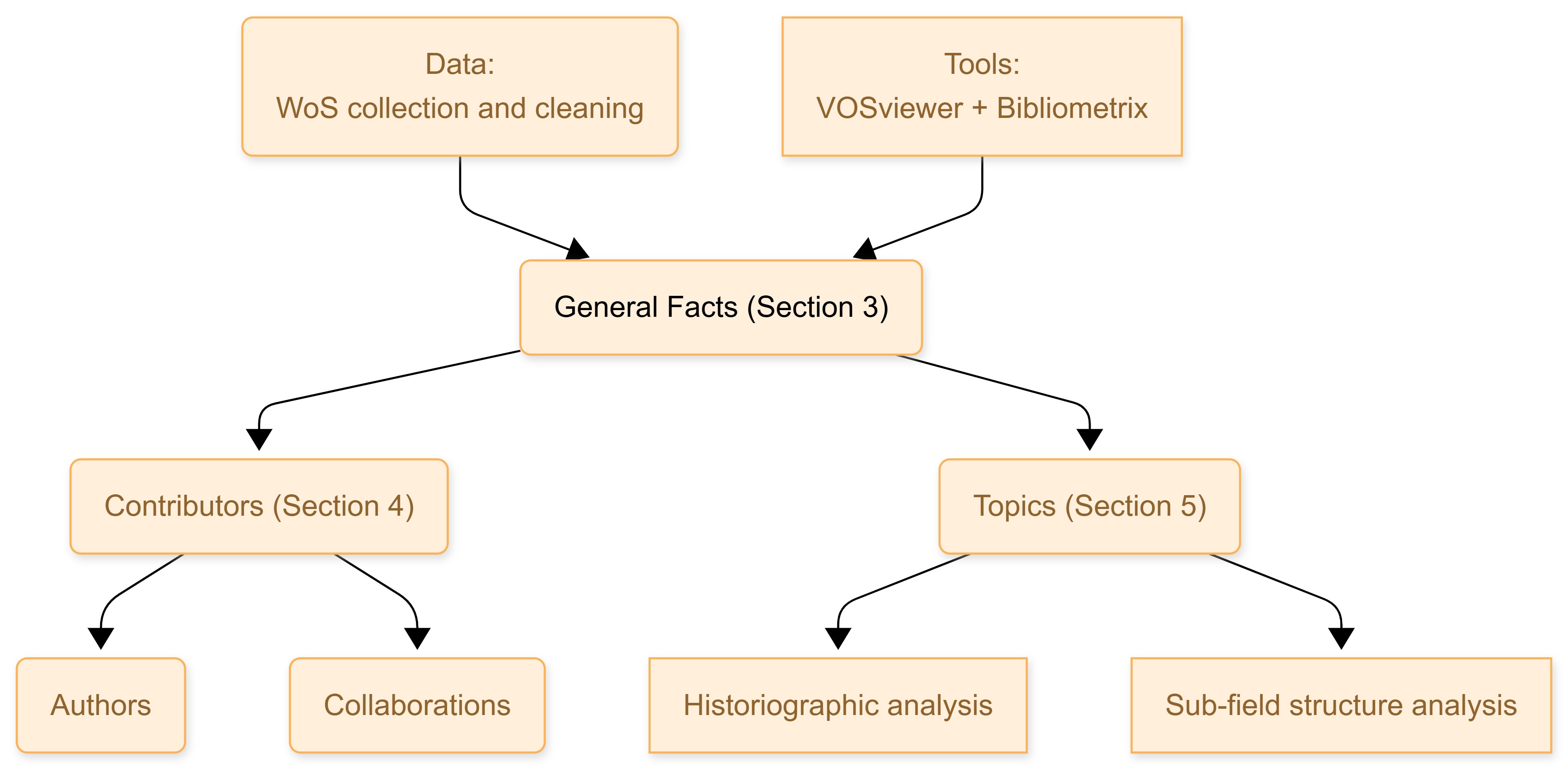}
    \caption{{Methodology overview for analyzing WoS collection. }}
    \label{fig:flowchart}
\end{figure}

{Our analysis is based on WoS Core Collection and bibliometrics tools. Figure}~\ref{fig:flowchart} {presents an overview of the methodology for analyzing the WoS data collection.}

\subsection{Tools for Bibliometric Analysis}

In this study, we employ bibliometrics tools such as VOSviewer~\citep{van2010software} and the bibliometrix package~\citep{aria2017bibliometrix} for conducting a quantitative statistical analysis of publications. VOSviewer is a software tool for constructing and visualizing bibliometric networks, while bibliometrix is an open-source tool implemented in the R language for quantitative research in scientometrics and bibliometrics. 
Both tools are used to aid in visualizing contributors, topics, and other key information in the sampled literature, facilitating the generation of knowledge graphs, such as keyword co-occurrence maps and co-authorship networks. 

Additionally, data cleaning process and other necessary analysis were implemented using Python.

\subsection{Literature Search and Data Collection}

To obtain research data for a bibliometrics study on OpenStreetMap-related research, data was sourced from the WoS Core Collection through our institute's database access. 
The search was conducted using the topic keyword ``OpenStreetMap" or the title keyword ``openstreetmap," with publication dates up to \textit{June 30, 2024}, to download full records and cited references. 
{The search was conducted on \textit{July 11, 2024}.}\footnote{{To addresses the time lag inherent in peer review and indexing systems, a supplementary search covering the period from July 1, 2024, to December 31, 2024, was subsequently performed and is analyzed in Appendix A.}}

Care was taken to exclude irrelevant results by avoiding abbreviations like ``OSM," which could refer to unrelated terms, such as ``open spatial modulation" or ``oncostatin M." 
This search yielded 1,926 records in total, comprising 1,220 articles and 706 proceedings.

\subsection{Data Cleaning and Preprocessing}\label{sec:data_clean}

We perform pre-processing to clean the data. 
This step involves tasks such as eliminating duplicate entries and resolving inconsistencies. 
Particular attention was paid to the disambiguation of author and affiliation names to ensure accurate mapping of co-authorship and other bibliometric analyses. 

In our study, we rely on full author names for both VOSviewer and bibliometrix analysis, but discrepancies often arise due to variations in how names are recorded. 
One common issue encountered was the inconsistent representation of the same author's name across different publications, e.g., some names are listed with full names, others are abbreviated, sometimes even abbreviated in multiple ways. 
We noticed that it was especially problematic with publications from ISPRS Annals and Archives, where names are often formatted as ``First Initial. Last Name." WoS treats this as a full name as the actual full name is not included in the document, leading to multiple variations of the same author's name in the dataset. 
To address this, the original WoS files were thoroughly reviewed. 
These were manually corrected to ensure that instances of the same author were unified under a consistent full name. 

For affiliation name disambiguation, 
several common issues were addressed. For instance, minor variations in institution names, such as ``University of Washington Seattle" and ``University of Washington," were standardized to a single form. 
In cases where different campuses or components of a university system were involved, such as ``University of Wisconsin System" versus ``University of Wisconsin–Madison," the university system was removed to avoid duplicating statistics. 
Similarly, institutions like ``University of California Berkeley" and ``University of California Santa Barbara" were distinguished from the overarching ``University of California System" by treating each campus individually. 
Redundant entries, such as those involving organizations like ``Helmholtz Association" and ``Swiss federal institutes of technology domain" which comprise multiple institutes, were identified and removed to prevent inflation of counts. 
These steps ensured that the dataset accurately reflected affiliations without overcounting or misrepresenting institutional contributions.

There are also instances of different usage of terms without introducing ambiguity. 
For example, we noticed that in the WoS dataset, records with ``German Aerospace Center (DLR)" in the affiliations are all recorded as ``German Aerospace Centre (DLR)." 
As these do not introduce ambiguity in the literature analysis, no modifications are made.

In addition, for analysis of the document sources, we performed disambiguation of conference names, as the same conference is often recorded differently across different years in WoS. 
The disambiguation ensures consistency in the dataset, allowing for more accurate analysis and comparison of conference-related contributions over time. 

\section{General Facts}\label{sec:3}

\subsection{Overview of the Data Collection}

The bibliographic data covers a timespan from 2008 to 2024 and includes 875 sources. 
Within this period, a total of 1,926 documents have been indexed, reflecting a robust annual growth rate of 30.87\%. 
The average age of the documents is 5.13 years, 
and on average, each document has received 18.35 citations. 
The dataset is supported by a comprehensive reference list, containing 54,370 citations. 
This data indicates a rapidly expanding field with a high level of engagement and citation activity. 

The document contents include 1,836 {``Keywords Plus"}\footnote{KeyWords Plus are words or phrases that frequently appear in the titles of an article's references, but do not appear in the title of the article itself (\url{https://support.clarivate.com/ScientificandAcademicResearch/s/article/KeyWords-Plus-generation-creation-and-changes?language=en_US}).} and 
4,884 {author's keywords}\footnote{Author Keywords consist of a list of terms that authors believe best represent the content of their paper.}, indicating a diverse range of topics and research areas. It features 5,527 authors, with 108 contributing to single-authored documents. A total of 122 documents are single-authored, while the average number of co-authors per document is 3.97, reflecting a strong trend toward collaborative research. Additionally, 27.62\% of the documents involve international co-authorship, highlighting a significant level of global collaboration within the research community.

\subsection{Annual Scientific Production}\label{sec:AnnualScientificProduction}

\begin{figure}[!]
    \centering
    \includegraphics[width=0.98\linewidth]{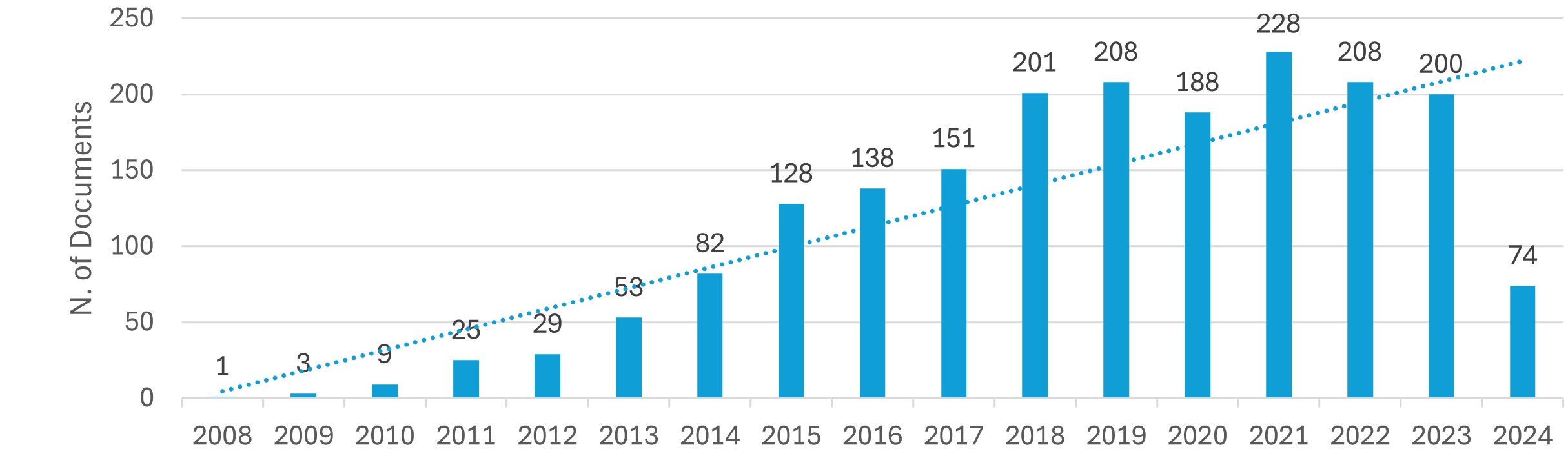}
    \caption{Annual scientific production of OSM research (WoS core collection) with the trending line from 2008 to June 2024. {The data for 2024 includes publications up to the study’s cutoff date of June 30, 2024. The search was conducted on July 11, 2024. \protect\footnotemark}}
    \label{fig:AnnualScientificProduction}
\end{figure}

\footnotetext{{The updated full-year 2024 values used in Appendix~A are shown in Figure}~\ref{fig:app_AnnualScientificProduction}.}

As shown in Figure~\ref{fig:AnnualScientificProduction}, {OSM research has shown steady growth since 2008, starting with a single publication. The output surpassed 50 in 2013, 100 in 2015, and 200 in 2018. A dip occurred in 2020, followed by a peak at 228 in 2021, then a slight decline in 2022 and 2023, however still exceeding 200 publications. By mid-2024, 74 articles had been recorded, reflecting ongoing research interest. }

\subsection{Sources}\label{sec:authors}

In the analysis of OSM-related research, we identified the main journals and conferences that serve as primary platforms for publishing and disseminating findings in this field. 

Table~\ref{tab:articles_by_source} highlights 21 journals that published more than 10 articles each.
Leading the category of publishing articles is \textit{ISPRS International Journal of Geo-Information}, followed by \textit{Remote Sensing} and \textit{Transactions in GIS}. 
{For conferences, Table}~\ref{tab:articles-conf} {lists 20 conferences with more than 5 published proceedings. 
Note that in compiling statistics for conferences, we aggregate related series conferences and list the frequency of these conferences. 
As can be seen, 
The \textit{ISPRS Congress} led with 26 proceedings, followed by \textit{SIGSPATIAL} with 21 proceedings. \textit{IGARSS} also featured prominently with 19 proceedings.} 

It is noticeable that OSM-related research spans a wide range of disciplines, reflecting its interdisciplinary nature and thematic diversity. Contributions originate from fields such as GIScience, cartography, remote sensing, urban studies, transportation, robotics, and sustainability sciences.
{While many OSM-related studies appear in GIScience journals (e.g., \textit{ISPRS International Journal of Geo-Information}, \textit{Transactions in GIS}), publications in interdisciplinary venues (\textit{Sustainability}, \textit{PLOS One}, \textit{IEEE Access}) highlight the broader impact of OSM research. Similarly, conference proceedings show its relevance beyond GIS, including AI, big data, and transportation (\textit{IEEE Big Data}, \textit{ICRA}, \textit{SIGSPATIAL}). This diversity shows that OSM research extends beyond GIScience and is gaining wider recognition, as exemplified by a recent publication in \textit{Nature Communications}}~\citep{Herfort2023}.

\renewcommand{\arraystretch}{1.1}

\begin{table}[!]
\scriptsize
    \centering
    \begin{tabular}{l|c}
        \hline
        \textbf{Name of Journal} & \textbf{\# Articles} \\
        \hline
        ISPRS International Journal of Geo-Information & 178 \\
        Remote Sensing & 75 \\
        Transactions in GIS & 67 \\
        International Journal of Geographical Information Science & 57 \\
        Computers, Environment and Urban Systems & 34 \\
        Sustainability & 32 \\
        IEEE Access & 22 \\
        ISPRS Journal of Photogrammetry and Remote Sensing & 21 \\
        IEEE Journal of Selected Topics in Applied Earth Observations and Remote Sensing & 20 \\
        PLOS One & 19 \\
        Environment and Planning B: Urban Analytics and City Science & 18 \\
        International Journal of Applied Earth Observation and Geoinformation & 18 \\
        Cartography and Geographic Information Science & 17 \\
        International Journal of Digital Earth & 16 \\
        Sensors & 15 \\
        IEEE Transactions on Geoscience and Remote Sensing & 14 \\
        International Journal of Health Geographics & 14 \\
        Geo-Spatial Information Science & 13 \\
        Transportation Research Record & 13 \\
        Applied Geography & 10 \\
        Geocarto International & 10 \\
        \hline
    \end{tabular}
    \caption{Most published journals and the number of articles by source (more than 10 articles).}
    \label{tab:articles_by_source}
\end{table}

\begin{table}[!]
\scriptsize
\centering
\begin{tabularx}{\textwidth}{>{\raggedright\arraybackslash}X|c|c}
\hline
\textbf{Name of Conference} & \textbf{\# Proceedings} & \textbf{Conference Frequency}\\
\hline
ISPRS Congress & 26 & Quadrennial\\ 
ACM SIGSPATIAL International Conference on Advances in Geographic Information Systems (SIGSPATIAL) & 21 & Annual\\ 
IEEE International Geoscience and Remote Sensing Symposium (IGARSS) & 19 & Annual\\ 
ISPRS Geospatial Week (GSW) & 11 & Biennial\\ 
International Conference on Cartography and GIS (ICC\&GIS) & 9 & Biennial\\ 
International Conference on Geographical Information Systems Theory, Applications and Management (GISTAM) & 9 & Annual\\  
Web and Wireless Geographical Information Systems (WGIS) & 9 & Annual\\  
IEEE Intelligent Vehicles Symposium (IV) & 7 & Annual\\  
IEEE International Conference on Big Data (Big Data) & 7  & Annual\\ 
Joint Urban Remote Sensing Event (JURSE) & 7 & Biennial\\
Semantic Web (ISWC) & 7 & Annual\\  
Computers Helping People with Special Needs (ICCHP) & 6 & Biennial\\
FOSSG - Academic Track & 6 & Annual\\  
IEEE International Conference on Data Engineering (ICDE) & 6 & Annual\\  
Conference of the Open Innovations Association (FRUCT) & 6  & Annual\\ 
IEEE International Conference on Mobile Data Management (MDM) & 5  & Annual\\ 
IEEE International Conference on Robotics and Automation (ICRA) & 5  & Annual\\ 
IEEE/RSJ International Conference on Intelligent Robots and Systems (IROS) & 5 & Annual\\  
International Conference on Geoinformatics (Geoinformatics) & 5 & Annual\\  
Hawaii International Conference on System Sciences (HICSS) & 5 & Annual\\  \hline
\end{tabularx}
\caption{Sources and the number of articles published  (more than 5 proceedings) 
}
\label{tab:articles-conf}
\end{table}

\section{Contributors}\label{sec:who}

This section focuses on the contributions, including influential authors, prominent institutions, and geographical distribution. 
We analyze the top authors and their evolving affiliations, examine the leading countries and institutions in OSM research, and then explore the collaborative efforts among authors, organizations, and countries.

\subsection{Authors}\label{sec:41authors} 

{We analyze contributions by both authors and institutions/countries, and examine the evolution of authors' affiliations over time.}

\subsubsection{Top Authors and the Contribution Over Time}\label{sec:411}

\begin{figure}[!]
  \centering
  \includegraphics[width=\linewidth]{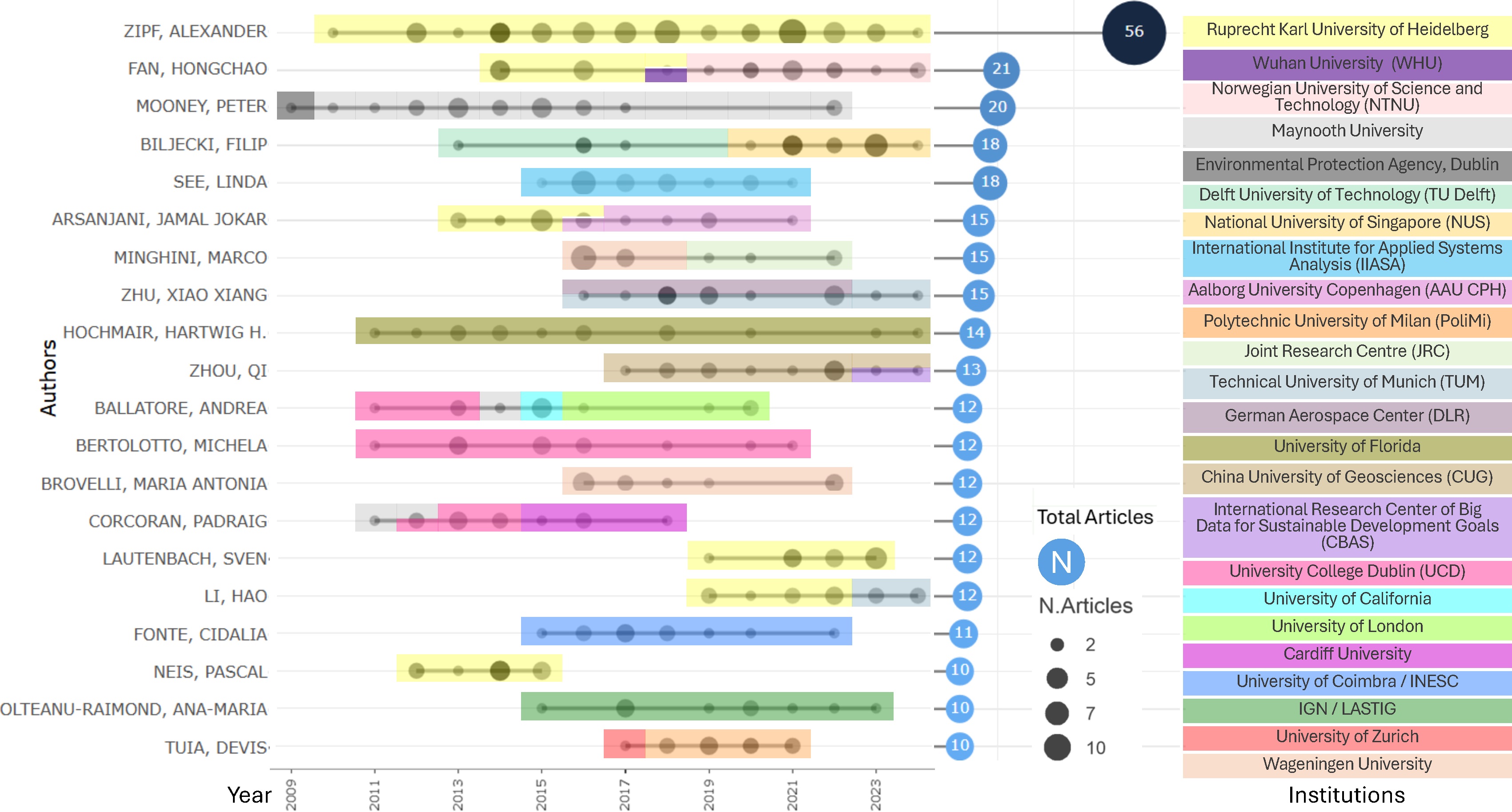}
  \caption{{Top-Authors' Productivity over Time, Number of Total Articles, and Authors' affiliations over time. Institutions are color-coded as shown on the right.}
  {The data for 2024 includes publications up to the study’s cutoff date of June 30, 2024. The search was conducted on July 11, 2024.}}
  \label{fig:author-overyear}
\end{figure}

\begin{table}[!]
\centering
\setlength{\tabcolsep}{0.7pt}
\renewcommand{\arraystretch}{1.7}
\fontsize{4.3pt}{4.3pt}\selectfont
\begin{tabular}{p{1.4cm}ccccccccccccccccr}
\hline
\textbf{Author} & {2009} & {2010} & {2011} & {2012} & {2013} & {2014} & {2015} & {2016} & {2017} & {2018} & {2019} & {2020} & {2021} & {2022} & {2023} & {\textbf{ALL}}\\
\hline
ZIPF, ALEXANDER & 0(3) & 33\%(3) & 0(2) & 44\%(9) & 17\%(6) & 57\%(7) & 44\%(9) & 38\%(13) & 42\%(12) & 44\%(18) & 29\%(7) & 27\%(11) & 77\%(13) & 71\%(7) & 50\%(6) &\textbf{ 42\%(126)} \\
FAN, HONGCHAO & 0(3) & - & - & 0(2) & - & 67\%(6) & 0(3) & 50\%(8) & 20\%(5) & 13\%(8) & 67\%(3) & 44\%(9) & 18\%(11) & 14\%(7) & 50\%(4) & \textbf{30\%(69)} \\
MOONEY, PETER & 100\%(1) & 50\%(2) & 10\%(10) & 29\%(7) & 100\%(4) & 67\%(3) & 67\%(6) & 33\%(6) & 100\%(1) & - & 0(1) & 0(2) & - & 67\%(3) & 0(1) & \textbf{41\%(47)} \\
BILJECKI, FILIP & - & - & - & - & 100\%(1) & 0(1) & 0(6) & 33\%(6) & 50\%(2) & 0(3) & 0(1) & 25\%(4) & 36\%(11) & 12\%(17) & 26\%(23) & \textbf{23\%(75)} \\
SEE, LINDA & - & 0(2) & 0(5) & 0(7) & 0(17) & 0(6) & 9\%(11) & 70\%(10) & 21\%(14) & 27\%(11) & 6\%(18) & 15\%(13) & 9\%(11) & 0(10) & 0(9) & \textbf{13\%(144)} \\
ARSANJANI, JAMAL JOKAR & - & - & - & - & 29\%(7) & 50\%(2) & 63\%(8) & 33\%(6) & 20\%(5) & 25\%(4) & 29\%(7) & 0(7) & 10\%(10) & 0(6) & 0(7) & \textbf{22\%(69)} \\
MINGHINI, MARCO & - & - & - & 0(1) & 0(2) & - & 0(1) & 80\%(10) & 60\%(5) & 0(1) & 50\%(2) & 100\%(1) & - & 100\%(2) & 0(1) & \textbf{58\%(26)} \\
ZHU, XIAO XIANG & 0(3) & 0(1) & 0(4) & 0(9) & 0(9) & 0(15) & 0(24) & 3\%(32) & 5\%(19) & 13\%(24) & 43\%(7) & 3\%(40) & 0(16) & 6\%(62) & 4\%(24) & \textbf{5\%(289)} \\
HOCHMAIR,  HARTWIG H. & 0(2) & 0(1) & 100\%(1) & 33\%(3) & 33\%(6) & 33\%(6) & 14\%(7) & 50\%(4) & 0(3) & 40\%(5) & 0(1) & 25\%(4) & 0(3) & 0(3) & 25\%(4) & \textbf{24\%(53)} \\
ZHOU, Qi & - & - & - & 0(2) & - & 0(1) & 0(2) & 0(2) & 50\%(2) & 67\%(3) & 100\%(2) & 33\%(3) & 33\%(3) & 67\%(6) & 50\%(2) & \textbf{43\%(28)} \\
BALLATORE, ANDREA & - & 0(1) & 100\%(1) & - & 100\%(2) & 33\%(3) & 57\%(7) & 25\%(4) & 0(1) & 0(3) & 20\%(5) & 33\%(6) & - & 0(2) & 0(5) & \textbf{30\%(40)} \\
BERTOLOTTO, MICHELA & 0(4) & 0(4) & 25\%(4) & 0(8) & 75\%(4) & 0(3) & 38\%(8) & 67\%(3) & 0(1) & 20\%(5) & 0(5) & 33\%(3) & 25\%(4) & 0(5) & - & \textbf{18\%(61)} \\
BROVELLI, MARIA ANTONIA & - & 0(2) & 0(4) & 0(3) & 0(5) & 0(2) & 0(4) & 45\%(11) & 29\%(7) & 14\%(7) & 10\%(10) & 0(10) & 0(2) & 21\%(14) & 0(20) & \textbf{12\%(101)} \\
CORCORAN, PADRAIG & - & 0(1) & 13\%(8) & 33\%(6) & 100\%(3) & 50\%(4) & 33\%(3) & 40\%(5) & 0(2) & 14\%(7) & 0(3) & 0(4) & 0(6) & 0(4) & 0(7) & \textbf{19\%(63)} \\
LAUTENBACH, SVEN & 0(1) & 0(4) & 0(5) & 0(4) & 0(4) & 0(2) & 0(1) & 0(5) & 0(8) & 0(3) & 11\%(9) & 0(5) & 100\%(3) & 43\%(7) & 63\%(8) & \textbf{17\%(69)} \\
LI, HAO & - & - & - & - & - & - & - & - & - & 0(1) & 100\%(2) & 25\%(4) & 100\%(2) & 75\%(4) & 29\%(7) & \textbf{50\%(20)} \\
FONTE, CIDALIA & 0(5) & 0(1) & - & 0(2) & 0(1) & 0(1) & 13\%(8) & 50\%(4) & 43\%(7) & 100\%(2) & 33\%(3) & 100\%(1) & 0(1) & 50\%(2) & 0(2) & \textbf{27\%(40)} \\
NEIS, PASCAL & - & - & - & 100\%(2) & 100\%(1) & 100\%(4) & 100\%(3) & - & - & - & - & - & - & - & 0(4) & \textbf{67\%(14)} \\
OLTEANU-RAIMOND, ANA-MARIA & - & - & - & 0(1) & 0(1) & - & 25\%(4) & 0(3) & 100\%(3) & 0(1) & 100\%(1) & 100\%(2) & 100\%(1) & 50\%(2) & 25\%(4) & \textbf{43\%(23)} \\
TUIA, DEVIS & 0(12) & 0(9) & 0(14) & 0(14) & 0(14) & 0(20) & 0(21) & 0(19) & 6\%(16) & 12\%(17) & 21\%(14) & 22\%(9) & 18\%(11) & 0(15) & 0(12) & \textbf{4\%(217)} \\
\hline
\end{tabular}
\caption{{Top-Authors' publications by year {(2009-2023)}: percentage of OSM-related publications (Total number of publications). Years without publications
are marked with ‘-’.}}
\label{tab:osm_all_pub}
\end{table}

In Figure~\ref{fig:author-overyear}, 
{the top 20 most published authors' total articles are shown in green circles and productivity over time is shown with the gray circles in the timeline. 
Each author has published more than ten OSM-related papers, with Alexander Zipf being the most prolific (56 papers). }

{To further determine the researchers' primarily focusing on OSM-related topics and researchers engaging in OSM as a secondary area of interest, we compared the number of OSM-related publications to the total publications per year for each author, as shown in  Table}~\ref{tab:osm_all_pub}{. We retrieved individual citation reports from WoS and analyzed records from 2009 - the year the first OSM-related article was published by top authors - to 2023.}

{It is important to note that the WoS Core Collection does not include all published articles and some works by top authors may not be indexed. As a result, individual data points for a given year have limited explanatory power. Therefore, our analysis focuses on identifying trends in WoS-indexed publications rather than interpreting isolated data points.}

{Key insights from the analysis include: }

\begin{itemize}
    \item 
    {\textit{Consistent Contribution to OSM Research:}  
    Authors such as Zipf and Mooney have consistently maintained high levels of OSM research activity over time, whereas others have shown intermittent or declining interest. }

    \item 
    {\textit{Proportion of OSM-Related Publications:}  
    The percentage of OSM-related publications reflects the degree of specialization of authors (c.f., the last column of Table}~\ref{tab:osm_all_pub}){. Some authors exhibit a strong focus on OSM research, with seven exceeding 40\%. For instance, Neis has the highest percentage at 67\%, followed by Minghini at 58\%. In contrast, authors such as Zhu and Tuia, despite having comparable absolute numbers of OSM-related articles, have much lower percentages—5\% and 4\%, respectively—indicating a broader research scope beyond OSM. }

    \item 
    {\textit{Shifting Research Interests:}  
    The percentage of OSM-related articles also highlights evolving research interests among authors. For example, Lautenbach had no OSM-related publications from 2009 to 2018 but shifted significantly toward OSM topics after 2019, with Zhou displaying a similar pattern. Conversely, Corcoran focused extensively on OSM-related research in earlier years but shifted away after 2019, while Arsanjani has contributed fewer OSM-related studies since 2020. }

    \item 
    {\textit{Sustained Research Output:}  
    Most authors have demonstrated continuous publication activity over the analyzed years. A notable exception is Neis, who had no publications between 2016 and 2022, possibly due to a change in professional focus or research interests. }

    \item 
    {\textit{Interdisciplinary Connections of OSM Research:}  
    Further investigation into the primary research domains of these authors reveals significant interdisciplinary connections. 
    For instance, Zhu and Tuia primarily focus on remote sensing; Lautenbach specializes in ecosystem services and GIScience; Zhou’s work spans road networks and map generalization; Corcoran is engaged in network science; and Arsanjani focuses on Earth observation and land use science. These findings underscore the intersection of OSM research with various scientific disciplines.} 
    
\end{itemize}

These insights illustrate the dynamic nature of research interests and productivity patterns within the OSM-related research community. 
These productivity and research-interest patterns may also provide insights into which authors are likely to continue contributing significantly to OSM-related research in the coming years.

\subsubsection{Institutions and Countries}\label{sec:412}

\definecolor{green(pigment)}{rgb}{0.0, 0.65, 0.31}
\definecolor{forestgreen(web)}{rgb}{0.13, 0.55, 0.13}
\definecolor{trueblue}{rgb}{0.0, 0.45, 0.81}
\definecolor{steelblue}{rgb}{0.27, 0.51, 0.71}
\definecolor{brass}{rgb}{0.71, 0.65, 0.26}
\definecolor{citrine}{rgb}{0.89, 0.82, 0.04}
\definecolor{richlavender}{rgb}{0.67, 0.38, 0.8}

\begin{table}[!]
\scriptsize
\centering
\setlength{\tabcolsep}{7pt}
\begin{tabular}{l|c|c|c}
\hline
{Organization} & {Documents} & {Citations} & {mean NCS} \\ \hline
Ruprecht Karls University Heidelberg (Heidelberg University) & \textbf{102} & \textbf{3644} & \textbf{1.67} \\ 
Wuhan University (WHU) & \textbf{79} & 1134 & 1.14 \\ 
Chinese Academy of Sciences (CAS) & \textbf{56} & 945 & 1.13 \\ 
German Aerospace Centre (DLR) & 53 & 979 & 1.17 \\ 
China University of Geosciences (CUG) & 44 & 625 & 1.23 \\ 
Technical University of Munich (TUM) & 42 & 950 & 0.98 \\ 
University of London & 39 & \textbf{3963} & 1.38 \\ 
Centre National de la Recherche Scientifique (CNRS) & 34 & 301 & 0.54 \\ 
University College London (UCL) & 28 & \textbf{3794} & \textbf{1.59} \\ 
National University of Singapore (NUS) & 26 & 605 & \textbf{2.82} \\ 
Maynooth University & 23 & 616 & 0.88 \\ 
Polytechnic University of Milan (PoliMi) & 23 & 392 & 1.00 \\ 
University College Dublin (UCD) & 22 & 533 & 0.79 \\[1pt]
George Mason University (GMU) & 21 & 450 & 0.96 \\ 
University of Chinese Academy of Sciences (UCAS) & 21 & 259 & 0.95 \\ 
International Institute for Applied Systems Analysis (IIASA) & 20 & 520 & 1.28 \\ 
\hline
\end{tabular}
\caption{{Documents, citations, and mean NCS by organization (16 organizations with more than 20 documents). Top 3 numbers in each column are highlighted with \textbf{bold}.}}
\label{tab:documents_citations}
\end{table}

{Approximately 1.3\% of organisations generate 34\% of all OSM publications. Table}~\ref{tab:documents_citations} {lists the 16 institutions with $\ge$ 20 documents and reports three metrics, documents, total citations, and mean Normalized Citation Score (NCS); the top three values in each column are shown in \textbf{bold}.} 
{NCS adjusts for publication age by dividing a paper’s citations by the average citations of all papers published in the same year}~\footnote{https://www.rdocumentation.org/packages/bibliometrix/versions/4.3.3/topics/normalizeCitationScore}. {A NCS of 1.0 equals the worldwide, year-specific average; values above 1 signal above-average impact. 
For all 1,926 OSM-related papers in WoS collection, we calculated NCS using Bibliometrix, and then averaged those scores for each institution.}  

{The results reveal marked differences in relative influence: 
\textit{National University of Singapore (NUS)} leads with a mean NCS of 2.82, followed by \textit{Heidelberg University} (1.67) and \textit{University College London (UCL)} (1.59). On the other hand, \textit{CNRS} (0.54) and \textit{University College Dublin (UCD)} (0.79) are below the global benchmark, as many of the most-cited papers were published early in the study period.}

\begin{figure}[!]
  \centering
  \includegraphics[width=\linewidth]{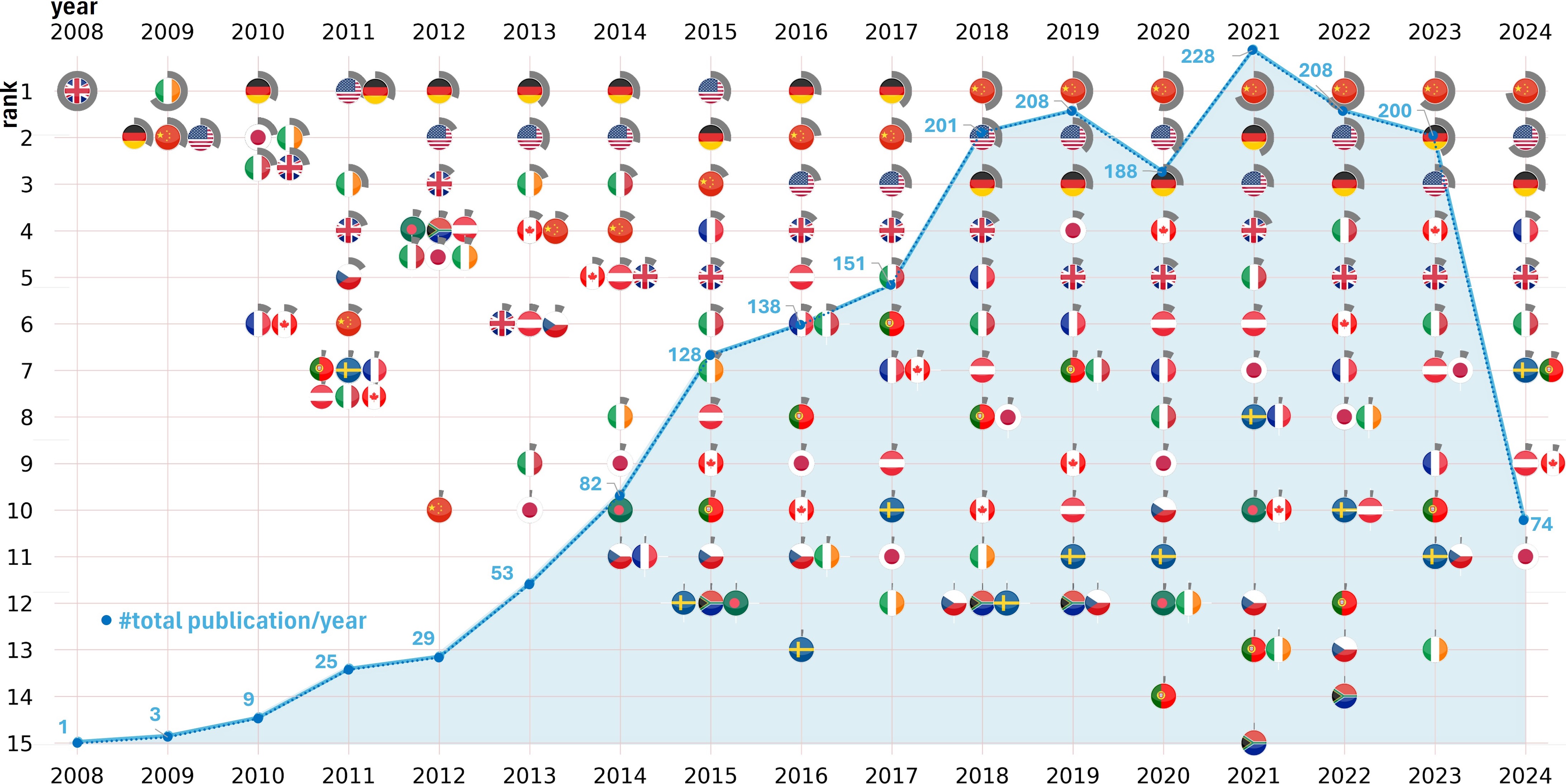}
  \caption{
  Annual rankings of countries by OSM research output from 2008 to June of 2024, featuring flags with circular charts indicating each country’s publication percentage of the total annual output. 
  The blue line chart in the background illustrates the total number of publications each year. }
  \label{fig:countries-rank}
\end{figure}

Figure~\ref{fig:countries-rank} presents the annual rankings of countries by OSM research output from 2008 to June 2024, highlighting the field's global expansion. The United Kingdom led in 2008, with Ireland, China, Germany, and the USA joining in 2009, followed by Japan, Italy, France, and Canada in 2010, strengthening the international research network. Since 2010, Germany has been a dominant contributor, with the USA and China emerging as key players, consistently ranking among the top three since 2015. 

South Africa became the first major African contributor in 2012, followed by Ghana and Nigeria, which saw growing participation in later years. Latin America entered in 2013 with Brazil, later joined by Argentina and Colombia. In Asia, Bangladesh marked South Asia's entry in 2012, with India, Japan, Singapore, and Nepal contributing in subsequent years. European research expanded with Austria, Portugal, and Spain alongside the UK, Italy, France, and the Netherlands. North America remained strong with Canada and the USA. By 2020, OSM research had achieved global representation, 
reflecting increasing international collaboration in the field.

\subsubsection{Top Authors' Affiliations Over Time}\label{sec:413}

Research is conducted by authors, and research topics often move with the authors as they move between different institutions. 
The analysis of the affiliations of top authors 
reveals that these topics are influenced by the authors' affiliations over time. 
Figure~\ref{fig:author-overyear} shows top 20 most published authors' productivity and the their corresponding affiliations over time. 
The analysis of top authors' affiliations reveals that 
24 different institutions have been instrumental in shaping this research landscape, as visualized in Figure~\ref{fig:author-overyear}.

We first analyze the data by distinguishing between {authors' affiliations}, if they are unchanged or changed: 

\begin{enumerate}
    \item[A.] \textit{Authors' Affiliation Unchanged}

    Several leading authors have maintained affiliations with a single institution throughout their OSM publication periods. These authors include Alexander Zipf, Linda See, Hartwig H. Hochmair, Michela Bertolotto, Maria Antonia Brovelli, Cidalia Fonte, and Ana-Maria Olteanu-Raimond.

    There are two distinct scenarios within this group:

    \begin{itemize}
    \item Scenario 1: Unique Affiliation with No Overlap with other Top Authors

     These authors are the only top contributors from their respective institutions, with no affiliation overlap with other top authors:

     \begin{itemize}
         \item Linda See: IIASA; 
         \item Hartwig H. Hochmair: University of Florida;
        \item Cidalia Fonte: University of Coimbra / INESC;
        \item Ana-Maria Olteanu-Raimond: IGN / LASTIG.

     \end{itemize}

    \item Scenario 2: Shared Affiliation with Multiple Top Authors 
    
     In these cases, institutions have been associated with multiple top authors, indicating that these institutions or their leading authors may have played a significant role in fostering other prolific contributors to the OSM field and these institutions have been pivotal in the diffusion of OSM topics:

    \begin{itemize}
         \item Alexander Zipf, affiliated with Heidelberg University, associated with other top authors like Hongchao Fan, Jamal Jokar Arsanjani, Sven Lautenbach, Hao Li, and Pascal Neis;
         
        \item Michela Bertolotto, affiliated with UCD, linked with top authors Andrea Ballatore and Padraig Corcoran;

        \item Maria Antonia Brovelli, from PoliMi, associated with Marco Minghini;

        \item Peter Mooney, affiliated with Maynooth University since 2010 (briefly with another institute before), associated with Andrea Ballatore and Padraig Corcoran.
     \end{itemize}

    \end{itemize}

    \item[B.] \textit{Authors' Affiliation Changed at Least Once}

    We observed two scenarios among authors who have changed their institutional affiliations:

    \begin{itemize}
        \item Continued Output After Affiliation Change
        
        In most cases, top authors continued to produce significant OSM-related research even after changing their institutions. This indicates a strong, enduring connection between the authors and their research topics, which they carried with them to their new affiliations. 
   
        For example: 
        Filip Biljecki continued his OSM research after moving from TU Delft to NUS in Singapore.
       
        Several authors who conducted research at Heidelberg University also followed this pattern:
        Hongchao Fan moved to NTNU in Norway; 
        Jamal Jokar Arsanjani relocated to AAU CPH in Denmark; 
        Hao Li transitioned to TUM.

    \item  No Output After Affiliation Change 
    
   In some cases, after changing institutions, authors no longer produced OSM-related research, suggesting their research was strongly tied to their original institution. This may indicate a shift in their research focus or job responsibilities. 
   An example of this is
   Pascal Neis. 

\end{itemize}
     
\end{enumerate}

\begin{figure}[!h]
  \centering
  \includegraphics[width=\linewidth]{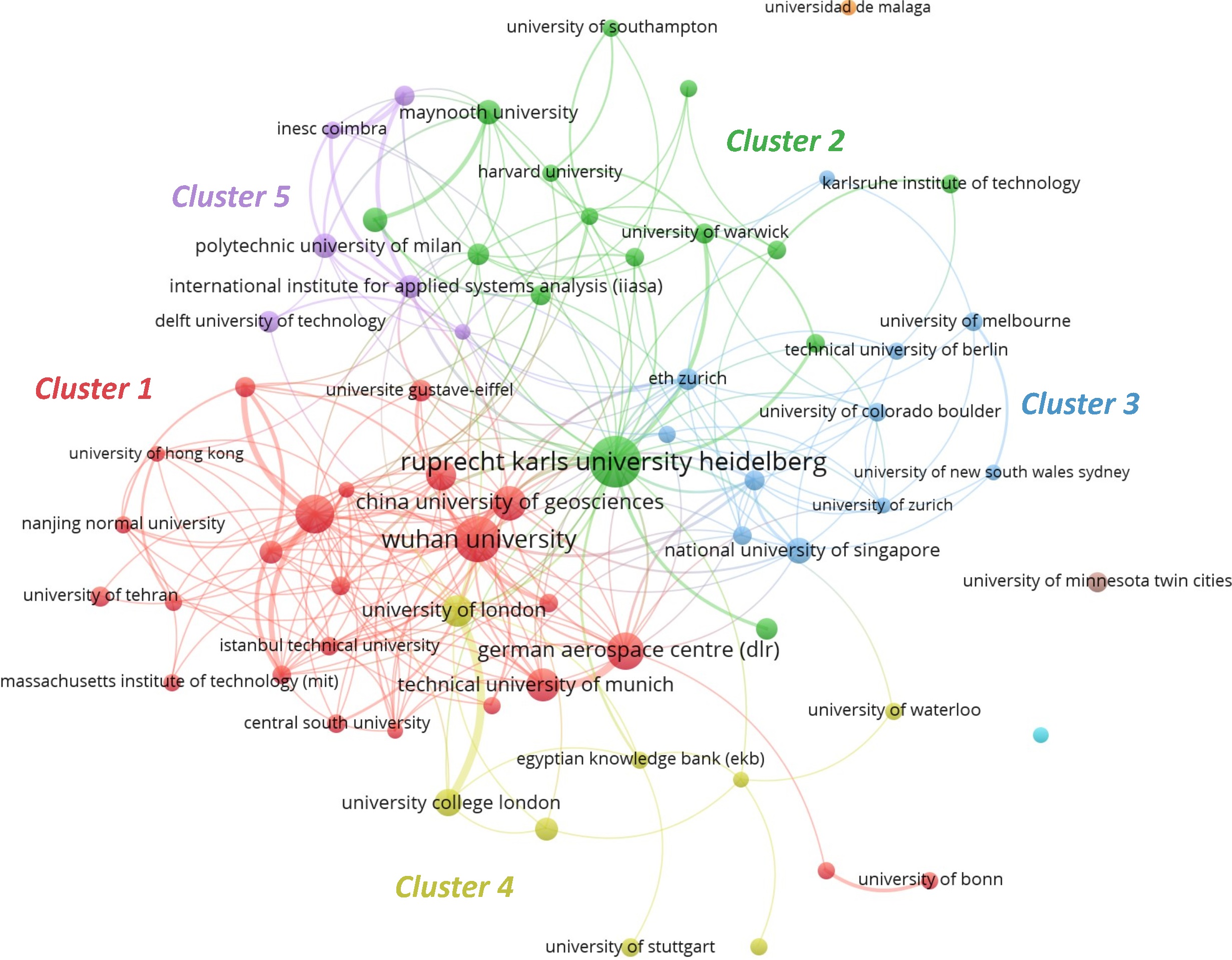}
  \caption{Organizations' Co-Authorship Network. 
  67 organizations are selected with a minimum threshold of 10 documents. The top 5 clusters are labeled. 
  Interactive version can be accessed here \protect\footnotemark.
}
  \label{fig:affi_net}
\end{figure}

Top authors in OSM research are distributed across different institutions and countries. This distribution highlights the global interest and investment in OSM research across regions. 
We further analyse the data by looking at the {institutions and countries over time, c.f., the timeline in Figure~\ref{fig:author-overyear}}. 
We observe three categories: 

\begin{enumerate}

    \item[A.] \textit{Institutions with Long-term Presence}\\
    Institutions and countries with a long-term presence     
    of top authors producing significant OSM research demonstrate a sustained interest and support for OSM-related studies. These include:
    \begin{itemize}
        \item  Germany, notably, two institutions - Ruprecht Karl University of Heidelberg and TUM.
        \item   Ireland, featuring two key institutions - Maynooth University and UCD.
        \item  Austria, represented by IIASA.
        \item   United States, specifically, the University of Florida.
        \item   China, with contributions from CUG.
        \item   Italy, highlighted by the PoliMi.
    \end{itemize} 
    \item[B.] \textit{Institutions on the Right Side of the Timeline (Relatively) }
    \\
    The institutions to which top authors have moved are positioned on the right side of the timeline, indicating that these institutions and countries typically reflect a growing support and interest in OSM research. 
    The authors have dispersed across various institutions and regions, including: 
        \begin{itemize}
        \item  Norway: NTNU
        \item  Denmark: AAU CPH
        \item  Singapore: NUS
        \item Italy: JRC
        \item  China: CBAS 
        \item UK: University of London
        \item  Netherlands: Wageningen University
        \end{itemize}
    \item[C.] \textit{Institutions on the Left Side of the Timeline (Relatively)} \\
        Before moving to new institutions, the original institutions of these top authors appear on the left side of the timeline. We observed that some institutions no longer appear on the map after these authors moved: 
    \begin{itemize}
        \item One possible reason is that OSM research may not be a primary focus at these institutions. 
        \\
        For instance, the \textit{University of Zurich} no longer appears after Devis Tuia moved to another institution, as shown in Figure~\ref{fig:author-overyear}. However, from Figure~\ref{fig:affi_net}, we can see that the University of Zurich has 10 OSM-related publications. Further investigating shows that 2 papers are contributed by Tuia, and other 8 papers involve different authors from the university spanning from 2015 to 2023. It indicates that OSM research may not be a primary focus of the university or other related authors from the university. 
        
        \item Another possibility is that OSM research continues at these institutions, but without a prominent leading author. 
        \\
        An example is \textit{WHU}, which appears only briefly in Figure~\ref{fig:author-overyear}. 
        However, as shown in Figure~\ref{fig:affi_net}, \textit{WHU} has produced 79 OSM-related documents. This suggests that there is strong research interest in OSM at \textit{WHU}, distributed among multiple authors who are not in the top 20 OSM-author list. \\
        Interested readers are referred to the interactive online view of Figure~\ref{fig:affi_net}, as indicated in the footnote, where more detailed information can be explored. 
    \end{itemize}
\end{enumerate}

{All of the situations described above illustrate that OSM research is largely driven by researchers, with their relocations playing a key role in the dissemination of related topics across institutions and countries. 
Some institutions have actively fostered OSM researchers, while others have benefited from incoming scholars who introduce OSM-related studies to new academic environments. These patterns also reflect, to some extent, the willingness of institutions and countries to support and engage with OSM research.  }

\subsection{Collaborations} 

In this section, we explore collaboration patterns in OSM research, focusing on both authors' partnerships and institutional and national co-authorship networks. 

\subsubsection{Authors' Collaboration}

To construct the co-authorship network between authors, a minimum threshold of five documents per author was used to select 92 authors. Figure~\ref{fig:author-net} shows the clustered network.  
Some authors are not interconnected, resulting in a network where the largest connected subset comprises 49 authors and many small sub-nets. Clusters with at least 3 authors are labeled in Figure~\ref{fig:author-net}.

{The \textit{largest subset} of the co-authorship network distributes across \textbf{five clusters}, each with distinct patterns of connectivity, underscoring the complex web of interrelationships among these clusters, with varying degrees of connectivity reflecting different collaborative dynamics within the largest subset of authors, shown in the center of Figure}~\ref{fig:author-net}{, and an overview of each cluster is presented in
Table}~\ref{tab:leading_authors}. 
{Notably, \textit{Alexander Zipf} and \textit{Peter Mooney} stand out as both the most published authors in their respective clusters and key collaborators across multiple clusters. \textit{Zipf} has extensive connections, linking to 4 authors in Cluster 1, 5 in Cluster 3, 2 in Cluster 5, and 1 in Cluster 4. Similarly, \textit{Mooney} is a central figure, connecting with authors in Clusters 1, 2, and 3, while also being linked to all other authors in Cluster 5. 
Cluster-wise, 
Cluster 4 has significantly fewer cross-cluster connections than other clusters, with only two authors establishing links beyond their group.}

\footnotetext{\label{fn:osmbib}\url{https://github.com/ya0-sun/OSMbib}}

{Several \textit{small clusters} are not connected to the largest subset of authors and are distributed along the edges of the network in Figure}~\ref{fig:author-net}.
{Cluster 6 includes 6 authors, Cluster 7 has 4 authors, and Cluster 8 features 3 authors, reflecting varying degrees of collaboration within these groups. Additionally, Clusters 9 to 17 each contain 2 authors, while clusters 18 to 29 each have a single author, indicating a broader distribution of individual contributions across the network. }

Some small clusters contain \textit{multiple authors}, suggesting that there are isolated groups with their focused research areas or collaborative networks that do not overlap with the larger connected subset. This indicates the presence of distinct, potentially specialized research communities within the network. Conversely, some small clusters consist of only \textit{a single author}, suggesting that these individuals may work independently or in highly niche fields that do not intersect with other prolific researchers, e.g., \textit{Boeing, Geoff}, the sole author of OSMnx~\citep{boeing2017osmnx}, as underlined in \textcolor{red}{red} in Figure~\ref{fig:author-net}. This separation highlights the diversity in research collaboration and the varied nature of academic and professional networks.


\begin{table}[!h]
    \centering
    \scriptsize
    \renewcommand{\arraystretch}{1.3}
    \begin{tabular}{l|p{3.4cm}|p{5.3cm}|p{2cm}}
        \hline
        \textbf{Cluster} & {Most Published Author (\#Documents)} & {Linking Author (Connected Clusters)} & {Not Connected Clusters} \\
        \hline
        \multirow{2}{*}{\textcolor{red}{Cluster 1} }  & \multirow{5}{*}{See, Linda (18)} & See, Linda (Clusters 2, 5) & \multirow{5}{*}{None} \\
        \multirow{3}{*}{14 authors} &  & Minghini, Marco (Clusters 3, 5) &  \\
        &  & Arsanjani, Jamal Jokar (Clusters 2, 4, 5) & \\
        & &  4 authors to Cluster 2 via Zipf, Alexander& \\
        & & 6 authors to Cluster 1 via Mooney, Peter & \\
        \hline
        \multirow{2}{*}{\textcolor{green(pigment)}{Cluster 2}}  & \multirow{4}{*}{Zipf, Alexander (56)} & Zipf, Alexander (Clusters 1, 3, 4, 5) & \multirow{4}{*}{None} \\
        \multirow{2}{*}{12 authors}& & Schultz, Michael (Cluster 4) &  \\
        & & Li, Hao (Clusters 3, 4) & \\
        & & 4 authors & \\
        \hline
        \textcolor{trueblue}{Cluster 3}  & \multirow{2}{*}{Fan, Hongchao (21)} & Juhasz, Levente (Clusters 1, 5) & \multirow{2}{*}{Cluster 4} \\
        9 authors& & 5 authors to Cluster 2 via Zipf, Alexander &  \\
        \hline
        \textcolor{citrine}{Cluster 4}  & \multirow{2}{*}{Zhu, Xiao Xiang (15)} & Taubenboeck, Hannes (Cluster 1) & \multirow{2}{*}{Cluster 3, 5}  \\
        8 authors& & Ghamisi, Pedram (Cluster 2) & \\
        \hline
        \textcolor{cyan}{Cluster 5} & \multirow{3}{*}{Mooney, Peter (20)} & Mooney, Peter (Clusters 1, 2, 3) & \multirow{3}{*}{Cluster 4} \\
        \multirow{2}{*}{6 authors}& & Ballatore, Andrea (Clusters 1, 2) &  \\
        & & 2 authors to Cluster 2 via Zipf, Alexander& \\
        \hline
    \end{tabular}
    \caption{{Most Published Author, Linking Authors, and Not Connected Clusters by Cluster}}
    \label{tab:leading_authors}
\end{table}

\subsubsection{Organizations' Co-authoships and Contributions by Countries} 

This section examines the collaborative networks among organizations and the contributions of different countries to OSM research. 

To construct the co-authorship network, organizations with at least 10 documents were selected. Of the 1,265 analyzed, 76 met this threshold, and after removing duplicate affiliations (Section~\ref{sec:data_clean}), 67 remained. 
Figure~\ref{fig:affi_net} visualizes the network, which comprises \textit{eight} clusters: three large clusters with over 15 organizations, two mid-sized clusters with 8 and 6 organizations, and three single-institute clusters:

\begin{figure}[!h]
  \centering
  \includegraphics[width=\linewidth]{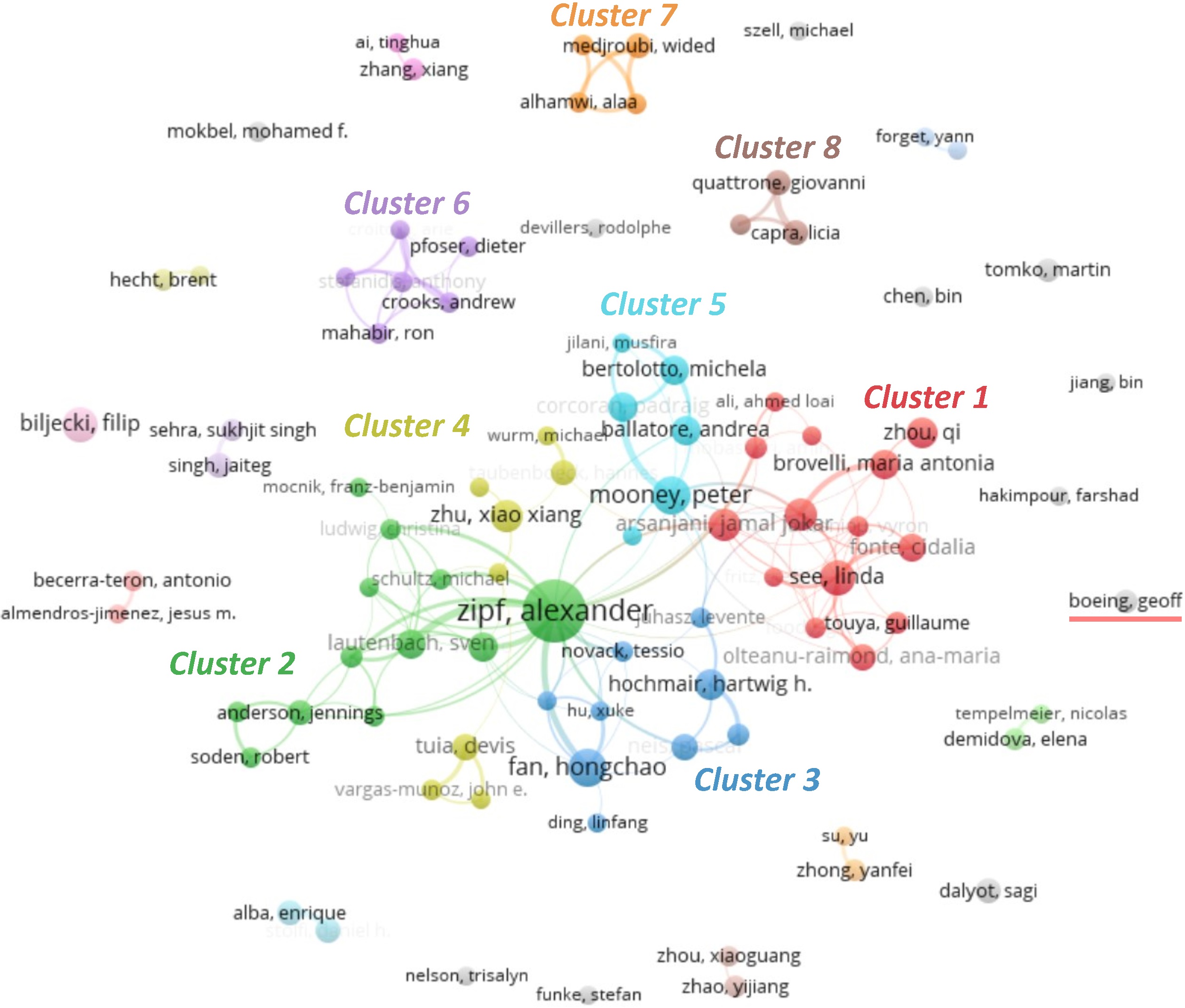}
  \caption{Co-authorship Network. 
  92 authors are selected with a minimum threshold of 5 documents per author. Clusters with at least 3 authors are labeled. 
Interactive version can be accessed here\fnref{fn:osmbib}.
}
  \label{fig:author-net}
\end{figure}

\begin{description}

\item [Cluster] 1, in \textcolor{red}{red}, {comprises 24 institutions, mainly from China, Germany, and France. \textit{WHU}, \textit{CAS}, and \textit{DLR} lead.}

\item [Cluster] 2, in \textcolor{green(pigment)}{green}, {includes 15 institutions from Europe and North America, led by \textit{Heidelberg University}, \textit{Maynooth University}, and \textit{UC Santa Barbara}, with a strong transatlantic presence.}

\item [Cluster] 3, in \textcolor{trueblue}{blue}, {covers 11 institutions across Europe, North America, Asia, and Australia. \textit{NUS}, \textit{ETH Zurich}, and \textit{NTNU} are key contributors.}

\item [Cluster] 4, in \textcolor{citrine}{citrine}, {includes eight institutions, mainly from the UK, USA, and Canada, led by \textit{UCL} and \textit{GMU}.}

\item [Cluster] 5, in \textcolor{richlavender}{purple}, {comprises six European institutions, led by \textit{PoliMi}, with strong representation from the Netherlands and Portugal.}

\item [Clusters] {6, 7, and 8 each feature a single institution: Israel's \textit{Technion}, Spain's \textit{Universidad de Málaga}, and the USA's \textit{University of Minnesota Twin Cities}.}

\end{description}

\begin{figure}[!t]
  \centering
  \includegraphics[width=.83\linewidth]{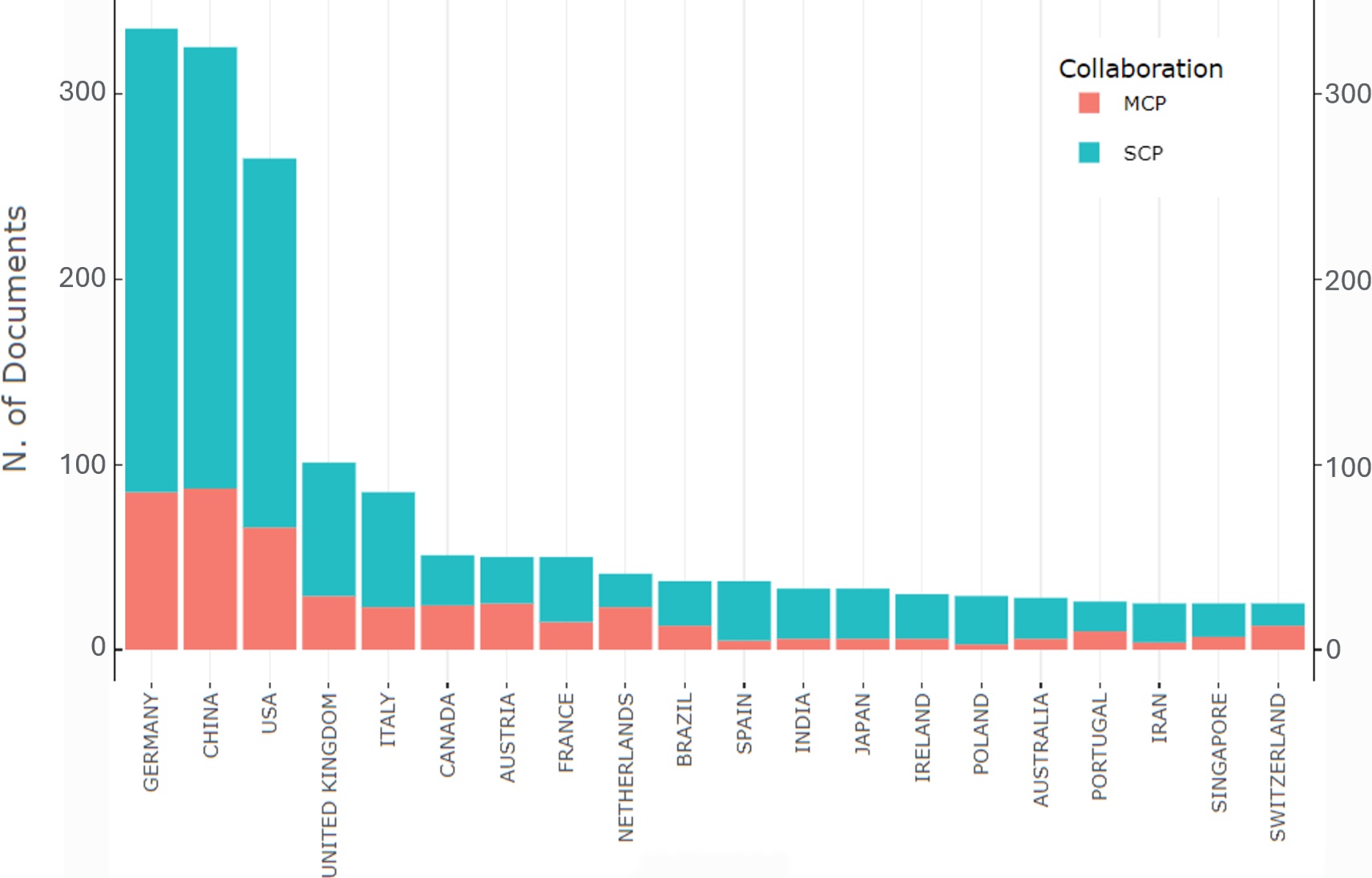}
  \caption{Number of Publications by Countries, including SCP and MCP.}
  \label{fig:CountryCorrespondingAuthor}
\end{figure}

{For the contributions by countries, we classify the publications as Single Country Publications (SCPs) or Multiple Country Publications (MCPs). SCPs involve authors from the same country, while MCPs indicate international collaboration. Figure}~\ref{fig:CountryCorrespondingAuthor} {presents the top corresponding author countries.}

{Germany leads in both national and international collaborations, followed by China and the USA, all of which show a balanced mix of SCPs and MCPs (around 25\%). 
The UK (28.7\%) and Italy (27.1\%) have a strong international presence, with a notable share of MCPs. Countries like Canada, Austria, and the Netherlands have nearly half or more of their publications as MCPs, demonstrating strong international collaboration. France and Brazil show a blend of national and international collaborations (MCP \> 30\%).
In contrast, Spain, India, Japan, Poland, and Iran focus more on domestic research, with MCPs less than 20\%. Meanwhile, Portugal (38.5\%) and Switzerland (52\%) stand out with a high MCP ratio.}

{The co-authorship network underscores the global nature of OSM research, with varying levels of collaboration and specialization across institutions and countries. Clusters containing institutions from multiple countries, alongside those featuring different institutions from the same country, such as Germany, highlight both international partnerships and diverse national strategies. 
These patterns reveal distinct institutional and national approaches—some prioritize domestic research, while others actively pursue international collaboration. Beyond institutional ties, individual researchers play a crucial role in shaping these networks, influencing the research strategies of their affiliated institutions and countries.}

\section{Research Topics}\label{sec:topics}

In this section, we analyze the range of research topics covered in OSM studies. 
We begin with key papers and a historiographic map to trace the field’s development, then explore the sub-field structure of OSM research and the research trends.

\subsection{Historiographic Map, Key Papers, and Key Research Interests}\label{sec:hist_map}

We identify the key papers in OSM research by building the historiographic map, which represents a historical development network of the most significant direct citations from a body of bibliographic records, illustrating the intellectual connections in chronological order~\citep{garfield2004historiographic}. 
The cited works of thousands of authors within a collection of articles are sufficient to map the historiographic structure of the field, identifying its key foundational works. 

\begin{figure}[H]
    \centering
    \includegraphics[width=1\linewidth]{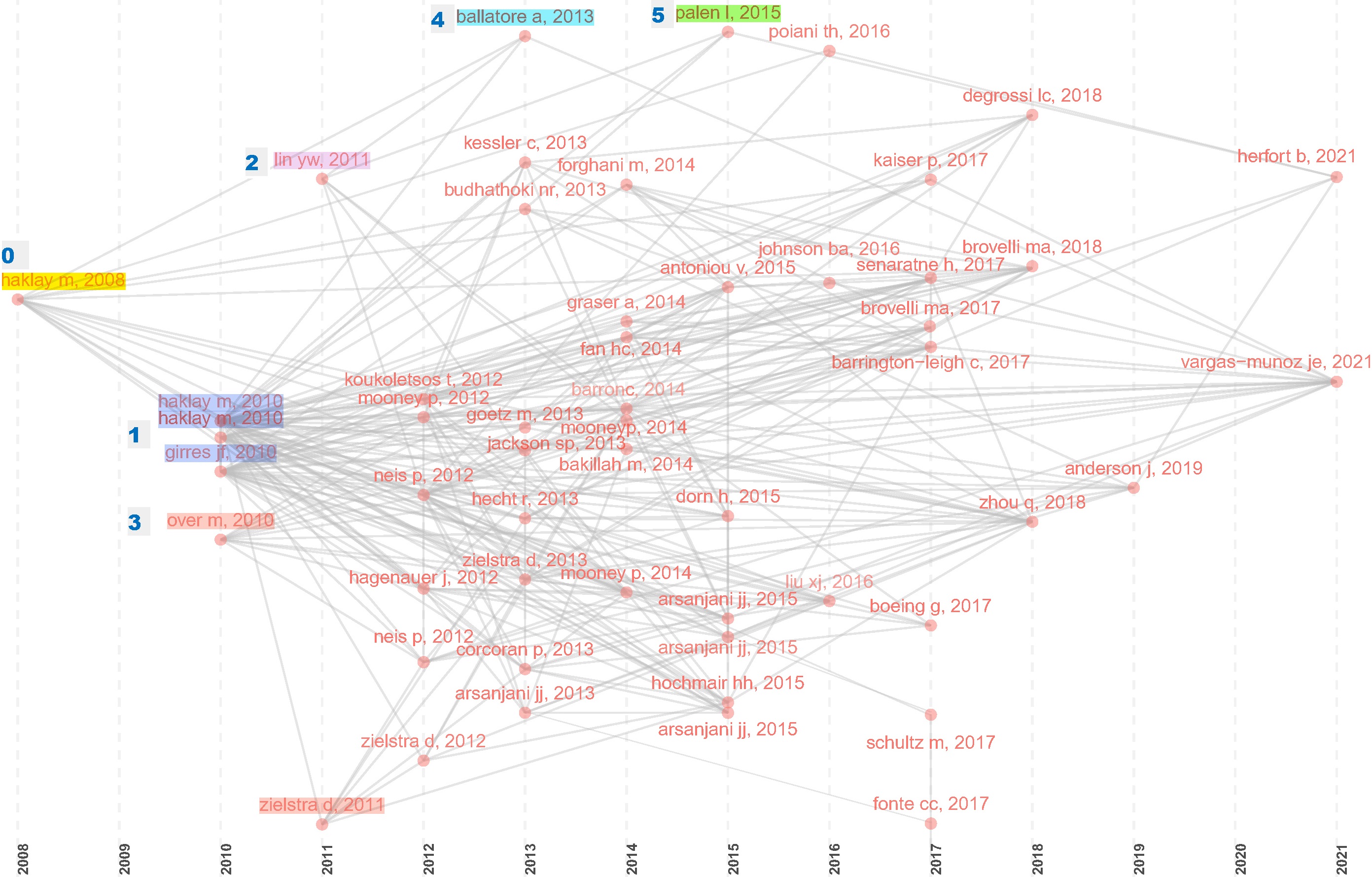}
    \caption{Bibliometric historiography graph of the most-cited work in our WoS collection. 
    The first paper is indexed as 0. The starting paper(s) of the five key research interests are highlighted and indexed. 
    }
    \label{fig:historical-year}
\end{figure}

In our study, we built the historiographic map of OSM research containing \textbf{\textit{52}} highly cited articles by setting Local Citation Score (LSC)$>=20$, representing the citation count of each paper by other papers in our WoS collection for OSM. 
The resulting historiographic map is as shown in Figure~\ref{fig:historical-year}, providing a comprehensive view of how research topics have evolved over time. 
We categorize the overall progression into four periods: 2008-2011, 2012-2014, 2015-2017, and 2018-2021. {As publications typically require time to accumulate citations, papers published after 2022, such as Herfort et al.'s 2023 study on the completeness and inequalities of global urban building data in OSM}~\citep{Herfort2023}{, have not yet appeared in these maps due to their recent publication date.}

From Figure~\ref{fig:historical-year}, 
we can see that it all started with an introduction, marked with ``0" in the figure - Haklay M, 2008 - \citep{haklay2008openstreetmap}, with LSC of 295. 
By examining the 52 highly cited papers, we identified {\textit{\textbf{Five Key Research Interests}}}. 
We find out that, three major research interests have already been revealed by 
seven highly cited papers published in the initial time period (2008-2011). 
Two other key research interests appeared in the second and third time periods, in year 2013 and 2015, respectively. 
In the later time period,  key papers can be found in all these areas of interest. 

The \textbf{\textit{five key research interests}} are listed below, with relevant highly cited papers shown in the corresponding table, and the starting paper(s) of each
key research interest are highlighted with numbers in Figure~\ref{fig:historical-year}:

\begin{enumerate}
    \item  {Quality Assessment and Validation of OSM Data} (c.f., Table~\ref{table:quality_assessment});
    \item  {Collaborative Contributing, Contributor Behavior, and Activity Analysis} (c.f., Table~\ref{table:collaborative_contributing});
    \item  {Mapping/Generating/Extracting Information using OSM and Other Data} (c.f., Table~\ref{table:mapping_generating_extraction});
    \item  {Tool Development} (c.f., Table~\ref{table:tool_development});
    \item  {OSM in Humanitarian and Disaster Response} (c.f., Table~\ref{table:osm_humanitarian_disaster_response}).
\end{enumerate}

\setlength{\tabcolsep}{3.7pt}

\begin{table}[H]
\centering
\scriptsize
\begin{tabular}{l|p{11cm}|c}
\hline
\textbf{Time Period} & \textbf{Papers:} paper - title & LSC \\ \hline
\multirow{5}{*}{{2008-2011}} & \citep{haklay2010good} - ``How Good is Volunteered Geographical Information? A Comparative Study of OpenStreetMap and Ordnance Survey Datasets" & 465 \\
    & \citep{girres2010quality} - ``Quality Assessment of the French OpenStreetMap Dataset"  & 270 \\
    & \citep{haklay2010many} - ``How Many Volunteers Does It Take to Map an Area Well? The Validity of Linus' Law to Volunteered Geographic Information" & 120 \\ \hline

\multirow{12}{*}{2012-2014} & \citep{koukoletsos2012assessing} - ``Assessing Data Completeness of VGI Through an Automated Matching Procedure for Linear Data" & 60 \\
    & \citep{hecht2013measuring} - ``Measuring Completeness of Building Footprints in OpenStreetMap Over Space and Time" & 86 \\
    & \citep{zielstra2013assessing} - ``Assessing the Effect of Data Imports on the Completeness of OpenStreetMap: A United States Case Study" & 67 \\
    & \citep{kessler2013trust} - ``Trust as a Proxy Measure for the Quality of Volunteered Geographic Information in the Case of OpenStreetMap" & 50 \\
    & \citep{jackson2013assessing} - ``Assessing Completeness and Spatial Error of Features in Volunteered Geographic Information" & 43 \\
    & \citep{fan2014quality} - ``Quality Assessment for Building Footprints Data on OpenStreetMap" & 148 \\
    & \citep{barron2014comprehensive} - ``A Comprehensive Framework for Intrinsic OpenStreetMap Quality Analysis" & 144 \\
    & \citep{forghani2014quality} - ``A Quality Study of the OpenStreetMap Dataset for Tehran" & 45 \\ \hline

\multirow{11}{*}{{2015-2017}} & \citep{brovelli2017towards} - ``Towards an Automated Comparison of OpenStreetMap With Authoritative Road Datasets" & 25 \\
    & \citep{hochmair2015assessing} - ``Assessing the Completeness of Bicycle Trail and Lane Features in OpenStreetMap for the United States" & 22 \\
    & \citep{barrington2017world} - ``The World's User-Generated Road Map Is More Than 80\% Complete" & 83 \\
    & \citep{antoniou2015measures} - ``Measures and Indicators of VGI Quality: An Overview" & 53 \\
    & \citep{dorn2015quality} - ``Quality Evaluation of VGI Using Authoritative Data - A Comparison With Land Use Data in Southern Germany" & 50 \\
    & \citep{arsanjani2015assessment} - ``An Assessment of a Collaborative Mapping Approach for Exploring Land Use Patterns for Several European Metropolises" & 23 \\
    & \citep{senaratne2017review} - ``A Review of Volunteered Geographic Information Quality Assessment Methods" & 100 \\ \hline

\multirow{6}{*}{{2018-2021}} & \citep{brovelli2018new} - ``A New Method for the Assessment of Spatial Accuracy and Completeness of OpenStreetMap Building Footprints" & 32 \\
    & \citep{degrossi2018taxonomy} - ``A Taxonomy of Quality Assessment Methods for Volunteered and Crowdsourced Geographic Information" & 22 \\
    & \citep{zhou2018exploring} - ``Exploring the Relationship Between Density and Completeness of Urban Building Data in OpenStreetMap for Quality Estimation" & 20 \\ \hline

\end{tabular}
\caption{\textbf{Key Research Interest 1}: Quality Assessment and Validation of OSM Data}
\label{table:quality_assessment}
\end{table}

\begin{table}[!h]
\centering
\scriptsize
\begin{tabular}{l|p{11cm}|c}
\hline
\textbf{Time Period} & \textbf{Papers:}  paper - title & LSC \\ \hline

\multirow{1}{*}{{2008-2011}} & \citep{lin2011qualitative} - ``A Qualitative Enquiry into OpenStreetMap Making" & 24 \\ \hline

\multirow{9}{*}{{2012-2014}} & \citep{neis2012analyzing} - ``Analyzing the Contributor Activity of a Volunteered Geographic Information Project - The Case of OpenStreetMap" & 127 \\
    & \citep{neis2012towards} - ``Towards Automatic Vandalism Detection in OpenStreetMap" & 31 \\
    & \citep{mooney2012annotation} - ``The Annotation Process in OpenStreetMap" & 58 \\
    & \citep{corcoran2013analysing} - ``Analysing the Growth of OpenStreetMap Networks" & 42 \\
    & \citep{mooney2014analysis} - ``Analysis of Interaction and Co-Editing Patterns Amongst OpenStreetMap Contributors" & 31 \\
    & \citep{budhathoki2013motivation} - ``Motivation for Open Collaboration: Crowd and Community Models and the Case of OpenStreetMap" & 47 \\
    & \citep{mooney2014has} - ``Has OpenStreetMap a Role in Digital Earth Applications?" & 20 \\ \hline

\multirow{3}{*}{{2015-2017}} & \citep{arsanjani2015exploration} - ``An Exploration of Future Patterns of the Contributions to OpenStreetMap and Development of a Contribution Index" & 23 \\
    & \citep{jokar2015emergence} - ``The Emergence and Evolution of OpenStreetMap: A Cellular Automata Approach" & 20 \\ \hline

\multirow{1}{*}{{2018-2021}} & \citep{anderson2019corporate} - ``Corporate Editors in the Evolving Landscape of OpenStreetMap" & 33 \\ \hline

\end{tabular}
\caption{\textbf{Key Research Interest 2}: Collaborative Contributing, Contributor Behavior, and Activity Analysis}
\label{table:collaborative_contributing}
\end{table}

\begin{table}[!h]
\centering
\scriptsize
\begin{tabular}{l|p{11cm}|c}
\hline
\textbf{Time Period} & \textbf{Papers:} short id - title & LSC \\ \hline

\multirow{4}{*}{{2008-2011}} & \citep{over2010generating} - ``Generating Web-Based 3D City Models from OpenStreetMap: The Current Situation in Germany" & 62 \\
    & \citep{zielstra2011comparative} - ``Comparative Study of Pedestrian Accessibility to Transit Stations Using Free and Proprietary Network Data" & 27 \\ \hline

\multirow{7}{*}{{2012-2014}} & \citep{hagenauer2012mining} - ``Mining Urban Land-Use Patterns from Volunteered Geographic Information by Means of Genetic Algorithms and Artificial Neural Networks" & 48 \\
    & \citep{zielstra2012using} - ``Using Free and Proprietary Data to Compare Shortest-Path Lengths for Effective Pedestrian Routing in Street Networks" & 20 \\
    & \citep{jokar2013toward} - ``Toward Mapping Land-Use Patterns from Volunteered Geographic Information" & 46 \\
    & \citep{goetz2013towards} - ``Towards Generating Highly Detailed 3D CityGML Models from OpenStreetMap" & 26 \\
    & \citep{bakillah2014fine} - ``Fine-Resolution Population Mapping Using OpenStreetMap Points-of-Interest" & 39 \\ \hline

\multirow{8}{*}{{2015-2017}} & \citep{Fonte2017} - ``Generating Up-To-Date and Detailed Land Use and Land Cover Maps Using OpenStreetMap and GlobeLand30" & 26 \\
    & \citep{johnson2016integrating} - ``Integrating OpenStreetMap Crowdsourced Data and Landsat Time Series Imagery for Rapid Land Use/Land Cover (LULC) Mapping: Case Study of the Laguna De Bay Area of the Philippines" & 34 \\
    & \citep{liu2016automated} - ``Automated Identification and Characterization of Parcels With OpenStreetMap and Points of Interest" & 56 \\
    & \citep{schultz2017open} - ``Open Land Cover from OpenStreetMap and Remote Sensing" & 53 \\
    & \citep{kaiser2017learning} - ``Learning Aerial Image Segmentation from Online Maps" & 25 \\ \hline

\multirow{2}{*}{{2018-2021}} & \citep{vargas2020openstreetmap} - ``OpenStreetMap: Challenges and Opportunities in Machine Learning and Remote Sensing" & 23 \\ \hline

\end{tabular}
\caption{\textbf{Key Research Interest 3}: Mapping, Generating, and Extracting Information Using OSM Data and Other Data}
\label{table:mapping_generating_extraction}
\end{table}

\begin{table}[!h]
\centering
\scriptsize
\begin{tabular}{l|p{11cm}|c}
\hline
\textbf{Time Period} & \textbf{Papers:} short id - title & LSC \\ \hline

\multirow{3}{*}{{2012-2014}} & \citep{ballatore2013geographic} - ``Geographic Knowledge Extraction and Semantic Similarity in OpenStreetMap" & 36 \\
    & \citep{graser2014towards} - ``Towards an Open Source Analysis Toolbox for Street Network Comparison: Indicators, Tools and Results of a Comparison of OSM and the Official Austrian Reference Graph" & 20 \\ \hline

\multirow{2}{*}{{2015-2017}} & \citep{boeing2017osmnx} - ``OSMnx: New Methods for Acquiring, Constructing, Analyzing, and Visualizing Complex Street Networks" & 86 \\ \hline

\end{tabular}
\caption{\textbf{Key Research Interest 4}: Tool Development}
\label{table:tool_development}
\end{table}

\begin{table}[!h]
\centering
\scriptsize
\begin{tabular}{l|p{11cm}|c}
\hline
\textbf{Time Period} & \textbf{Papers:} short id - title & LSC \\ \hline

\multirow{4}{*}{{2015-2017}} 
& \citep{palen2015success} - ``Success \& Scale in a Data-Producing Organization: The Socio-Technical Evolution of OpenStreetMap in Response to Humanitarian Events" & 25 \\
& \citep{poiani2016potential} - ``Potential of Collaborative Mapping for Disaster Relief: A Case Study of OpenStreetMap in the Nepal Earthquake, 2015" & 21 \\ \hline

\multirow{1}{*}{{2018-2021}} 
& \citep{herfort2021evolution} - ``The Evolution of Humanitarian Mapping Within the OpenStreetMap Community" & 26 \\ \hline

\end{tabular}
\caption{\textbf{Key Research Interest 5}: OSM in Humanitarian and Disaster Response}
\label{table:osm_humanitarian_disaster_response}
\end{table}

{The first two key interests of highly cited papers are relevant to the crowdsourced and multi-source nature of OSM. }

The first focuses on data quality, while the second studies the contributors. 
21 highly cited papers study and discuss Quality Assessment and Validation of OSM Data (c.f., Table~\ref{table:quality_assessment}), with methods varying from comparative Analysis with Authoritative Datasets, Automated Matching and Feature Comparison, Temporal Analysis, Taxonomy and Theoretical Frameworks, to Qualitative Methods, and Review and Synthesis of Quality Indicators. 

{The second key research interest focuses on Collaborative Contributing and Contributor Behavior, reflected in 11 highly cited papers. 
The analysis evolved from individual contributors at the beginning to Corporate Editors in 2021}~\citep{anderson2019corporate}{, revealing a shift of contributors. 
Notably, 
studies in the social domain also align with this research interest, exploring questions related to contributors, such as who has the authority to map and whether participation is an empowering act. These studies delve into conflicting understandings of reality, mapping disputes, and challenges to official and commercial cartography, thereby intersecting with geopolitical issues, as well as social, political, legal, and governance aspects}~\citep{Bittner2017, Jackson2018, Lin2019, Scassa2013}.

The third key interest focuses on how to utilize OSM data, often combined with other data sources.
The application areas include generating 3D city models, assessing pedestrian accessibility, mapping land-use patterns, creating land use and land cover maps, and performing aerial image segmentation. In 2021, a review paper by Vargas-Munoz JE summarized the challenges and opportunities of using OSM in machine learning and remote sensing~\citep{vargas2020openstreetmap}. 

The fourth key interest is Tool Development. 
This includes:
A semantic tool for geographic knowledge extraction in OSM~\citep{ballatore2013geographic}; 
An open-source toolkit for comparing street networks~\citep{graser2014towards}; 
and the renowned OSMnx, a Python package that simplifies acquiring, modeling, analyzing, and visualizing geospatial features from OSM~\citep{boeing2017osmnx}.

The final key research interest is Humanitarian Mapping and Disaster Relief. 
It covers the evolution and impact of OSM in humanitarian contexts, including its role in the 2015 Nepal earthquake and its broader socio-technical development in response to emergencies~\citep{palen2015success,poiani2016potential}. The latest research highlights the ongoing evolution of humanitarian mapping practices within the OSM community~\citep{herfort2021evolution}.

\subsection{Sub-field Structure of OSM Research and Research Trends}\label{sec:wos-trend}

\definecolor{green(pigment)}{rgb}{0.0, 0.65, 0.31}
\definecolor{forestgreen(web)}{rgb}{0.13, 0.55, 0.13}
\definecolor{trueblue}{rgb}{0.0, 0.45, 0.81}
\definecolor{steelblue}{rgb}{0.27, 0.51, 0.71}
\definecolor{brass}{rgb}{0.71, 0.65, 0.26}
\definecolor{citrine}{rgb}{0.89, 0.82, 0.04}
\definecolor{richlavender}{rgb}{0.67, 0.38, 0.8}

Key papers are highly cited, however do not represent all areas of research, nor do they capture the quantity and scope of studies within different topics. Therefore, we analyze the topics across the entire database.

We conduct a co-occurrence analysis of keywords to extract the sub-field structure of OSM research. 
A minimum of \textit{five} occurrences was set as the threshold. Out of 6,264 keywords (both author keywords and keywords plus), 416 met this criterion, and after excluding the two most frequent keywords - ``openstreetmap" with 816 occurrences and ``volunteered geographic information" with 225 occurrences - to avoid overshadowing other keywords, a total of 414 keywords were analyzed. 

\begin{figure}[!h]
    \centering
    \subfigure[Network clusters of the keywords.]{
        \includegraphics[width=.83\linewidth]{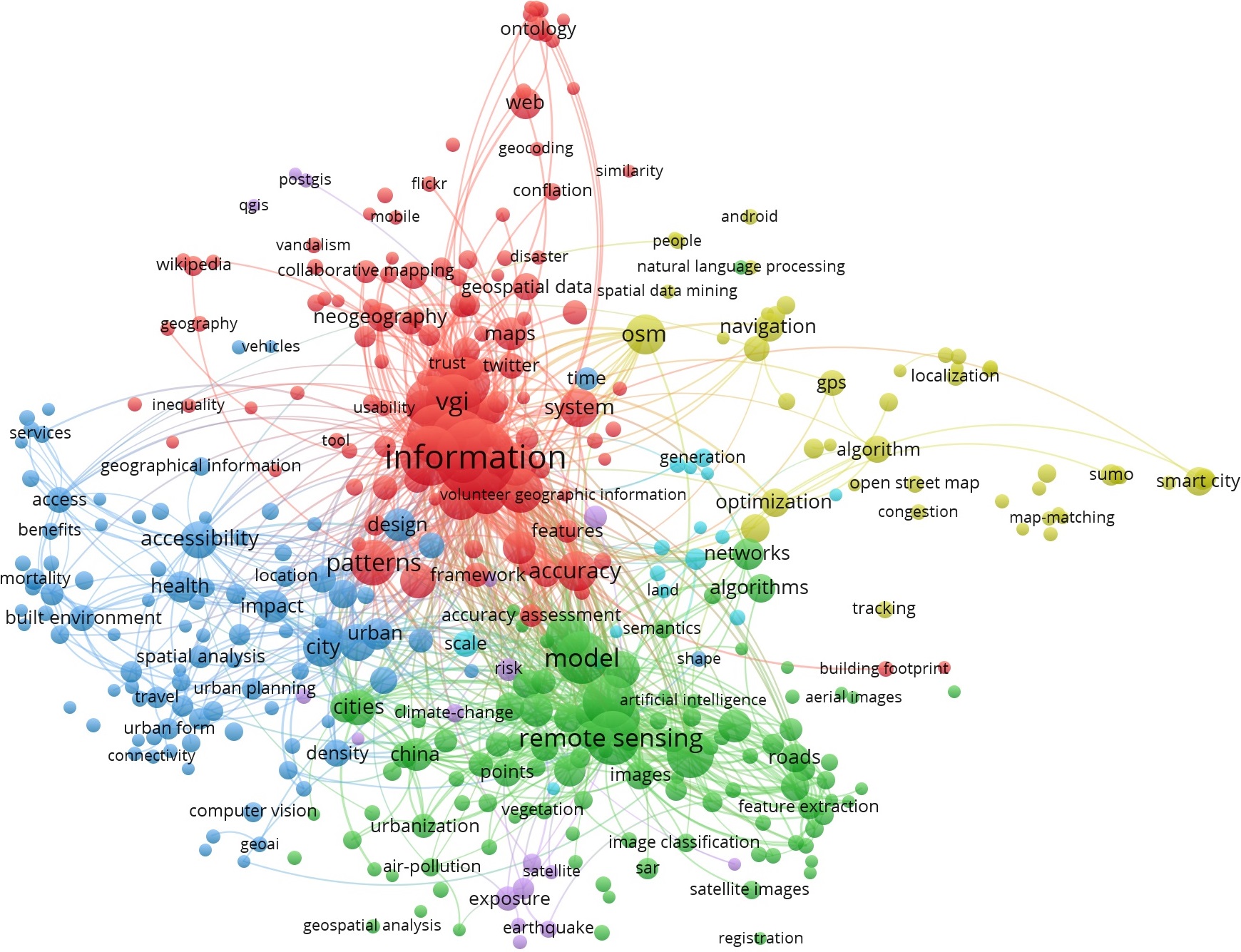}
    }\\
    \subfigure[Overlay visualization of the keywords.]{
        \includegraphics[width=.83\linewidth]{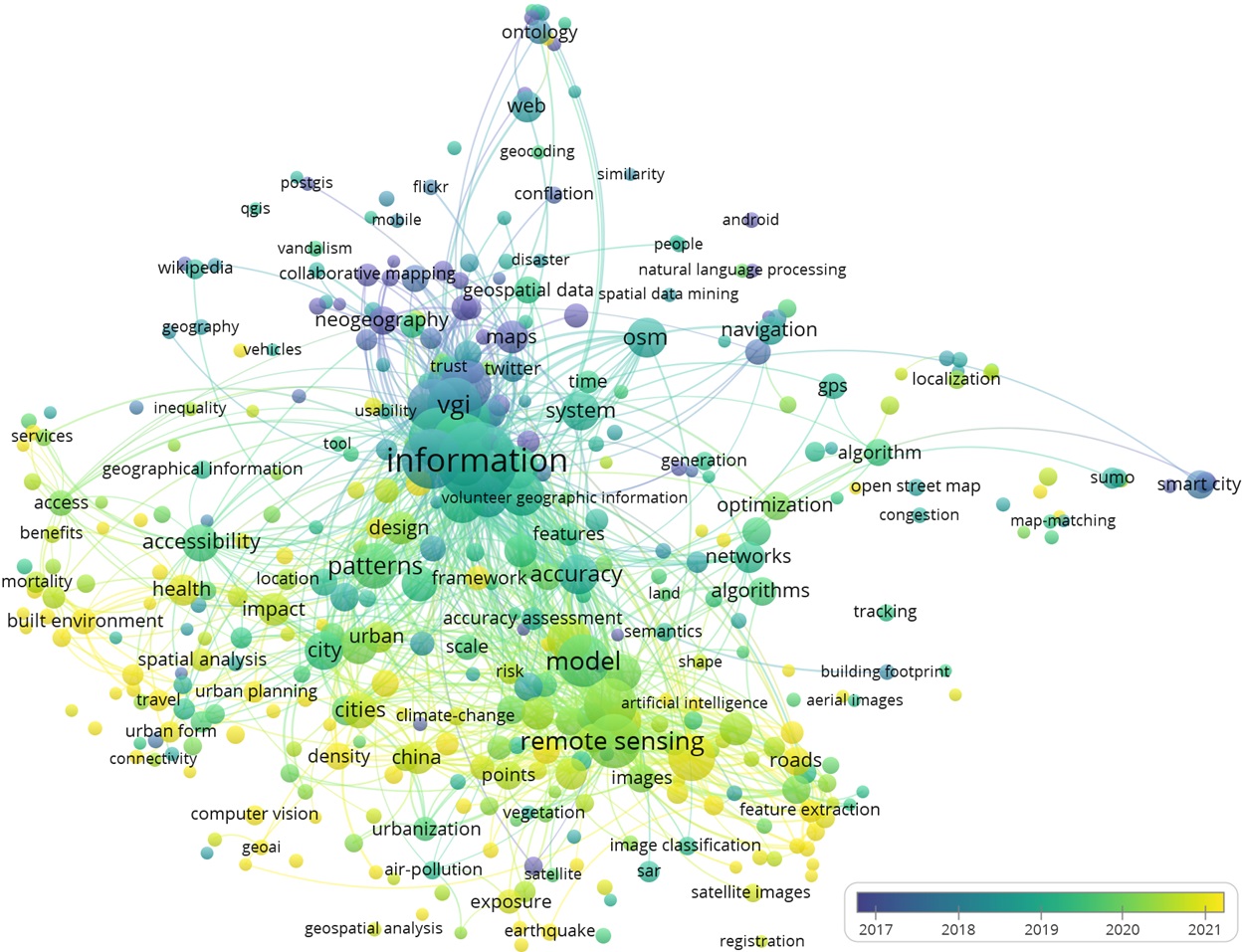}
    }
    \caption{Network clusters and overlay visualization of keywords. 
    (a) Network clusters of the keywords (Interactive version available here\fnref{fn:osmbib}); 
    (b) Overlay visualization. 
    }
    \label{fig:topic-cluster}
\end{figure}

As visualized in Figure~\ref{fig:topic-cluster} (a), these keywords formed six distinct clusters, indicating a diverse and well-structured landscape of research sub-fields, which are summarized below. Table~\ref{tab:topic_clusters} lists the key themes and the keywords with these themes for each topic cluster.

\begin{description}

\item[{Topic cluster}]\textbf{1: Geo Information}  (c.f., \textcolor{red}{red cluster} in Figure~\ref{fig:topic-cluster} (a).)  
{The cluster focuses on information quality and management in geographic data systems, emphasizing open data, crowdsourced contributions, and strategies for improving data accuracy and integration.}

\item[\textbf{Topic cluster}]\textbf{2: Remote Sensing} (c.f., \textcolor{green(pigment)}{\textbf{green cluster}} in Figure~\ref{fig:topic-cluster} (a).)  
{The cluster focuses on remote sensing and AI-driven data analysis for urban and environmental monitoring, emphasizing practical applications and real-world impact.}

\item[\textbf{Topic cluster}]\textbf{3: Urban Planning} (c.f., \textcolor{trueblue}{\textbf{blue cluster}} in Figure~\ref{fig:topic-cluster} (a).)  
{The cluster focuses on urban planning and its impact on health, accessibility, and infrastructure, examining how design and data-driven analysis enhance urban development.}

\item[\textbf{Topic cluster}]\textbf{4: Navigation and Transportation} (c.f., \textcolor{citrine}{\textbf{citrine cluster}} in Figure~\ref{fig:topic-cluster} (a).)  
{The cluster explores advanced navigation and transportation technologies, focusing on “optimization” and “simulation” for traffic modeling and efficiency.}

\item[\textbf{Topic cluster}]\textbf{5: Risk and Vulnerability} (c.f., \textcolor{richlavender}{\textbf{purple cluster}} in Figure~\ref{fig:topic-cluster} (a).)  
{The cluster examines risk and vulnerability assessment in environmental and climate-related hazards. }

\item[\textbf{Topic cluster}]\textbf{6: Geospatial Modeling} (c.f., \textcolor{cyan}{\textbf{cyan cluster}} in Figure~\ref{fig:topic-cluster} (a).)  
{The cluster examines geospatial modeling and analysis, focusing on data acquisition, modeling techniques, and spatial analysis.}

\end{description}

\begin{table}[!]
    \centering
    \scriptsize
    \begin{tabularx}{\textwidth}{l|X}
        \hline
        \textbf{Topic Cluster} & \textbf{Key Themes and keywords} \\
        \hline
        \multirow{3}{*}{\textbf{\textcolor{red}{Geo}} } 
        & \textbf{Geographic data quality and management:} “Information”, “Quality Assessment”, “Data Quality”, “Accuracy” \\
        {\textbf{\textcolor{red}{Information}} }& \textbf{Participatory mapping:} “GIS”, “Volunteered Geographic Information (VGI)”, “Completeness”, “Crowdsourcing” \\
         & \textbf{Spatial modeling:} “Patterns”, “Models”, “OpenStreetMap (OSM)”, “Validation” \\
        \hline
        \multirow{3}{*}{\textbf{\textcolor{green(pigment)}{Remote}}}  
        & \textbf{AI-driven remote sensing:} “Classification”, “Deep Learning”, “Machine Learning” \\
        & \textbf{Advanced data processing:} “Data Fusion”, “Feature Extraction”, “Model” \\
        {\textbf{\textcolor{green(pigment)}{Sensing}}}  & \textbf{Satellite-based applications:} “Google Earth Engine”, “Sentinel-2”, “Urban Areas”, “Land Cover”, “Roads” \\
        \hline
        \multirow{3}{*}{\textbf{\textcolor{trueblue}{Urban Planning}}}  
        & \textbf{Urban design and infrastructure:} “City”, “Land Use”, “Impact”, “Accessibility” \\
        & \textbf{Spatial organization and mobility:} “Infrastructure”, “Road Networks”, “Built Environment”, “Density”, “Urban Form” \\
        & \textbf{Analytical tools:} “Spatial Analysis”, “Geographic Information Systems (GIS)” \\
        \hline
        \multirow{3}{*}{\textbf{\textcolor{citrine}{Navigation \& }}}  
        & \textbf{Mobility optimization: }“Optimization”, “Simulation”, “Routing”, “GPS”, “Smart Mobility” \\
        {\textbf{\textcolor{citrine}{Transportation}}}&\textbf{Intelligent transportation:} “Smart City”, “Algorithms”, “Autonomous Vehicles” \\
        & \textbf{Emerging navigation tools: }“Traffic Simulation”, “Augmented Reality”, “3D Reconstruction” \\
        \hline
        \multirow{3}{*}{\textbf{\textcolor{richlavender}{Risk \&}}}  
        & \textbf{Environmental hazard assessment:} “Exposure”, “Risk”, “Uncertainty” \\
        & \textbf{Disaster risk analysis:} “Climate Change”, “Earthquakes” \\
        {\textbf{\textcolor{richlavender}{Vulnerability}}}& \textbf{Geospatial resilience strategies:} “Crowdsourcing”, “Satellite”, “PostGIS”, “QGIS”, “Sustainable Development”, “Resilience Planning” \\
        \hline
        \multirow{3}{*}{\textbf{\textcolor{cyan}{Geospatial}}}  
        & \textbf{Data acquisition:} “LiDAR”, “Photogrammetry” \\
        &\textbf{3D spatial modeling:} “Reconstruction”, “3D City Models” \\
        {\textbf{\textcolor{cyan}{Modeling}}} & \textbf{Land use monitoring:} “Change Detection”, “Cover Change”, “Urban Analysis” \\
        \hline
    \end{tabularx}
    \caption{{Topic clusters and key themes in geospatial research. Topic clusters are color-coded with the colors in Figure}~\ref{fig:topic-cluster} (a).}
    \label{tab:topic_clusters}
\end{table}

{To analyze trends in OSM research, we use VOSviewer’s overlay visualization, which color-codes items by average publication year (Figure}~\ref{fig:topic-cluster}(b)). {This highlights the evolution of topics within each cluster, distinguishing older from emerging research areas.
As can be seen in Figure}~\ref{fig:topic-cluster}{(b), older topics are concentrated in the top-right, corresponding to Topic clusters 1 “Geo Information” and 4 “Navigation and Transportation”. Newer topics are more prominent in the bottom-left, where Topic clusters 2 “Remote Sensing”, 3 “Urban Planning”, and 5 “Risk and Vulnerability” indicate growing research interest. Cluster 6 “Geospatial Modeling” lies in between but trends toward recent developments.}

Table~\ref{tab:topic_evolution_clusters} illustrates the evolution of research topics across different clusters.
For most studies, OSM serves as a data source, leading to overlaps with various research domains. Nevertheless, the identified keywords reflect the progression of these OSM-related fields. 
For instance, the topic in the sub-field Geo-Information shifted from digital earth in 2015 to knowledge graph in recent years. In Remote Sensing, research was initially centered around data and topics in 2017–2018 but later incorporated advanced tools such as Google Earth Engine and Convolutional Neural Networks (CNN) by 2021–2022. In Urban Planning, early studies explored network-based approaches such as Complex Networks in 2017, which later transitioned to AI-driven methodologies such as GeoAI in 2022. Similarly, in Navigation and Transportation, the focus moved from optimization techniques such as Evolutionary Algorithms in 2016 to intelligent mobility solutions such as Autonomous Vehicles in 2021. In the Risk and Vulnerability domain, research initially leveraged spatial technologies such as Crowd-Sourcing and PostGIS in 2016, while more recent studies emphasize sustainability-driven approaches such as Sustainable Development in 2022. This evolution also highlights the widespread application of machine learning across various domains, demonstrating its increasing role in different aspects of OSM-related research.

\begin{table}[!h]
    \centering
    \scriptsize
    \begin{tabular}{l|lc}
        \hline
        \textbf{Topic Cluster} & \textbf{Topic} & \textbf{Average Publication Year} \\
        \hline
        \multirow{6}{*}{\textbf{\textcolor{red}{Geo Information}}}
        & Community & 2015 \\
        & Digital Earth & 2015 \\
        & Semantic Similarity & 2015 \\
        & GIS & 2018 \\
        & Framework & 2021 \\
        & Management & 2021 \\
        & Knowledge Graph & 2021 \\
        \hline
        \multirow{7}{*}{\textbf{\textcolor{green(pigment)}{Remote Sensing}}}
        & Biodiversity & 2017 \\
        & Lidar Data & 2018 \\
        & Modis & 2018 \\
        & Cloud Computing & 2018 \\
        & Landscape & 2018 \\
        & Urban Land Use & 2021 \\
        & Deep Learning & 2021 \\
        & Google Earth Engine & 2021 \\
        & Feature Extraction & 2021 \\
        & Semantic Segmentation & 2022 \\
        & Earth Observation & 2022 \\
        & Convolutional Neural Network (CNN) & 2022 \\
        \hline
        \multirow{6}{*}{\textbf{\textcolor{trueblue}{Urban Planning}}}
        & Urban Streets & 2017 \\
        & Complex Networks & 2017 \\
        & City & 2019 \\
        & Land-Use & 2020 \\
        & Health & 2020 \\
        & GeoAI & 2022 \\
        & Spatial Autocorrelation & 2022 \\
        & Spaces & 2022 \\
        \hline
        \multirow{6}{*}{\textbf{\textcolor{citrine}{Navigation \& Transportation}}}
        & Smart Mobility & 2016 \\
        & Evolutionary Algorithm & 2016 \\
        & Optimization & 2020 \\
        & Recognition & 2020 \\
        & 3D Reconstruction & 2021 \\
        & Trajectory & 2021 \\
        & Autonomous Vehicles & 2021 \\
        & Calibration & 2022 \\
        \hline
        \multirow{6}{*}{\textbf{\textcolor{richlavender}{Risk \& Vulnerability}}}
        & Crowd-Sourcing & 2016 \\
        & PostGIS & 2017 \\
        & Risk & 2020 \\
        & Exposure & 2021 \\
        & Climate Change & 2021 \\
        & Earthquake & 2021\\
        & Natural Hazards & 2021 \\
        & Sustainable Development & 2022 \\
        \hline
        \multirow{6}{*}{\textbf{\textcolor{cyan}{Geospatial Modeling}}}
        & Population Estimation & 2017 \\
        & 3D City Models & 2017 \\
        & LiDAR & 2018 \\
        & Change Detection & 2018 \\
        & Digital Elevation Model & 2021 \\
        & Flow & 2022 \\
        \hline
    \end{tabular}
    \caption{{Topic evolution across different clusters, ordered from older to newer research trends. Topic clusters are color-coded with the colors in Figure}~\ref{fig:topic-cluster} (a).}
    \label{tab:topic_evolution_clusters}
\end{table}

\fboxsep=10mm
\fboxrule=4pt

\section{Community Priorities: the SotM Conference as an Example}\label{sec:sotm}

OSM has a large and diverse community of users and contributors, most of whom are not involved in academic research.  
This community actively discusses and shares ideas through various platforms, such as community forums, mailing lists, meetups, and the SotM conferences. 
Many of the issues and innovations raised in these discussions often become topics of academic research. Community discussion plays a vital role in identifying practical challenges and shaping the broader research agenda. 
Therefore, in analyzing OSM research, it is crucial to consider the community's voices and perspectives. 

To study this issue, we analyzed the topics presented at SotM conferences as the examples of community presented research. 
SotM is the annual global conference organized by the OpenStreetMap Foundation (OSMF) and has served as a key gathering since 2007 (with exceptions in 2015 and 2023), where contributors, developers, and stakeholders worldwide come together to share insights and innovations. 
We identify the top contributors and trending topics within SotM. 

\subsection{Data Acquisition}

To analyze the presentations at SotM conferences, we carefully collected and selected data based on specific criteria to ensure the relevance and quality of our analysis. We collected the data from the conference website\footnote{\url{https://stateofthemap.org/}}by implementing the following selection criteria:

\begin{enumerate}
    \item Conference Scope:
   We focused exclusively on the State of the Map conferences organized by the OSMF. 
   Notably, regional and local conferences, such as SotM-Asia, SotM-Africa, SotM-US, and others, also carry the ``State of the Map" name. These regional SotMs are independently organized by local teams, separate from the OSMF. 
   However, they were excluded for two reasons:
   1) These regional conferences are not organized by the OSMF, and therefore may not fully align with the standards and objectives of the international SotM.
   2) Many regional conferences lack comprehensive documentation, making it challenging to gather reliable data for analysis.

 \item Academic Track:
   Since 2018, the SotM conference has included an academic track, renamed OSM Science in 2023\footnote{For convenience, we refer to both the Academic Track and OSM Science simply as the Academic Track in the rest of the text.}, and published proceedings. However, upon investigation, we found that these proceedings were not indexed in the WoS Core Collection, limiting their accessibility and visibility in academic research. Additionally, the number of presentations in the academic track was relatively small. Therefore, we decided not to focus exclusively on this track.

 \item Exclusions:
   We excluded several types of presentations from our analysis to maintain a clear focus on research-oriented talks. 
   Panels, Workshops, and Tutorials were not included as they are not traditional research talks but rather interactive sessions or instructional content.
   Lightning Talks are very short, 5-minute presentations, often with incomplete records, making them less suitable for detailed analysis.
\end{enumerate}

Based on the above criteria, we compiled a list and cleaned the data (i.e., author name disambiguation), resulting in a dataset consisting of a total of \textit{782}
records, which included the \textit{titles} and \textit{authors} of the presentations, forming the basis for the subsequent analysis.

\begin{table}[H]
\centering
\scriptsize
\begin{tabular}{lcc}
\hline
\textbf{Authors}    & \textbf{\#SotM Talks} & {\textbf{\#Academic Track}} \\ \hline
RAMM, FREDERIK      & 13  & - \\
TOPF, JOCHEN        & 11  & - \\
VAN EXEL, MARTIJN   & 9   & - \\
\textbf{MOONEY, PETER}  & 8   & 6\\
\textbf{ZIPF, ALEXANDER}  & 8   & 8\\
COAST, STEVE        & 7   & - \\
QUEST, CHRISTIAN    & 7   & - \\
WOOD, HARRY         & 7   & - \\
ZVEREV, ILYA        & 7   & - \\
MARON, MIKEL        & 6   & - \\
OLBRICHT, ROLAND    & 6   & - \\
WHITELEGG, NICK     & 6   & - \\
\textbf{ANDERSON, JENNINGS }  & 6   & 6\\
CHAPMAN, KATE       & 6   & - \\
ABELSHAUSEN, BEN    & 5   & 1\\
HOFF, HENK          & 5   & - \\
HOFFMANN, SARAH     & 5   & - \\
KNERR, TOBIAS       & 5   & - \\
MILLER, PETER       & 5   & - \\
MIURA, HIROSHI      & 5   & - \\
PAVIE, ADRIEN       & 5   & - \\
WATERS, TIM         & 5   & - \\
WEAIT, RICHARD      & 5   & - \\ \hline
\end{tabular}
\caption{{Authors with at least 5 talks at SotM, the number of their Talks and the number of their Talks in the Academic Track. Authors with at least 5 articles in WoS collection are highlight in \textbf{bold}.}}
\label{tab:authors_articles}
\end{table}

\begin{table}[H]
\scriptsize
\centering
\begin{tabular}{lccc}
\hline
\textbf{Authors}  & \textbf{\#WoS Articles} & \textbf{\#SotM Talks} & {\textbf{\#Academic Track}} \\ \hline
ZIPF, ALEXANDER         & 56  & 8    & 8\\
MOONEY, PETER           & 20  & 8    & 6 \\
MINGHINI, MARCO         & 15  & 5    & 4\\
BROVELLI, MARIA ANTONIA & 12  & 3    & 3\\
CORCORAN, PADRAIG       & 12  & 1    & -\\
LAUTENBACH, SVEN        & 12  & {2}    & 2\\
ANDERSON, JENNINGS      & 9   & 6    & 6\\
DALYOT, SAGI            & 9   & 1    & 1\\
DITTUS, MARTIN          & 6   & 2    & - \\
LUDWIG, CHRISTINA       & 6   & {2}    & 2\\
NOVACK, TESSIO          & 6   & 1    & 1 \\
PALEN, LEYSIA           & 6   & 1    & 1\\
SODEN, ROBERT           & 6   & 1    & - \\
YEBOAH, GODWIN          & 5   & 4    & 4\\ \hline
\end{tabular}
\caption{{Top Authors in the WoS collection with SotM contributions, and the number of articles/talks, and the number of their talks in the Academic Track.}}
\label{tab:wos_sotm_comparison}
\end{table} 

\subsection{Key Authors}

Table~\ref{tab:authors_articles} lists authors who have at least 5 talks in SotM. 
To determin the authors' fields of contribution, we reviewed the authors' OpenStreetMap profiles, and gathered the information from the OpenStreetMap website. 
For further information, please refer to the webpages linked in the research data 
and their profiles available on the OSM website. 

These contributors represent the vast OSM community that have significantly shaped OSM through software development, community leadership, humanitarian efforts, and educational outreach. 
The contributions of these OSM community members span several areas. Some major activities and contributions include:

\begin{itemize}

    \item 
    \textit{{OSM Founder}}: 
    {Steve Coast started OSM and started SotM. He developed early OSM versions, served as Chairman of the OSM Foundation, co-founded CloudMade to support OSM. }
    
    \item \textit{Geofabrik}: Frederik Ramm and Jochen Topf co-founded Geofabrik, a consultancy specializing in OSM services and software development.

    \item \textit{JOSM}: Several contributors, including Frederik Ramm and Tim Waters, have developed or maintained tools and plugins related to the JOSM editor, a key software for OSM mapping.

    \item \textit{Overpass API}: Roland Olbricht is notable for developing and maintaining the Overpass API, a powerful tool for querying OSM data.

    \item \textit{OSMF Board Members}: Individuals like Frederik Ramm, Martijn van Exel, Mikel Maron, Kate Chapman, and Sarah Hoffmann have served on the board, guiding the strategic direction of OSM.

    \item \textit{OSM Working Groups}: Members like Tobias Knerr, Harry Wood, and Mikel Maron have contributed to various OSM working groups, focusing on governance, data management, and communications.

    \item \textit{HOT (Humanitarian OpenStreetMap Team)}: Contributors such as Kate Chapman, Harry Wood, Ben Abelshausen, and Mikel Maron have been deeply involved with HOT, focusing on disaster response and humanitarian mapping efforts.

    \item \textit{Blogging and Writing}: Authors like Steve Coast, Jochen Topf, and Ilya Zverev have written extensively on OSM, sharing insights, updates, and technical knowledge through blogs and books.

    \item \textit{Software Development}: Contributors like Martijn van Exel, Nick Whitelegg, Tim Waters, and Roland Olbricht have developed various OSM-related software, tools, and platforms that enhance the OSM ecosystem.

\end{itemize}

Notably, top SotM contributors like Peter Mooney, Alexander Zipf, and Jennings Anderson (c.f., highlight in \textbf{bold} in Table~\ref{tab:authors_articles}) have both published more than five academic articles in the WoS dataset, underscoring their dual engagement in both community and academic spheres. 
Multiple researchers in academia, as listed in Table~\ref{tab:wos_sotm_comparison}, have presented in SotM.
{In addition, 
Marco Minghini, Alexander Zipf, Martin Dittus, and Godwin Yeboah, have made multiple appearances in SotM. 
The overlap between SotM conference contributions and academic research, particularly within the WoS collection, suggests a strong connection between community-driven discussions and scholarly work in the OSM ecosystem. }

{Moreover, Table 14 reveals that academic participation is largely concentrated within the Academic Track. This concentration highlights a dual dynamic: while the track successfully attracts scholars who might not otherwise attend SotM, the high specificity of engagement suggests it often functions as a ``conference within a conference." This separation risks limiting the exposure of research to the broader contributor community. Nevertheless, the Academic Track continues to serve as a vital bridge for knowledge exchange, allowing scholars to establish contact with the community and researchers like Peter Mooney to actively foster cross-community dialogue}~\citep{mooney2018coordinating, grinberger2019bridging}.

\subsection{Trend Topics Comparing to the WoS Collection}\label{sec:somt-trend}

To analyze trend topics from SotM talks, we start by extracting topics from the titles, which are key for identifying themes. We then process the text using Unigrams (single words) to pinpoint individual keywords and Bigrams (pairs of consecutive words) to uncover common two-word phrases. During this process, we remove stop words like ``based", ``study", and ``project" that do not contribute to trend identification, and merge synonyms such as ``osm" with ``openstreetmap" and ``map" with ``maps" to maintain consistency in topic representation.
Then we employ bibliometrix to generate and visualize trend topics from the processed data. 

For a fair comparative analysis, we conduct the same analysis for titles of documents in the WoS collection to identify trends and discrepancies. Table~\ref{tab:word_freq} lists the parameters used in extracting topics in both Unigrams and Bigrams from SotM and WoS. The resulting trend topics are shown in Figure~\ref{fig:trend-topics}, including the term frequency, the time span, and the median year.

\begin{table}[!]
\centering
\scriptsize
\caption{Annual Word Frequency and Word Count in SotM and WoS Publications}
\begin{tabular}{lcccc}
\toprule
\textbf{} & \textbf{Type} & \textbf{Threshold of Frequency} & \textbf{Number of Words/Year} \\
\midrule
\textbf{SotM} & Unigrams & 5 & 3 \\
           & Bigrams  & 2 & 3 \\
\midrule
\textbf{WoS} & Unigrams & 5 & 3 \\
           & Bigrams  & 3 & 3 \\
\bottomrule
\end{tabular}
\label{tab:word_freq}
\end{table}

\definecolor{sotm}{rgb}{0.643, 0.439, 0.729}
\definecolor{wos}{rgb}{0.169, 0.588, 0.957}

\begin{figure}[!]
    \centering
    \hspace{.9cm}\subfigure{\includegraphics[height=.6\linewidth]{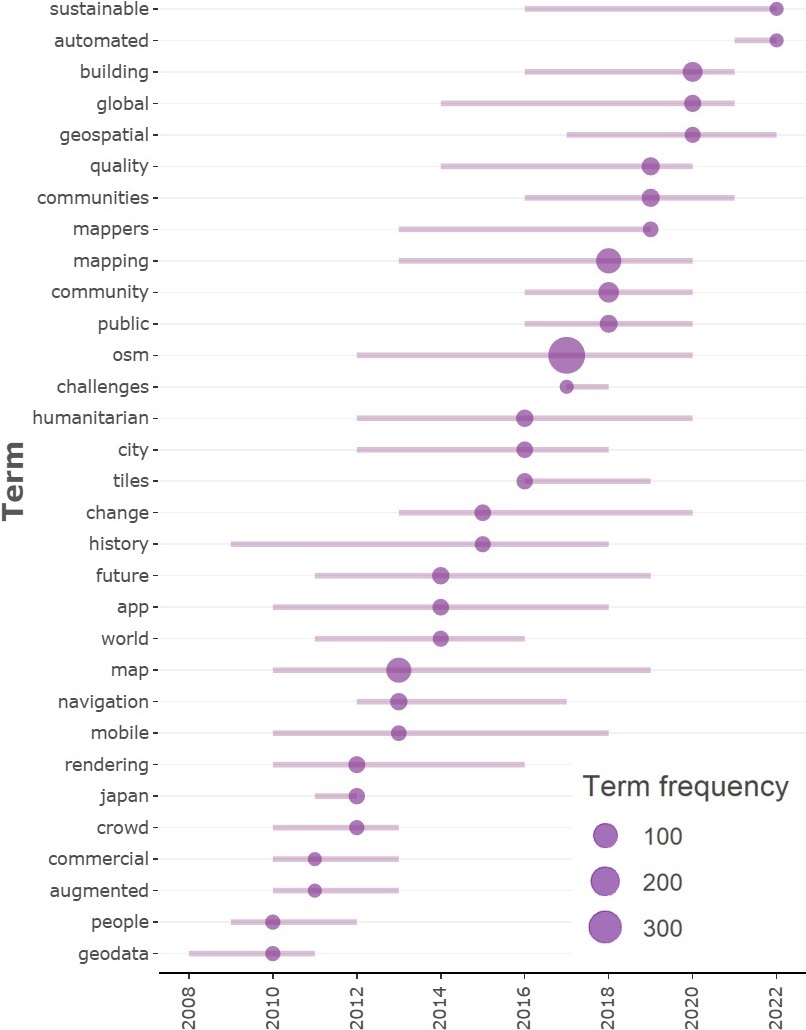}}\hfill
    \subfigure{\includegraphics[height=.599\linewidth]{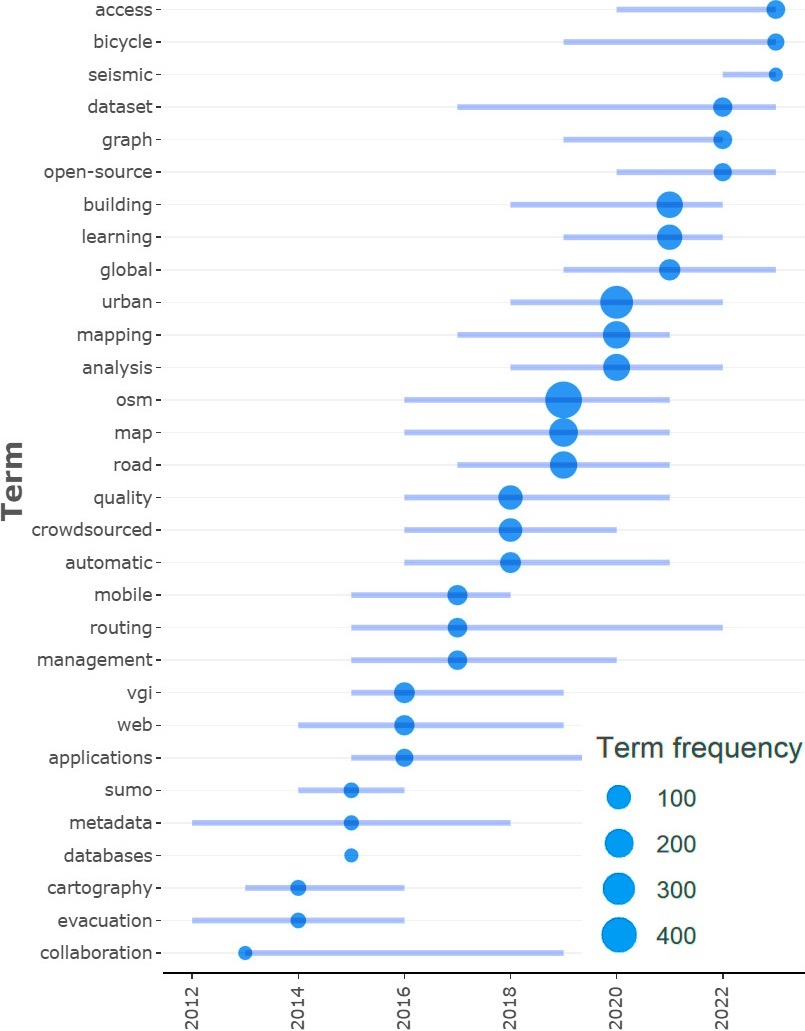}}\hspace{1.3cm}
    \\
    \hspace{0cm} \subfigure{\includegraphics[height=.66\linewidth]{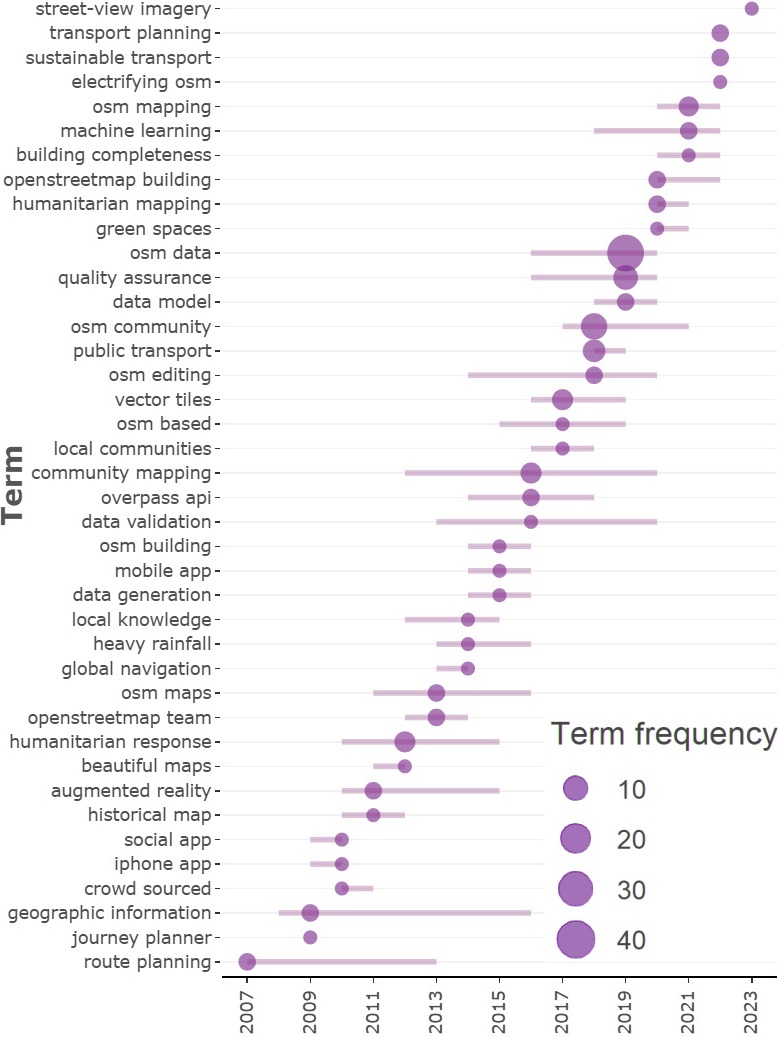}}\hfill
    \subfigure{\includegraphics[height=.66\linewidth]{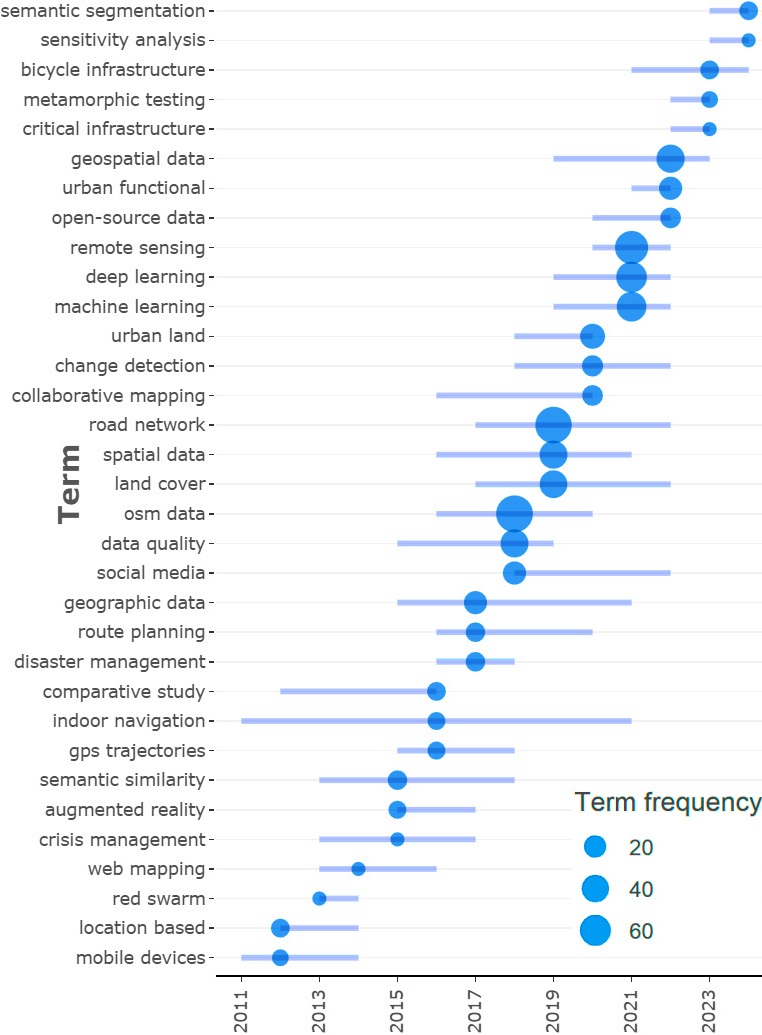}} \hspace{0.5cm}  
    \caption{Trend topics based on titles of \textcolor{sotm}{\textbf{SotM}} and \textcolor{wos}{\textbf{WoS}} documents: (up) Unigrams; (down) Bigrams.}
    \label{fig:trend-topics}
\end{figure}

\subsubsection{Common Terms: Earlier Appearance in SotM Followed by WoS}

Common terms appearing in both WoS and SotM, as detailed in Table~\ref{tab:term_comparison}, reveal their frequency and median year of occurrence. 

Comparing the common terms, we observe a trend where many terms have earlier median years in SotM compared to WoS. 
\textit{Route planning} peaked a full decade earlier at SotM (2007 vs 2017), and \textit{map} / \textit{mapping} did so 6 and 2 years earlier.
Emerging technologies followed the same pattern: \textit{augmented reality} appeared in 2011 at SotM but only in 2015 in WoS. 
Practical topics, \textit{building}, \textit{road}, and \textit{global} issues, all show SotM medians one to three years ahead of WoS, suggesting that practitioners highlighted these concerns before they became mainstream in research papers. 
This suggests that these topics topics emerged and developed earlier at SotM, gaining attention there before becoming more prominent in WoS. 
Only \textit{machine learning} shows identical median years (2021), indicating that both communities converged on this theme at roughly the same time. 
On average, the median year in WoS is about \textit{\textbf{2.8 years}} later than in SotM, showing a lag in the academic recognition of trends initially identified in the SotM community.

\begin{table}[!]
\centering
\scriptsize
\begin{tabular}{lcccc}
\hline
\textbf{Term } & {WoS Frequency} & {SotM Frequency } & {WoS Year(Median)} & {SotM Year(Median)} \\
\hline
osm         & 427           & 378 & 2019 & 2017 \\
mapping     & 150       & 94  & 2020 & 2018 \\
map         & 179           & 91  & 2019 & 2013 \\
building    & 130      & 30  & 2021 & 2020 \\
global      & 49        & 12  & 2021 & 2020 \\
road        & 148          & 9   & 2019 & 2017 \\
mobile      & 36          & 7   & 2017 & 2013 \\
\hline
route planning      & 7          & 3   & 2017 & 2007 \\
augmented reality   & 5         & 3   & 2015 & 2011 \\
osm data            & 73        & 43  & 2018 & 2019 \\
machine learning    & 36	   & 3    &2021		&2021\\ 
\hline
\end{tabular}
\caption{Comparison of frequency and median year of common terms between WoS and SotM.}
\label{tab:term_comparison}
\end{table}

\subsubsection{Unique Terms: Distinct Scopes in SotM and WoS}

Each collection contains unique terms that reflect differing research foci and stages in OSM studies. While the {WoS} collection emphasizes modern, data-driven approaches, the {SotM} collection captures the formative stages of GIS technology and community mapping.

\begin{itemize}
    \item Terms Unique to \textbf{SotM}: These terms highlight elements related to user interaction, humanitarian efforts, and the foundational aspects of geospatial technologies. 
    For example, terms such as \textit{people}, \textit{augmented}, \textit{commercial}, \textit{humanitarian}, and \textit{community} underscore a human-centric approach with an emphasis on practical applications and community-driven mapping projects. Additionally, phrases like \textit{OSM maps}, \textit{OSM community}, and \textit{local communities} illustrate the significant role of OSM and its community, while the term \textit{OSM team} hints at earlier organizational efforts. Further, \textit{public transport}, \textit{quality assurance}, and \textit{vector tiles} suggest applied solutions in urban planning and public services, with terms such as \textit{iphone app} and \textit{social app} reflecting early interest in mobile and social applications.
    
    \item Terms unique to \textbf{WoS}: These terms focus on modern technological innovations, advanced data analysis, and community-driven initiatives. Notable examples include \textit{deep learning}, \textit{remote sensing}, and \textit{semantic segmentation}, which indicate the integration of artificial intelligence and sophisticated data analytics in geospatial research. Additionally, terms like \textit{disaster management}, \textit{crisis management}, \textit{seismic}, and \textit{critical infrastructure} highlight a focus on leveraging geospatial technologies for emergency response and infrastructure protection.

\end{itemize}

\subsubsection{Timeliness in Addressing Recent Events}

We also observed that topics at SotM conferences are highly timely and often address significant events happened recently. 

For instance, the term \textit{Japan} shows as a trend topic in Figure~\ref{fig:trend-topics}(upper left), with the starting year of 2011. 
Notably, 
the SotM talks such as 
\citep{Kastl2011} \citep{Miura2011} and \citep{Ikiya2012} 
highlight the immediate and evolving response to the Japan tsunami and earthquake happend in 2011. 
The median year is 2012, might be related to the fact that SotM 2012 took place in Japan and there were several discussions likely connected to Japan. 

Another example shows up in the preliminary program of 2024 SotM\footnote{As of the writing of this paper, the 2024 SotM conference has not yet taken place.}: a talk by Yair Grinberger and Marco Minghini
~\citep{Grinberger2024} which discusses the recent challenges posed to the VGI community by international events and the subsequent changes in management practices.

In contrast, academic research often experiences a delay due to publication cycles, meaning that studies on such events are typically published some time after the events have occurred.

\section{Future Trends}\label{sec:future}

{As OSM enters its third decade, the research community, the contributor base, and external stakeholders are all undergoing transitions. This chapter outlines key trends shaping the future of OSM from two complementary perspectives: the trends of academic research, and the evolving roles and partnerships among the academia, community, and industry.}

\subsection{Research Trends}\label{sec:futureResearch}

Future trends have always been a focal point in research. In an excellent editorial by Grinberger et al.~\citep{grinberger2023openstreetmap}, three key topics were identified that represent the data-oriented approach within OSM science: data quality, applications, and machine learning. In Sections \ref{sec:wos-trend} and \ref{sec:somt-trend}, we analyzed the evolution of research topics. This section builds on that analysis by exploring future research topics and trends, drawing on both WoS topic trends and SotM discussions to predict emerging topics.

On one hand, we consider emerging topics within different clusters identified in WoS. On the other hand, from our analysis in Section \ref{sec:somt-trend}, we observed that common terms tend to appear 2.8 years earlier in SotM discussions compared to academic research. Therefore, we have identified newer topics in SotM, such as ``pedestrian," ``green space," ``editing," and ``map editors," which may gain prominence in the near future.

Since the 2023 global SotM conference did not take place, we also considered discussions from regional SotM events in 2023, including SotM Africa, Europe, US, France, and the proceedings of OSM Science 2023. 
This combined approach allows us to predict topics likely to become significant in upcoming OSM research.

{We have identified the following \textbf{six} research directions with significant potential, as listed in Table}~\ref{tab:future_osm_trends}. {The first three directions pertain to OSM data: research on OSM data itself, the data sources used to generate OSM data, and the integration of OSM with 3D data structures. They align with the first and fourth Key Research Interests identified in Section}~\ref{sec:hist_map}, {i.e., Quality Assessment and Validation of OSM Data, and Tool Development. 
The fourth and fifth directions focus on application fields that utilize OSM data and on emerging hot study areas for OSM research, respectively. They align with the third and fifth key research interests, i.e., 
Mapping/Generating/Extracting Information using OSM and Other Data, and OSM in Humanitarian and Disaster Response. 
The last direction examines contributors and the influence of their participation, which also aligns with the second Key Research Interest identified in Section}~\ref{sec:hist_map}, {Collaborative Contributing, Contributor Behavior, and Activity Analysis. 
Notably, they also align with the data quality aspect and application aspect pointed out by Grinberger et al.}~\citep{grinberger2023openstreetmap}. 
{We detail these directions in the following sections. }

\begin{table}[h]
    \centering
    \scriptsize
    \renewcommand{\arraystretch}{1.3}
    \begin{tabular}{l|l}
        \hline
        \# & {\textbf{Research directions}}  \\
        \hline
        1& {Research centered on OSM Data} \\
        \hline
        2& {Data sources and Multi-Source Integration for generating OSM data} \\
        \hline
        3& {Integration of OSM with 3D Data Structures} \\
        \hline
        4& {Applications of OSM data}  \\
        \hline
        5& {Study Arears in OSM research} \\
        \hline
        6& {Gender Topics: Potential changes driven by OSM contributors} \\
        \hline
    \end{tabular}
    \caption{{Predicted Research Directions in OSM.}}
    \label{tab:future_osm_trends}
\end{table}

    \subsubsection{{Research centered on OSM Data}}

    This research direction centers on the data itself, including aspects such as data quality, data sources, data generation processes, and the potential for automation. 
    Recent studies have explored the various data elements in OSM, such as Points of Interest (PoIs), buildings, and roads, as demonstrated in OSM Science 2023~\citep{oostwegel2023global, andorful2023exploring, melanda2023openstreetmap, li2023beyond}. The research goes beyond geometric data, addressing the missing attribute data, which has recently gained attention~\citep{biljecki2023quality, sun2023flickrstr}. These topics are also extensively discussed within the OSM community. 
    Additionally, contributions from organizations and companies like Overture, Mapillary, and TomTom are gaining significance in this area.

    Key areas of focus that have been discussed in regional SotM include:

    \begin{itemize}
        \item Data Organization:
        \textit{Tags} and the \textit{Taginfo} tool are crucial, with discussions highlighted such as 
        \citep{Zimmermann2023} at SotM France, and \citep{Topf2023} at SotM EU. 

   \item Data Generation Processes and Tools:
        The automation and semi-automation of data generation, such as the \textit{Rapid Editor}, are essential topics. These were discussed by \citep{Housel2023} at SotM US 
        and by \citep{Turksever2023} at SotM Africa.

   \item Research and Data Processing Tools:
        Tools like \textit{ohsome} and \textit{Overpass} are crucial for analyzing OSM data. Presentations \citep{Reinmuth2023} at SotM Africa 
        and \citep{Reinmuth2023US} at SotM US 
        illustrate this focus. Additionally, Overpass was covered in multiple sessions at SotM France, including \citep{Riche20231}, \citep{Riche20232} and \citep{Buckel2023}.

   \item Contributions from Organizations and Companies:
        The involvement of organizations and companies in OSM is evident in discussions from SotM 2023 regional conferences. For example, \textit{Overture} and \textit{TomTom}'s open data initiatives were discussed by \citep{Clauss2023} and \citep{Baidoun2023} at SotM France, 
        while Mapillary's contributions were highlighted by \citep{Neerhut20231} at SotM EU, \citep{Neerhut20232} at SotM US, and \citep{Turksever20231} at SotM Africa. 

    \end{itemize}

It is foreseeable that achieving high-precision, globally covered, up-to-date data that includes both geometric and semantic information will remain a persistent hot topic. 

\subsubsection{{ Data Sources and Multi-Source Integration}}

Data sources used for OSM research are predominantly based on remote sensing, though there is growing incorporation of social media data and other sources~\citep{hoffmann2023using, li2020instance, sun2020cgnet}. The use of multi-source data is expected to increase, with emerging trends pointing towards the integration of street-level imagery, texts, and other diverse datasets~\citep{hu2023location, hu2024dlrgeotweet,chen2022mining, wang2023insights, lim2024integration, hou2024global, sun2025slifloor}. 
{The trend in data sources is clearly towards multi-source registration and integration~\citep{zhao2019exploring, ding2020framework, sun2020auto, ding2021consistency, leitenstern2024flexmap}.}

Key tools and methods associated with this trend, showing as keywords in Figure~\ref{fig:topic-cluster}, include:

\begin{itemize}
    \item Data and Tools: ``google earth engine", ``GeoAI", ``street view", ``calibration", ``dataset", ``multi-source data".

    \item Methods: ``multiscale analysis", ``graph theory", ``adaptation techniques", ``CNN", ``artificial intelligence", and ``computer vision".
\end{itemize}

\subsubsection{{Integration with 3D Data Structures}}

This topic aligns with ``Topic Cluster 6: Geospatial Modeling" discussed in Section~\ref{sec:wos-trend}. The integration of 3D modeling into OSM research is becoming increasingly relevant. While LoD1 models, which often treat height as an attribute~\citep{chen2021maskH, li20233dcentripetalnet,sun2021bbox}, are relatively simple, LoD2, LoD3, and even LoD4 models offer more detailed representations and are often linked with urban digital twins~\citep{gui2021automated, dehbi2021optimal, biljecki2019raise, pantoja2022generating, wang2024framework,li2024review}. These detailed models typically require LiDAR or photogrammetry data for accurate reconstruction~\citep{xu2021voxel, pan2022enriching, hoegner2018mobile, xu2019pairwise}. Given the current complexity of fine-grained 3D models and the high barriers to editing, such models are not yet widely contributed by volunteers. 

Future predictions suggest that research may focus on developing data structures and frameworks to integrate various data types and address these challenges. This will likely result in new studies on standards, data structures, data transformation, tools, and 3D editing software as the fields of 3D modeling and OSM increasingly converge~\citep{kolbe2021semantic, gilbert2020built, donkers2016automatic, boeters2015automatically, floros2018investigating,biljecki2021extending, heeramaglore2022semantically}.

\subsubsection{{ Applications of OSM Data }}\label{sec:appOSM}

The applications focus on addressing human needs particularly in urban science. 
As OSM data is utilized across various domains, we can expect to see more research exploring new application areas.

A prominent emerging topic is accessibility, which has been widely discussed in the 2023 regional SotM conferences. For instance:

\begin{itemize}
    \item SotM France: Discussions on improving walkability data by \citep{Gervais2023} and mapping accessibility without the wheelchair tag by \citep{Lainez2023}.
    \item SotM EU: Discussion on accessibility for wheelchair users by \citep{Julien2023}. 
    \item OSM Science 2023: Discussion on bike-transit accessibility by \citep{Passmore2023}. 
\item SotM US: Discussion on hospital accessibility by \citep{Edmisten2023}. 
 
    \item SotM Africa: Discussion on footways accessibility by \citep{Vestena2023}. 
\end{itemize}

In urban science, OSM data is being increasingly used to explore topics related to sustainability and the United Nations' Sustainable Development Goals (SDGs). Figure~\ref{fig:topic-cluster} reveals emerging keywords such as health (e.g., ``obesity", ``public health"), environmental comfort (e.g., ``CO2 emissions", ``land-surface temperature"), accessibility (e.g., ``walking", ``cycling", ``spatial accessibility"), and urban morphology (e.g., ``urban forms", ``space syntax"). 

As data resolution and accuracy improve, new research possibilities will emerge, such as enhancing models related to buildings and road networks. For example, improving earthquake risk estimates and detecting earthquake-damaged buildings with OSM building data has been demonstrated in recent studies~\citep{sun2023qqb,zadeh2023improving}. 

After all, OSM, made by people, aims to serve people by addressing a wide range of societal needs, including safety, health, mobility, comfort, and more. Future research will continue to focus on addressing and solving these human-centric challenges.

\subsubsection{{ Study Areas of OSM Research}}

Emerging research areas are expected to keep focusing on regions like China and Sub-Saharan Africa, as highlighted in the network analysis of Figure~\ref{fig:topic-cluster}, which shows significant mentions of \textit{China} (33 times) and \textit{Africa/Sub-Saharan Africa} (13 times).

\textit{China}, with its rapid urbanization and population growth, is poised to remain a key area of study. As China's population stabilizes and begins to decline, new research needs will arise, focusing on how human demands and infrastructure, such as road networks and buildings, adapt to these changes. 
Similarly, \textit{Sub-Saharan Africa}, where urbanization is accelerating and data is often scarce, presents a critical area for future research. This region’s ongoing urban and population growth will require extensive study to address various developmental challenges.

In summary, \textit{China} and \textit{Sub-Saharan Africa} are likely to remain important research areas, with studies focusing on the unique challenges posed by urbanization and demographic changes.

\subsubsection{{ Gender Topics}}\label{sec:genderTopics}

We observed that at SotM Africa, several talks focused on women's participation. 
While gender issues have been explored to some extent in the existing research landscape, such as in studies like Gardner's work on gender representation in OSM~\citep{gardner2020quantifying}, this area remains under-researched. For instance, Figure~\ref{fig:topic-cluster} shows that ``gender" appears only 6 times, with an average publication year of 2018, indicating limited recent focus.

However, gender-related topics are expected to gain more prominence in OSM research, particularly with a focus on contributors and their narratives. 
The discussions at regional SotM conferences, especially in Africa, highlight a growing emphasis on women's participation in OSM. Talks at SotM Africa 2023, such as \citep{Likiliwike2023}, \citep{Chilufya2023}, \citep{Hopeful2023} and \citep{Makuate2023}, 
indicate an emerging interest in empowering women in the geospatial field.

This trend suggests that Africa could be a key driver in advancing gender-focused research in OSM. As women increasingly contribute to geospatial data and infrastructure development in data-scarce regions like Africa, their narratives and needs are likely to become more visible. 

We anticipate that as more women become involved, 
there will be an increase in research from two aspects: 
on one hand, this may lead to research on gender-related editing behaviors; on the other hand, it could result in more studies addressing women's needs within the geospatial framework. A notable, yet still rare, example of such work is the study by Karlekar and Bansal~\citep{karlekar-bansal-2018-safecity}, which explores diverse forms of sexual harassment based on personal stories from the map application ``Safecity". This type of research, though currently limited, is likely to expand as gender issues gain more attention in the OSM community.

\subsection{Evolving Roles and Partnerships}\label{sec:trends_system}

{OSM is undergoing transitions in both its internal research development and external influences, driven by technological advancements and growing corporate involvement. 
OSM's future will also be shaped by how different stakeholders adapt their roles and foster new forms of collaboration. This section discusses these evolving dynamics.}

\titleformat{\paragraph}[block]{\bfseries}{ }{0pt}{}  

\subsubsection{Academic Engagement} 

{In academic research of OSM, we observe a centralization and dispersion pattern among contributors, 
and also a strong diversity across publication venues, author profiles, and research topics.}
\paragraph*{\normalfont\textit{{A. Coexisting Patterns among Contributors: Centralization and Dispersion}}}

{The analysis of contributors and their collaboration networks, conducted in Section~\ref{sec:who}, reveals a dual dynamic of \textbf{centralization} and \textbf{dispersion} that has shaped the development of OSM-related research over the past two decades. 

On the one hand, the field remains highly \textit{centralized}. A small number of prolific authors (e.g., Alexander Zipf, c.f., Section~\ref{sec:411}) and institutions (e.g., Heidelberg University, Wuhan University, c.f., Section~\ref{sec:412}) have contributed disproportionately to the literature. 
The author collaboration network is dominated by a large connected component, in which key figures such as Zipf and Mooney act as bridges between otherwise separate clusters~(c.f., Figure~\ref{fig:author-net}). 
These observations point to the presence of strong research centers that not only accumulate resources and talent but also help set research agendas for the broader community. 

On the other hand, there is clear evidence of \textit{dispersion}. Geographically, contributors have expanded from a European focus to include institutions across North America, Asia, Africa, and South America~(c.f., Figure~\ref{fig:countries-rank}). 
Researcher mobility has played a key role in transferring expertise and establishing new research sites~(c.f., Section~\ref{sec:413}). 
Thematically, while many authors are anchored in GIScience, others have published in multiple disciplines, indicating a diversification of research directions~(c.f., Table~\ref{tab:osm_all_pub}). 
The collaboration network also includes many isolated clusters~(c.f., Figure~\ref{fig:author-net}), suggesting the presence of smaller, independent groups working on more specialized topics. 

These two patterns, \textit{centralization and dispersion}, coexist and carry different implications. \textit{Centralization} brings strength through concentration of expertise, visibility, and collaborative density, but it also introduces \textbf{vulnerabilities}. The field may become overly reliant on a few individuals or institutions, making it susceptible to disruption if they shift priorities or withdraw. Furthermore, strong centralization may reinforce intellectual path-dependencies, making it harder for novel ideas or methods to emerge. 
\textit{Dispersion}, in contrast, contributes to the \textbf{resilience} of the research landscape. The presence of globally distributed contributors, diverse thematic interests, and researcher mobility ensures that innovation continues even if core actors disengage.

Recognizing this duality helps clarify both the historical trajectory and future directions of OSM research. A balanced strategy that strengthens collaboration while supporting new contributors and topics may help sustain the field’s long-term vitality.}

\paragraph*{\normalfont\textit{B. Diversity across publication venues, author profiles, and research topics}}

{We observe strong diversity in OSM research across publication venues, author profiles, and research topics: the field spans multiple disciplines, engages both core and applied researchers, and covers themes ranging from data quality to broad applications (cf.~Sections}~\ref{sec:3},~\ref{sec:who},~\ref{sec:topics}). 
{This diversity as structural fragmentation; as quantitatively visible in the Co-authorship Network~(Figure}~\ref{fig:author-net}), {small, isolated clusters persist alongside the main connected component, reflecting distinct research silos. }
{These developments indicate a bifurcation in the academic landscape:}
\begin{itemize}
\item
{\textit{OSM-centric research}: it examines OSM as a socio-technical system, focusing on data quality, contributor dynamics, platform governance, and tool development.}
\item
{\textit{OSM-aided research}: it uses OSM as a readily available spatial data source to address domain-specific problems, often without engaging with the OSM community.}
\end{itemize}

{\textit{OSM-aided research}' growth reflects OSM's data maturity and impact, indicating that OSM has achieved the scale and quality necessary for broad reuse. 
However, this success may introduce tensions: 
some researchers use OSM as a free annotated dataset, extracting value without engaging with or contributing back to the community }~\citep{grinberger2022bridges}. 
{Such one-sided use can erode trust of the community that produces and maintains the data, reinforce perceptions of academic exploitation, and widen the gap between academia and the OSM community. 

In the coming years, the continued rise of \textit{application-oriented research}, driven by the accessibility of OSM data and advances in AI, may further increase the risk of disciplinary detachment and community disengagement. Meanwhile, certain \textit{OSM-centric} topics, such as geometric accuracy assessments in the context of large corporate contributions}~\citep{microsoft2025globalml}, {may decline in relevance. 
To address these challenges, we propose two complementary strategies: }

\begin{itemize}

\item 

{In \textit{OSM-centric research}, we propose redirecting focus toward toward emerging challenges at the intersection of technology, semantics, and ethics. First, human–AI collaboration should be advanced through intuitive validation tools, mitigation of automation bias, and support for community learning. Second, greater attention is needed to semantic and subjective data quality, including attribute completeness, semantic richness, and handling of conflicting or subjective tags. Third, data justice and map ethics must be prioritized to address coverage inequality, ensure fair representation, and foster inclusive mapping practices.}

\item 

{In \textit{OSM-aided research}, we call for more reciprocal practices, including open-sourcing trained models, sharing derived datasets, and developing tools that benefit the OSM community.}

\end{itemize}

{Together, these directions aim to maintain a healthy balance between application-driven utility and community-oriented stewardship—ensuring that OSM research remains both impactful and sustainable.}

\subsubsection{Community Participation}

{In parallel, 
AI-assisted mapping, rising corporate contributions, and stabilized geometric data growth have increased data volume}~\citep{sirko2021continental,gonzales2023building,microsoft2025globalml} 
{and are also reshaping volunteer engagement. 
SotM US 2022 talk on the RapiD editor explicitly discussed how AI handles large-scale geometry while humans shift toward validation and supervision, illustrating this role transition in practice}~\citep{housel2022rapidx,Housel2023,Turksever2023}. 
{As automation and large-scale uploads reduce the visibility of individual edits, contributors may feel less impactful, risking fatigue and role disorientation if new, meaningful participating roles are not established. 
Meanwhile, providing data for 
OSM-aided research, contributors may feel reduced to data annotators and zero-cost labor, 
raising a critical question: what defines meaningful participation in the AI era?

{With machines increasingly handling routine tasks, human contributors are uniquely positioned to focus on qualitative enrichment~}{(c.f., Section~}\ref{sec:appOSM}, \ref{sec:genderTopics}). This signals a role evolution from manual laborers of the map to designers and knowledge curators of the map. 
This shift points to three critical, human-centered contributions: }
\begin{itemize}
    \item 

{\textit{Knowledge enrichment}: Adding machine-hard-to-capture information such as place attributes, temporal context, and local knowledge.}
    \item 
{\textit{Subjective experience mapping}: Describing atmospheres, perceived safety, aesthetics, and other qualitative spatial experiences that enrich user understanding.}
    \item 
{\textit{Defining innovative use cases}: Initiating thematic maps and hyper-local applications that reflect community needs and creativity.}

\end{itemize}

{Ultimately, the goal is not to resist automation, but to redefine human contribution to focus more on tasks that require judgment, empathy, lived experience, and creativity. 
This transition can elevate OSM from a representation of the physical world to a richer, more inclusive social knowledge system.

More broadly, OSM’s evolution offers a lens into a fundamental challenge of the AI age: how to preserve and foster human motivation and creativity in collaborative data ecosystems. As a flagship of VGI and citizen science, OSM remains a critical site for experimentation and reflection on this future.}

\subsubsection{Strengthening Collaboration}

A key trend shaping the future of OSM is the growing emphasis on collaboration between the community, academia, and industry. {The increasingly intertwined relationship between the community and academia is exemplified by the formal inclusion of the Academic Track at SotM since 2018 and the prominence of Heidelberg as a major hub for OSM research} ({see also Table}~\ref{tab:documents_citations} {and  Figure}~\ref{fig:affi_net}). On an individual level, many researchers are now active contributors within the community~{(c.f., Table}~\ref{tab:authors_articles}, {Table}~\ref{tab:wos_sotm_comparison}); this dual involvement actively fosters mutual understanding and co-evolution. 
{The critical need for such collaboration is underscored by the trend analysis in Section 6.3.1, which revealed an average lag of 2.8 years between a topic's emergence in SotM discussions and its widespread adoption in academic literature (Figure~}\ref{fig:trend-topics}).

Central to this process are the “link scholars” who straddle both academic and community spheres~{(c.f.,  Table~}\ref{tab:wos_sotm_comparison}). As OSM research enters a critical juncture, these scholars play a pivotal role in three capacities:
\begin{itemize} 
    \item {\textit{Translation}: Framing community challenges (e.g., tool usability, tag conflicts) as research questions, and translating academic advances into practical tools and guidelines.}
    \item {\textit{Coordination:} Facilitating joint events where contributors and researchers co-design tools and studies aligned with community needs.}
    \item {\textit{Advocacy:} Using their institutional visibility to represent community interests in academic, policy, and industry forums to ensure their values are reflected in OSM’s future. }
\end{itemize}

\textit{Industry participation} in OSM is growing, with companies contributing data and participating under the umbrella of community engagement like SotM~\citep{Clauss2023,Baidoun2023,Neerhut20231,Neerhut20232,Turksever20231}
While this brings resources and technical capacity, it also raises concerns about corporate influence over standards and priorities, potentially misaligned with community values~\citep{overture_forum_discussion,osm_ethics_forum2023}. 
For example, commercial imperatives may deprioritize non-profitable regions or introduce standards that conflict with grassroots practices. 
Sustaining OSM’s future requires balanced collaboration among academia, community, and industry: 
academia contributes theory and tools; community provides local knowledge and ensures data relevance; industry offers infrastructure and scale. 
Together, through transparent and sustained dialogue, they can transform OSM into a more equitable, intelligent, and human-centered global knowledge infrastructure.
\section{Conclusion}\label{sec:conclude}

Over the past two decades, OSM has made significant strides in geographic information science and mapping. 
This review provides an in-depth analysis of OSM research till 2024, 
combining insights from academic publications and community discussions. 
Our study involves a detailed statistical examination of OSM-related research using the WoS Core Collection, and a thorough review of presentations from the SoTM conferences. We aim to address the OSM research landscape by exploring who is involved, what research topics are prominent, and how these areas are evolving. 

We investigate the key contributors - identifying influential countries, institutions, and individuals - and analyze their impact and collaborations, underscoring the driving force of the research, the authors. We analyse the key topics and research sub-fields, highlighting the evolution of core themes and emerging trends in OSM research. Additionally, we bridge the gap between academic research and community-driven initiatives by comparing academic findings with insights from SoTM talks.
Furthermore, we forecast six future research trends by examining evolving topics in both WoS and SoTM 
and highlight shifting roles and collaborations shaping OSM’s future. 
These findings can help researchers and practitioners navigate future directions and foster more balanced, impactful, and collaborative developments.

Despite the findings, we acknowledge this study has several limitations. First, while WoS is one of the best available sources, it has inherent data constraints. Some articles, particularly from newer journals and conferences, may not yet be indexed. Additionally, inconsistencies in indexing, keyword variations, and multiple author name formats can affect research discoverability and impact, underscoring the need for better transparency, accurate translation, and enhanced searchability. Second, our bibliometric approach offers a structured overview of the research landscape but does not explore individual studies in depth. As a result, some less-studied yet important subfields, such as social studies of OSM, briefly mentioned in Section~\ref{sec:hist_map} under the Key research interest 2 \textit{Collaborative Contributing, Contributor Behavior, and Activity Analysis}, are inevitably omitted. 
Finally, our analysis has focused on mapping the overall research landscape rather than providing an exhaustive examination of all individual research works. More detailed analyses of specific research areas would require considerably more space and beyond the scope of this work.

As Steve Coast recently wrote \citep{stevecoast2024}, the notion of enabling volunteers to edit maps was deemed unthinkable in 2004, with map data traditionally controlled by a select few. In just 20 years, OSM has transformed this paradigm, creating a global, freely accessible map supported by a vast and engaged community, alongside substantial contributions from various organizations and companies. This shift has not only reshaped the way we interact with geographic data but has also spurred a rich body of research. OSM stands as a benchmark in VGI, driving extensive academic and practical exploration across diverse fields. 
As we look forward, the continued evolution of OSM's research landscape promises to yield further insights and innovations, setting the stage for the next 20 years of groundbreaking advancements in geographic information science; 
{the rise of AI, expanding corporate participation, and changing contributor dynamics the OSM community with both challenges and unprecedented opportunities to reimagine how we build, use, and share geospatial knowledge. 
To this end, our collective efforts matter more than ever. 
Academic researchers are uniquely positioned and entrusted to collaborate closely with volunteers, communities, and industry, helping to guide OSM’s evolution toward a future that is open, innovative, and resilient.}

\appendix 
\section{\\Supplementary Analysis of OSM Research for Year 2024}

{We conducted a supplementary search of the WoS Core Collection covering \textit{1 July--31 December 2024}. This analysis complements the primary dataset (\textit{2008--June 2024}) and mitigates the time lag inherent in peer review and indexing, providing an updated view of research trends in late 2024. 
This update does not change the main analysis. } 

\subsection*{Data Description}

The main analysis in this work is based on a WoS search conducted on \textit{11 July 2024}, covering publications up to \textit{30 June 2024} and yielding 1,926 records, including 74 items assigned to $PY\footnote{PY: Publication Year} = 2024$. To obtain a complete view of the calendar year and preserve comparability with previous years, we repeated the search on \textit{11 August 2025} and downloaded all records indexed as $PY = 2024$\footnote{We observed inconsistencies between the mid-2024 and updated WoS records, that are typically caused by delayed indexing, metadata updates, and \textit{Early Access} year reassignments. To ensure complete coverage, we retrieved all items indexed as $PY = 2024$ using the same query and inclusion criteria as the main analysis and removed duplicates.}. 

This updated snapshot contains 191 OSM-related publications (118 journal articles, 64 proceedings papers, and 9 other document types).

Comparing the mid-2024 and full-year 2024 snapshots shows:
\begin{itemize}
\item  57 records appear in both snapshots;
\item  17 mid-year records are no longer assigned to 2024 (3 reassigned to 2025; 14 removed due to WoS metadata updates);
\item  134 records are newly identified OSM-related publications for 2024.
\end{itemize}

We refer to this difference set, i.e., 134 records, as subset $S2$. Because both $S2$ and the full-year 2024 dataset (191 records) are substantially smaller than the main dataset (1,926 records), the statistical thresholds used in the primary analysis (e.g., $\geq 10$ papers per author or $\geq 5$ papers per institution) are no longer applicable. 
Consequently, this appendix provides a lightweight verification using simplified indicators and a qualitative trend assessment, focusing on whether research patterns in the second half of 2024 follow earlier trajectories or show any notable deviations.

\begin{figure}[!]
    \centering
    \includegraphics[width=0.98\linewidth]{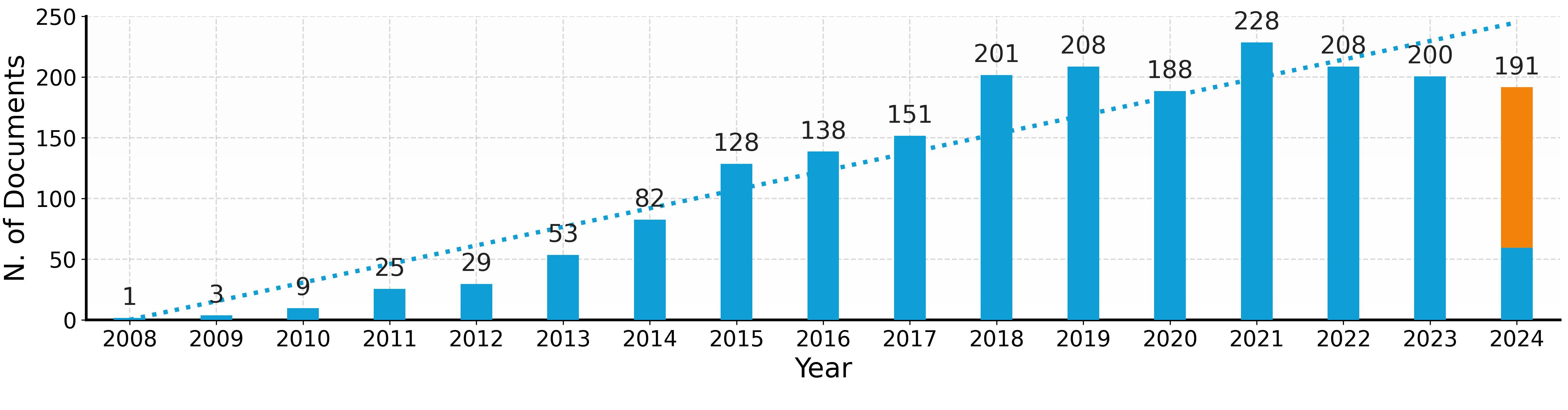}
    \caption{Annual scientific production of OSM research (WoS Core Collection). For 2024, the blue bar (57) is the corrected value of  Figure~\ref{fig:AnnualScientificProduction} (publications indexed up to June 30), while the orange segment shows the additional records found in the updated WoS snapshot, bringing the full-year total to 191. The dotted line indicates the long-term trend.}
    \label{fig:app_AnnualScientificProduction}
\end{figure}

\subsection*{Publication Trend}

\subsubsection*{Publication Volume}  
The full-year 2024 output of 191 WoS-indexed OSM publications confirms the stabilization trend observed in the main analysis (Section~\ref{sec:AnnualScientificProduction}). After peaking in 2021 (228 papers) and maintaining high publication volumes in 2022–2023 ($>$200 papers), the 2024 total indicates a moderate consolidation. Nevertheless, it remains close to the 200-item level, suggesting a plateau rather than a structural decline. This supports the interpretation that OSM research has transitioned from an expansionary phase into a mature period of sustained productivity. Figure~\ref{fig:app_AnnualScientificProduction} shows the Annual scientific production of OSM research from 2008 to 2024.

\subsubsection*{Key Sources}
The source analysis for 2024 reaffirms the prominence of technical journals. The \textit{\textbf{International Journal of Digital Earth}} emerged as the leading venue (7 papers), alongside established journals such as \textit{ISPRS International Journal of Geo-Information} (6 papers). Earth observation journals, including \textit{IEEE J-STARS}, \textit{Sensors}, \textit{Remote Sensing of Environment}, and \textit{Remote Sensing}, also contributed multiple papers (3 each), consistent with long-term patterns (Table~\ref{tab:articles_by_source}).

A noteworthy development in late 2024 is the increased visibility of OSM research in computer science and AI venues. For example, the top-tier robotics journal, \textit{IEEE Robotics and Automation Letters}, published deep-learning-based OSM map generation~\citep{jiang2024pmapnet}, marking a step toward high-level technical integration. Additional interdisciplinary contributions appeared in CVPR and ACM SIGSPATIAL workshops~\citep{sastry2024geosynth, friedsam2024osm, wang2024cycletrajectory, kapp2024surfaceai}. This underscores a migration of OSM research from pure cartography toward the fields of computer vision, robotics, and artificial intelligence.

\subsubsection*{Contributors}

The 2024 geographical distribution confirms the ``tri-polar'' dominance of China, the USA, and Europe identified in Section~\ref{sec:412}. China leads (46 articles), followed by the USA (37) and Germany (30). Europe remains the main hub: the combined output of Germany, the UK (14), and the Netherlands (13) exceeds any other region, and international collaborations still account for a large share of publications.

The July--December 2024 subset shows a partly different institutional landscape from Table~4 and Section~\ref{sec:412}. The Chinese Academy of Sciences (CAS), Wuhan University, and the Technical University of Munich remain major contributors, but institutions that were not prominent in the 2008--June 2024 dataset now rank among the leading publishers, such as Tsinghua University and UT Southwestern Medical Center. 
 The most active authors in the 2024 subset are absent from the long-term top list (Table~\ref{tab:osm_all_pub}), including \textit{Chen Hongruixuan} and \textit{Song Jian} (land-cover change detection~\citep{chen2024objformer, chen2024change}), \textit{Ana Basiri} (contributor behaviour and toponym identification~\citep{solomon2024evaluating, shingleton2024enhancing}), \textit{Kiyun Yu} (knowledge-base QA and spatial indexing~\citep{yang2024geographic, jeong2024generating}), and \textit{Hanli Liu} and \textit{Carlos J.~Hellin} (3D urban reconstruction~\citep{liu2024vertex, liu20243d}). 
As this appendix uses only newly indexed 2024 publications, these authors are not directly comparable to the long-term contributors in Table~\ref{tab:osm_all_pub}. 
Nevertheless, this pattern may indicate that recent momentum in OSM research is increasingly driven by new groups, particularly from Earth observation, computer vision, and AI communities.

\subsection*{Topics} 

The topic trends predicted in the main analysis (Section~\ref{sec:futureResearch}) are reflected in the newly indexed 2024 publications. 
Research on OSM data itself continued through studies on API standardisation~\citep{arnold2024varying}. 
Multi-source integration remained strong, with OSM incorporated into multimodal foundation models~\citep{balsebre2024city}, used as structural priors for HD-map generation~\citep{jiang2024pmapnet}, and evaluated in relation to open aerial imagery~\citep{mandourah2024analysing}. Integration with LiDAR and aerial imagery continued for grid modelling~\citep{weber2024open}, urban functional zone classification~\citep{mo2024urban}, and polyhedral building reconstruction~\citep{liu2024vertex}. Applications further diversified across environmental modelling~\citep{ding2024spatial}, conflict-zone assessment~\citep{bachmann2024cultural}, accessibility routing~\citep{nguyen2024mypath}, and green-space perception~\citep{teeuwen2024how}. Geographic coverage includes studies in rural China~\citep{wang2024exploring} and expanded further into the Global South, including  Kenya~\citep{zhou2024mapping}, and Africa-wide SDG assessment~\citep{cardenas2024automated}. Gender-related research also continued through analyses of demographic participation patterns in OSM editing behaviour~\citep{solomon2024evaluating}. Overall, these works indicate continuity in major thematic trajectories.

The new 2024 publications highlight rapid growth in generative AI and large language model (LLM) integration within OSM research. The emergence of ``GeoAI" has materialized into specific applications of foundation models, exemplified by satellite image synthesis \citep{sastry2024geosynth} and City Foundation Models for learning general-purpose representations from OSM \citep{balsebre2024city}. Furthermore, the integration of LLMs became prominent in late 2024, evidenced by the use of embeddings to enrich building function classification \citep{memduhoglu2024enriching} and application of LLM-informed POI classification for semantic trajectory mining \citep{liu2024semantic}. These studies confirm that OSM research is actively adopting cutting-edge generative AI technologies. 

\subsection*{Summary} 

The supplementary dataset for the second half of 2024 corroborates the structural and thematic trends identified in the main analysis. No notable shifts in contributor demographics, publication venues, or core research topics were detected—an expected outcome given that a six-month window is typically insufficient for major bibliometric changes to materialise.

Overall, the full-year 2024 WoS snapshot shows that OSM research remains highly active, with publication output stabilising at a high level and research directions increasingly aligned with geospatial foundation models, LLM, and corporate geodata ecosystems such as Overture Maps. These observations reinforce the future directions outlined in Section~\ref{sec:future} and suggest that OSM research is entering a new phase 
of deeper integration with multi-modal AI and large-scale geodata infrastructures. 

\section*{Data and codes availability statement}

The data and code that support the findings of this study are available at
\url{https://doi.org/10.6084/m9.figshare.27636516}. 
The same repository also provides access to the interactive VOSviewer visualizations.

\section*{Acknowledgement(s)}

We would like to acknowledge Jianming Zhou for his preliminary investigation of Vosviewer.

\section*{Author contribution} 
Yao Sun: Conceptualization, Methodology, Data curation, Formal analysis, Writing - original draft, Writing - review \& editing; 
Liqiu Meng: Conceptualization, Writing - review \& editing, Funding acquisition; 
Andrés Camero: Methodology, Writing - review \& editing;
Stefan Auer: Writing - review \& editing; 
Xiao Xiang Zhu: Writing - review \& editing, Funding acquisition.

\section*{Disclosure statement}
No potential conflict of interest was reported by the author(s). 
The authors employed ChatGPT to polish the language.

\section*{Funding}

The work is jointly supported 
by the Deutsche Forschungsgemeinschaft (DFG, German Research Foundation) -  499168241 for the project OpenStreetMap Boosting using Simulation-Based Remote Sensing Data Fusion (Acronym: OSMSim) and 
by the Technical University of Munich (TUM) Georg Nemetschek Institute under the project Artificial Intelligence for the automated creation of multi-scale digital twins of the built world (Acronym: AI4TWINNING).






\bibliographystyle{tfv}
\bibliography{mybibfile}

@article{chen2024objformer,
  author    = {Chen, Hongruixuan and Lan, Cuiling and Song, Jian and Broni-Bediako, Clifford and Xia, Junshi and Yokoya, Naoto},
  title     = {ObjFormer: Learning Land-Cover Changes From Paired OSM Data and Optical High-Resolution Imagery via Object-Guided Transformer},
  journal   = {IEEE Transactions on Geoscience and Remote Sensing},
  year      = {2024},
  doi       = {10.1109/TGRS.2024.3410389}
}

@inproceedings{chen2024change,
  author    = {Chen, Hongruixuan and Song, Jian and Yokoya, Naoto},
  title     = {Change Detection Between Optical Remote Sensing Imagery and Map Data Via Segment Anything Model (SAM)},
  booktitle = {2024 IEEE International Geoscience and Remote Sensing Symposium (IGARSS)},
  year      = {2024},
  doi       = {10.1109/IGARSS53475.2024.10642789}
}

@article{solomon2024evaluating,
  author    = {Solomon, Guy and Sutton, Dominick and Kayali, Merve Polat and Yuan, Xinyi and Gardner, Zoe and Basiri, Ana},
  title     = {Evaluating geotemporal behaviours of OpenStreetMap contributors},
  journal   = {AGILE: GIScience Series},
  year      = {2024},
  doi       = {10.5194/agile-giss-5-14-2024}
}

@inproceedings{shingleton2024enhancing,
  author    = {Shingleton, Joseph and Basiri, Ana},
  title     = {Enhancing Toponym Identification: Leveraging Topo-BERT and Open-Source Data to Differentiate Between Toponyms and Extract Spatial Relationships},
  booktitle = {27th AGILE Conference on Geographic Information Science},
  year      = {2024},
  doi       = {10.5194/agile-giss-5-12-2024}
}

@article{yang2024geographic,
  author    = {Yang, Jonghyeon and Jang, Hanme and Yu, Kiyun},
  title     = {Geographic Knowledge Base Question Answering over OpenStreetMap},
  journal   = {ISPRS International Journal of Geo-Information},
  year      = {2024},
  doi       = {10.3390/ijgi13010010}
}

@inproceedings{jeong2024generating,
  author    = {Jeong, Evelyn Hyeji and Yang, Taejoo and Yang, Jonghyeon and Yu, Kiyun},
  title     = {Generating a Semantic Parsing Dataset for GeoKBQA over OpenStreetMap},
  booktitle = {2024 IEEE International Conference on Big Data and Smart Computing (BigComp)},
  year      = {2024},
  doi       = {10.1109/BigComp60711.2024.00073}
}

@article{liu2024vertex,
  author    = {Liu, Hanli and Hellin, Carlos J. and Tayebi, Abdelhamid and Calles, Francisco and Gomez, Josefa},
  title     = {Vertex-Oriented Method for Polyhedral Reconstruction of 3D Buildings Using OpenStreetMap},
  journal   = {Sensors},
  year      = {2024},
  doi       = {10.3390/s24247992}
}

@article{liu20243d,
  author    = {Liu, Hanli and Hellin, Carlos J. and Tayebi, Abdelhamid and Delgado, Carlos and Gomez, Josefa},
  title     = {3D Reconstruction of Geometries for Urban Areas Supported by Computer Vision or Procedural Generations},
  journal   = {Mathematics},
  year      = {2024},
  doi       = {10.3390/math12213331}
}

@article{jiang2024pmapnet,
  author    = {Jiang, Zhou and Zhu, Zhenxin and Li, Pengfei and Gao, Huan-ang and Yuan, Tianyuan and Shi, Yongliang and Zhao, Hang and Zhao, Hao},
  title     = {P-MapNet: Far-Seeing Map Generator Enhanced by Both SDMap and HDMap Priors},
  journal   = {IEEE Robotics and Automation Letters},
  year      = {2024},
  publisher = {IEEE},
  doi       = {10.1109/LRA.2024.3447450}
}

@inproceedings{sastry2024geosynth,
  author    = {Sastry, Srikumar and Khanal, Subash and Dhakal, Aayush and Jacobs, Nathan},
  title     = {GeoSynth: Contextually-Aware High-Resolution Satellite Image Synthesis},
  booktitle = {2024 IEEE/CVF Conference on Computer Vision and Pattern Recognition Workshops (CVPRW)},
  year      = {2024},
  doi       = {10.1109/CVPRW63382.2024.00051}
}

@inproceedings{friedsam2024osm,
  author    = {Friedsam, Wenzel and Rupp, Tobias},
  title     = {OSM Ticket to Ride},
  booktitle = {Proceedings of the 17th ACM SIGSPATIAL International Workshop on Computational Transportation Science (IWCTS 2024)},
  year      = {2024},
  doi       = {10.1145/3681772.3698211}
}

@inproceedings{wang2024cycletrajectory,
  author    = {Wang, Meihui and Haworth, James and Ilyankou, Ilya and Christie, Nicola},
  title     = {CycleTrajectory: An End-to-End Pipeline for Enriching and Analyzing GPS Trajectories to Understand Cycling Behavior and Environment},
  booktitle = {Proceedings of the 2nd ACM SIGSPATIAL International Workshop on Sustainable Urban Mobility (SUMOB 2024)},
  year      = {2024},
  doi       = {10.1145/3681779.3696838}
}

@inproceedings{kapp2024surfaceai,
  author    = {Kapp, Alexandra and Hoffmann, Edith and Weigmann, Esther and Mihaljevic, Helena},
  title     = {SurfaceAI: Automated Creation of Cohesive Road Surface Quality Datasets Based on Open Street-Level Imagery},
  booktitle = {Proceedings of the 2nd ACM SIGSPATIAL International Workshop on Advances in Urban-AI (Urban-AI 2024)},
  year      = {2024},
  doi       = {10.1145/3681780.3697277}
}

@inproceedings{balsebre2024city,
  author    = {Balsebre, Pasquale and Huang, Weiming and Cong, Gao and Li, Yi},
  title     = {City Foundation Models for Learning General Purpose Representations from OpenStreetMap},
  booktitle = {Proceedings of the 33rd ACM International Conference on Information and Knowledge Management (CIKM 2024)},
  year      = {2024}
}

@article{memduhoglu2024enriching,
  author    = {Memduhoglu, Abdulkadir and Fulman, Nir and Zipf, Alexander},
  title     = {Enriching Building Function Classification Using Large Language Model Embeddings of OpenStreetMap Tags},
  journal   = {Earth Science Informatics},
  year      = {2024}
}

@inproceedings{liu2024semantic,
  author    = {Liu, Yifan and Kuai, Chenchen and Liao, Xishun and Ma, Haoxuan and He, Brian YUESHUAI and Ma, Jiaqi},
  title     = {Semantic Trajectory Data Mining with LLM-Informed POI Classification},
  booktitle = {2024 IEEE 27th International Conference on Intelligent Transportation Systems (ITSC)},
  year      = {2024}
}

@article{arnold2024varying,
  author = {Arnold, Laurin and Hukal, Philipp},
  title = {The varying effects of standardisation on digital platform innovation: evidence from OpenStreetmap},
  journal = {Innovation: Organization \& Management},
  year = {2024}
}

@article{bachmann2024cultural,
  title={Cultural Heritage in Times of Crisis: Damage Assessment in Urban Areas of Ukraine Using Sentinel-1 SAR Data},
  author={Bachmann-Gigl, Ute and Dabiri, Zahra},
  journal={ISPRS International Journal of Geo-Information},
  volume={13},
  number={9},
  pages={319},
  year={2024},
  publisher={MDPI}
}

@article{teeuwen2024how,
  author = {Teeuwen, Roos and Milias, Vasileios and Bozzon, Alessandro and Psyllidis, Achilleas},
  title = {How well do NDVI and OpenStreetMap data capture people’s visual perceptions of urban greenspace?},
  journal = {Landscape and Urban Planning},
  year = {2024}
}

@article{mo2024urban,
  author = {Mo, Y. and others},
  title = {Urban Functional Zone Classification Using Light-Detection-and-Ranging Point Clouds, Aerial Images, and Point-of-Interest Data},
  journal = {Remote Sensing},
  year = {2024}
}

@article{weber2024open,
  title={Open data-driven automation of residential distribution grid modeling with minimal data requirements},
  author={Weber, Moritz and Janecke, Luc and {\c{C}}akmak, H{\"u}seyin K and Hagenmeyer, Veit},
  journal={IEEE Transactions on Smart Grid},
  volume={15},
  number={6},
  pages={5721--5732},
  year={2024},
  publisher={IEEE}
}

@article{mandourah2024analysing,
  author = {Mandourah, Ammar and Hochmair, Hartwig H.},
  title = {Analysing the use of OpenAerialMap images for OpenStreetMap edits},
  journal = {Geo-spatial Information Science},
  year = {2024}
}

@article{wang2024exploring,
  title={Exploring the potential of OpenStreetMap Data in regional economic development evaluation modeling},
  author={Wang, Zhe and Zheng, Jianghua and Han, Chuqiao and Lu, Binbin and Yu, Danlin and Yang, Juan and Han, Linzhi},
  journal={Remote sensing},
  volume={16},
  number={2},
  pages={239},
  year={2024},
  publisher={MDPI}
}

@article{ding2024spatial,
  title={Spatial and temporal urban air pollution patterns based on limited data of monitoring stations},
  author={Ding, Junwei and Ren, Chen and Wang, Junqi and Feng, Zhuangbo and Cao, Shi-Jie},
  journal={Journal of Cleaner Production},
  volume={434},
  pages={140359},
  year={2024},
  publisher={Elsevier}
}

@incollection{nguyen2024mypath,
  author = {Nguyen, Thomas and others},
  title = {MyPath: Accessible Route Generation Using Crowd-Sensed Surface Information},
  booktitle = {Lecture Notes of the Institute for Computer Sciences, Social Informatics and Telecommunications Engineering},
  year = {2024}
}

@article{zhou2024mapping,
  author = {Zhou, Qi and Liu, Zixian and Huang, Zesheng},
  title = {Mapping Road Surface Type of Kenya Using OpenStreetMap and High-resolution Google Satellite Imagery},
  journal = {Scientific Data},
  year = {2024}
}

@article{cardenas2024automated,
  author = {Cardenas-Ritzert, Orion S. E. and others},
  title = {Automated Geospatial Approach for Assessing SDG Indicator 11.3.1: A Multi-Level Evaluation of Urban Land Use Expansion across Africa},
  journal = {ISPRS International Journal of Geo-Information},
  year = {2024}
}

@misc{overture_forum_discussion,
author       = {{OpenStreetMap community discussion}},
  title        = {Community discussion on Overture Maps Foundation concerns},
  howpublished = {OpenStreetMap Community Forum thread “Overturemaps\.org – big‑businesses OSMF alternative”},
  year         = {2022},
  note         = {“The tensions between the corporate players and the OSMF … allow overture\.org to cut the OSMF … control how and with what data contributions to OSM are made.”}
}

@misc{osm_ethics_forum2023,
author       = {{OpenStreetMap community discussion}},
  title        = {Community discussion on ethics, human rights, and corporate mapping activities in OSM communities},
  howpublished = {OpenStreetMap Community Forum thread “Sobre la ética, los Derechos Humanos y la acción de empresas con las comunidades OSM”},
  year         = {2023},
}

@misc{housel2022rapidx,
  author       = {Housel, Bryan and Clark, Ben},
  title        = {{RapiD}: AI‑assisted OpenStreetMap editor (RapiD/Rapid 2.0)},
  howpublished = {Presented at State of the Map US 2022 by OpenStreetMap US},
  year         = {2022},
  doi          = {10.5446/58201},
  note         = {Describes RapiD extensions to iD with AI‑generated roads, buildings, sidewalks :contentReference[oaicite:1]{index=1}.}
}

@article{sirko2021continental,
  author    = {Sirko, W. and Kashubin, S. and Ritter, M. and Annkah, A. and Bouchareb, Y. S. E. and Dauphin, Y. and Keysers, D. and Neumann, M. and Cisse, M. and Quinn, J. A.},
  title     = {Continental‑scale building detection from high resolution satellite imagery},
  journal   = {arXiv preprint arXiv:2107.12283},
  year      = {2021},
  note      = {Google Open Buildings: 1.8billion footprints across Africa, SouthSoutheast Asia, Latin America, Caribbean}
}

@misc{gonzales2023building,
  author       = {Gonzales, Jack Joseph},
  title        = {Building‑Level Comparison of Microsoft and Google Open Building Footprints Datasets},
  howpublished = {{InProceedings of the 12th International Conference on Geographic Information Science (GIScience 2023), LIPIcs 277:35:1–35:6}},
  year         = {2023},
  doi          = {10.4230/LIPIcs.GIScience.2023.35},
  note         = {Discusses matching and quality assessment of Microsoft and Google building footprints :contentReference[oaicite:2]{index=2}}
}

@misc{microsoft2025globalml,
  author       = {{Microsoft}},
  title        = {Global ML Building Footprints},
  howpublished = {GitHub repository},
  year         = {2025},
  note         = {1.4billion+ global building footprints detected from Bing Maps imagery (2014–2024), ODbL licence :contentReference[oaicite:3]{index=3}}
}

@article{sun2025slifloor,
  title={Building Floor Number Estimation from Crowdsourced Street-Level Images: Munich Dataset and Baseline Method},
  author={Sun, Yao and Chen, Sining and Tian, Yifan and Zhu, Xiao Xiang},
  journal={arXiv preprint arXiv:2505.18021},
  year={2025}
}

@article{Yan01092020,
author = {Yingwei Yan and Chen-Chieh Feng and Wei Huang and Hongchao Fan and Yi-Chen Wang and Alexander Zipf and},
title = {Volunteered geographic information research in the first decade: a narrative review of selected journal articles in GIScience},
journal = {International Journal of Geographical Information Science},
volume = {34},
number = {9},
pages = {1765--1791},
year = {2020},
publisher = {Taylor \& Francis},
}

@article{Herfort2023,
  author    = {Benjamin Herfort and Sven Lautenbach and Jo{\~a}o Porto de Albuquerque and Jennings Anderson and Alexander Zipf},
  title     = {A spatio-temporal analysis investigating completeness and inequalities of global urban building data in OpenStreetMap},
  journal   = {Nature Communications},
  year      = {2023},
  volume    = {14},
  number    = {1},
  pages     = {3985},
  doi       = {10.1038/s41467-023-39698-6},
  url       = {https://www.nature.com/articles/s41467-023-39698-6}
}

@article{Bittner2017,
  author    = {Christian Bittner},
  title     = {Participation and Marginality on the Geoweb: The Politics of Non-Mapping on OpenStreetMap Jerusalem},
  journal   = {GeoJournal},
  year      = {2017},
  volume    = {82},
  number    = {5},
  pages     = {887--906},
  doi       = {10.1007/s10708-016-9723-7},
  url       = {https://link.springer.com/article/10.1007/s10708-016-9723-7}
}

@article{Jackson2018,
  author    = {Steven J. Jackson and Alex Pompe and Gabriel Mugar},
  title     = {Plotting Practices and Politics: (Im)mutable Narratives in OpenStreetMap},
  journal   = {New Media \& Society},
  year      = {2018},
  volume    = {20},
  number    = {4},
  pages     = {1415--1433},
  doi       = {10.1177/1461444817697322},
  url       = {https://journals.sagepub.com/doi/10.1177/1461444817697322}
}

@article{Lin2019,
  author    = {Wen Lin},
  title     = {Editing Worlds: Participatory Mapping and a Minor Geopolitics},
  journal   = {Transactions of the Institute of British Geographers},
  year      = {2019},
  volume    = {44},
  number    = {2},
  pages     = {297--311},
  doi       = {10.1111/tran.12280},
  url       = {https://rgs-ibg.onlinelibrary.wiley.com/doi/10.1111/tran.12280}
}

@article{Scassa2013,
  author    = {Teresa Scassa},
  title     = {Prod-Users of Geospatial Information: Some Legal Perspectives},
  journal   = {GeoJournal},
  year      = {2013},
  volume    = {78},
  number    = {6},
  pages     = {935--946},
  doi       = {10.1007/s10708-013-9473-3},
  url       = {https://link.springer.com/article/10.1007/s10708-013-9473-3}
}

@misc{Likiliwike2023,
  author = {Carolina Likiliwike},
  title = {Motivate Girls in Geospatial Technologies},
  year = 2023,
  url  = {https://pretalx.com/sotm-africa-2023/talk/LJY7EH/},
  lastchecked = {2024-08-30}
}

@misc{Chilufya2023,
  author = {Charles Chilufya},
  title = {Bridging Gender Gaps in OSM},
  year = 2023,
  url  = {https://pretalx.com/sotm-africa-2023/talk/WUH9R3/},
  lastchecked = {2024-08-30}
}

@misc{Hopeful2023,
  author = {Bafamodei Hopeful},
  title = {Campaigning and Engaging Girls and Women Towards Positive Change using Open Mapping Tech in Nigeria},
  year = 2023,
  url  = {https://pretalx.com/sotm-africa-2023/talk/SEWJ8D/},
  lastchecked = {2024-08-30}
}

@misc{Makuate2023,
  author = {Marie Makuate and Shazmane Mandjee Rehamtula and Maimouna Ndao},
  title = {Women Centered Disaster Resilience in Saloum Islands (Sénégal)},
  year = 2023,
  url  = {https://pretalx.com/sotm-africa-2023/talk/QWENXA/},
  lastchecked = {2024-08-30}
}

@misc{Julien2023,
  author = {Robin Julien},
  title = {Accessibility information for wheelchair users in OSM and the On Wheels app},
  year = 2023,
  url  = {https://2023.stateofthemap.eu/program/accessibility-information-for-wheelchair-users-in-osm-and-the-on-wheels-app},
  lastchecked = {2024-08-30}
}

@article{Passmore2023,
  title={Assessing bike-transit accessibility with OpenStreetMap},
  author={Reid Passmore and Randall Guensler and Kari Watkins},
  journal={Proceedings of the OSM Science 2023},
  pages={53--56},
  year={2023}
}

@misc{Edmisten2023,
  author = {William Edmisten},
  title = {Visualizing Hospital Accessibility with OpenStreetMap},
  year = 2023,
  url  = {https://openstreetmap.us/events/state-of-the-map-us/2023/visualizing-hospital-accessibility-with-openstreetmap/},
  lastchecked = {2024-08-30}
}

@misc{Vestena2023,
  author = {Kauê de Moraes Vestena},
  title = {Acessibility Mapping of Footways using Openstreetmap},
  year = 2023,
  url  = {https://pretalx.com/sotm-africa-2023/talk/BRWGWQ/},
  lastchecked = {2024-08-30}
}

@misc{Gervais2023,
  author = {Bertrand Gervais and Florent Morel},
  title = {Pourquoi et comment améliorer les données de marchabilité dans OSM?},
  year = 2023,
  url  = {https://peertube.openstreetmap.fr/w/5xmC2MaFR5ZfwV3t4ujbF4},
  lastchecked = {2024-08-30}
}

@misc{Lainez2023,
  author = {Florian Lainez and Jean-Luc Chirpaz},
  title = {Projet Eazyway : mapper l'accessibilité… sans le tag wheelchair},
  year = 2023,
  url  = {https://peertube.openstreetmap.fr/w/o5AWXYerrMs7Us7v1YLdCF},
  lastchecked = {2024-08-30}
}

@misc{Clauss2023,
  author = {Hervé Clauss and Simon Hughes and Priscilla Zachée},
  title = {Questions-réponses sur Overture},
  year = 2023,
  url  = {https://peertube.openstreetmap.fr/w/q2q5tcVHFMoRBRnWg8U5je},
  lastchecked = {2024-08-30}
}

@misc{Baidoun2023,
  author = {Salim Baidoun and Hervé Clauss and Priscilla Zachée},
  title = {TomTom s’ouvre au monde, TomTom s’ouvre à OSM},
  year = 2023,
  url  = {https://peertube.openstreetmap.fr/w/iPmgfN9nyLV6TtGcgfBhAb},
  lastchecked = {2024-08-30}
}

@misc{Neerhut20231,
  author = {Edoardo Neerhut and Said Turksever},
  title = {Mapillary: 2 billion images and beyond},
  year = 2023,
  url  = {https://2023.stateofthemap.eu/program/mapillary-2-billion-images-and-beyond},
  lastchecked = {2024-08-30}
}

@misc{Neerhut20232,
  author = {Edoardo Neerhut},
  title = {Mapillary Camera Grant Program},
  year = 2023,
  url  = {https://openstreetmap.us/events/state-of-the-map-us/2023/mapillary-camera-grant-program/},
  lastchecked = {2024-08-30}
}

@misc{Turksever20231,
  author = {Said Turksever},
  title = {A Journey to 2 Billion Open Street-Level Imagery with Mapillary},
  year = 2023,
  url  = {https://pretalx.com/sotm-africa-2023/talk/TKAEYF/},
  lastchecked = {2024-08-30}
}

@misc{Riche20231,
  author = {Antoine Riche},
  title = {Overpass Turbo : le couteau suisse des données OSM},
  year = 2023,
  url  = {https://peertube.openstreetmap.fr/w/tzPeb3w8W7TKmCm4g7eiEF},
  lastchecked = {2024-08-30}
}

@misc{Riche20232,
  author = {Antoine Riche},
  title = {Requêtes Overpass : faites mieux que l'assistant d'Overpass Turbo},
  year = 2023,
  url  = {https://peertube.openstreetmap.fr/w/t2nyzAJ3xrF6LF7YEfVxPQ},
  lastchecked = {2024-08-30}
}

@misc{Buckel2023,
  author = {Alizée Buckel},
  title = {OSMOverpassConnector, un Transformer FME permettant l’accès aux données de l’API Overpass d’OSM},
  year = 2023,
  url  = {https://peertube.openstreetmap.fr/w/jrpmGrW8un2d8if6Pb3Ys7},
  lastchecked = {2024-08-30}
}

@misc{Reinmuth2023US,
  author = {Marcel Reinmuth},
  title = {Tracking OpenStreetMap History and Quality with the ohsome stack},
  year = 2023,
  url  = {https://openstreetmap.us/events/state-of-the-map-us/2023/tracking-openstreetmap-history-and-quality-with-the-ohsome-stack/},
  lastchecked = {2024-08-30}
}

@misc{Reinmuth2023,
  author = {Marcel Reinmuth},
  title = {History based quality measures of OpenStreetMap now in the ohsome dashboard},
  year = 2023,
  url  = {https://pretalx.com/sotm-africa-2023/talk/Q3NQMR/},
  lastchecked = {2024-08-30}
}

@misc{Housel2023,
  author = {Bryan Housel and Ben Clark},
  title = {Getting to know the new Rapid v2 Editor},
  year = 2023,
  url  = {https://openstreetmap.us/events/state-of-the-map-us/2023/getting-to-know-the-new-rapid-v2-editor/},
  lastchecked = {2024-08-30}
}

@misc{Turksever2023,
  author = {Said Turksever},
  title = {How to Map with Rapid},
  year = 2023,
  url  = {https://pretalx.com/sotm-africa-2023/talk/NRPXQZ/},
  lastchecked = {2024-08-30}
}

@misc{Topf2023,
  author = {Jochen Topf},
  title = {Exploring taginfo},
  year = 2023,
  url  = {https://2023.stateofthemap.eu/program/exploring-taginfo},
  lastchecked = {2024-08-30}
}

@misc{Zimmermann2023,
  author = {Jean-Louis Zimmermann and François Lacombe},
  title = {Tags d’OSM: sortir du yes-n},
  year = 2023,
  url  = {https://peertube.openstreetmap.fr/w/iXSK2NyRgkVadQapnZsZu2},
  lastchecked = {2024-08-30}
}

@misc{Grinberger2024,
  author = {Yair Grinberger and Marco Minghin},
  title = {What Happens When VGI is Threatened? A Systems Perspective Analysis of the Events Behind the Introduction of Rate Limiting in OpenStreetMap},
  year = 2024,
  url  = {https://2024.stateofthemap.org/sessions/FSYR9S/},
  lastchecked = {2024-08-30}
}

@misc{Kastl2011,
  author = {Daniel Kastl},
  title = {OSM in Japan before and after the tsunami},
  year = 2011,
  url  = {http://www.slideshare.net/kastl/osm-japan-before-and-after-the-tsunami},
  lastchecked = {2024-08-30}
}

@misc{Miura2011,
  author = {Hiroshi Miura},
  title = {Crisis mapping and disaster response},
  year = 2011,
  url  = {http://www.slideshare.net/miurahr/sotm2011-crisis-mapping-and-sinsaiinfo},
  lastchecked = {2024-08-30}
}

@misc{Ikiya2012,
  author = {Kinya Inoue and Tomomichi Hayakawa},
  title = {Fukushima mapping Before and after the disaster},
  year = 2012,
  url  = {http://www.slideshare.net/ikiya_OSM/fukushima-mapping-before-and-after-the-disaster},
  lastchecked = {2024-08-30}
}

@misc{stevecoast2024,
  author = {Steve Coast},
  title = {The Days Are Long but the Years Are Short},
  year = 2024,
  url  = {https://stevecoast.substack.com/p/the-days-are-long-but-the-years-are},
  lastchecked = {2024-08-30}
}

@article{leitenstern2024flexmap,
  title={FlexMap Fusion: Georeferencing and Automated Conflation of HD\~{} Maps with OpenStreetMap},
  author={Leitenstern, Maximilian and Sauerbeck, Florian and Kulmer, Dominik and Betz, Johannes},
  journal={arXiv preprint arXiv:2404.10879},
  year={2024}
}

@article{zhao2019exploring,
  title={Exploring semantic elements for urban scene recognition: Deep integration of high-resolution imagery and OpenStreetMap (OSM)},
  author={Zhao, Wenzhi and Bo, Yanchen and Chen, Jiage and Tiede, Dirk and Blaschke, Thomas and Emery, William J},
  journal={ISPRS Journal of Photogrammetry and Remote Sensing},
  volume={151},
  pages={237--250},
  year={2019},
  publisher={Elsevier}
}

@article{ding2021consistency,
  title={Consistency assessment for open geodata integration: An ontology-based approach},
  author={Ding, Linfang and Xiao, Guohui and Calvanese, Diego and Meng, Liqiu},
  journal={Geoinformatica},
  volume={25},
  number={4},
  pages={733--758},
  year={2021},
  publisher={Springer}
}

@article{ding2020framework,
  title={A framework uniting ontology-based geodata integration and geovisual analytics},
  author={Ding, Linfang and Xiao, Guohui and Calvanese, Diego and Meng, Liqiu},
  journal={ISPRS International Journal of Geo-Information},
  volume={9},
  number={8},
  pages={474},
  year={2020},
  publisher={MDPI}
}

@article{biljecki2021street,
  title={Street view imagery in urban analytics and GIS: A review},
  author={Biljecki, Filip and Ito, Koichi},
  journal={Landscape and Urban Planning},
  volume={215},
  pages={104217},
  year={2021},
  publisher={Elsevier}
}

@article{kang2018building,
  title={Building instance classification using street view images},
  author={Kang, Jian and K{\"o}rner, Marco and Wang, Yuanyuan and Taubenb{\"o}ck, Hannes and Zhu, Xiao Xiang},
  journal={ISPRS journal of photogrammetry and remote sensing},
  volume={145},
  pages={44--59},
  year={2018},
  publisher={Elsevier}
}

@article{bagheri2019fusion,
  title={Fusion of multi-sensor-derived heights and OSM-derived building footprints for urban 3D reconstruction},
  author={Bagheri, Hossein and Schmitt, Michael and Zhu, Xiaoxiang},
  journal={ISPRS International Journal of Geo-Information},
  volume={8},
  number={4},
  pages={193},
  year={2019},
  publisher={MDPI}
}

@article{zhuo2018optimization,
  title={Optimization of OpenStreetMap building footprints based on semantic information of oblique UAV images},
  author={Zhuo, Xiangyu and Fraundorfer, Friedrich and Kurz, Franz and Reinartz, Peter},
  journal={Remote Sensing},
  volume={10},
  number={4},
  pages={624},
  year={2018},
  publisher={MDPI}
}

@article{hoffmann2023using,
  title={Using social media images for building function classification},
  author={Hoffmann, Eike Jens and Abdulahhad, Karam and Zhu, Xiao Xiang},
  journal={Cities},
  volume={133},
  pages={104-107},
  year={2023}}

@article{hou2024global,
  title={Global Streetscapes—A comprehensive dataset of 10 million street-level images across 688 cities for urban science and analytics},
  author={Hou, Yujun and Quintana, Matias and Khomiakov, Maxim and Yap, Winston and Ouyang, Jiani and Ito, Koichi and Wang, Zeyu and Zhao, Tianhong and Biljecki, Filip},
  journal={ISPRS Journal of Photogrammetry and Remote Sensing},
  volume={215},
  pages={216--238},
  year={2024},
  publisher={Elsevier}
}

@article{lim2024integration,
  title={Integration of movement data into 3d gis},
  author={Lim, Joie and Biljecki, Filip and Stouffs, Rudi},
  journal={ISPRS Annals of the Photogrammetry, Remote Sensing and Spatial Information Sciences},
  volume={10},
  pages={219--227},
  year={2024},
  publisher={Copernicus Publications G{\"o}ttingen, Germany}
}

@article{wang2023insights,
  title={Insights in a city through the eyes of Airbnb reviews: Sensing urban characteristics from homestay guest experiences},
  author={Wang, Jiaxuan and Chow, Yoong Shin and Biljecki, Filip},
  journal={Cities},
  volume={140},
  pages={104399},
  year={2023},
  publisher={Elsevier}
}

@article{chen2022mining,
  title={Mining real estate ads and property transactions for building and amenity data acquisition},
  author={Chen, Xinyu and Biljecki, Filip},
  journal={Urban Informatics},
  volume={1},
  number={1},
  pages={12},
  year={2022},
  publisher={Springer}
}

@article{hu2024dlrgeotweet,
  title={DLRGeoTweet: A comprehensive social media geocoding corpus featuring fine-grained places},
  author={Hu, Xuke and El{\ss}ner, Tobias and Zheng, Shiyu and Serere, Helen Ngonidzashe and Kersten, Jens and Klan, Friederike and Qiu, Qinjun},
  journal={Information Processing \& Management},
  volume={61},
  number={4},
  pages={103742},
  year={2024},
  publisher={Elsevier}
}

@article{heeramaglore2022semantically,
  title={Semantically enriched voxels as a common representation for comparison and evaluation of 3D building models},
  author={Heeramaglore, Medhini and Kolbe, Thomas H},
  journal={ISPRS Annals of the Photogrammetry, Remote Sensing and Spatial Information Sciences},
  volume={10},
  pages={89--96},
  year={2022},
  publisher={Copernicus Publications G{\"o}ttingen, Germany}
}

@article{li20233dcentripetalnet,
  title={3DCentripetalNet: Building height retrieval from monocular remote sensing imagery},
  author={Li, Qingyu and Mou, Lichao and Hua, Yuansheng and Shi, Yilei and Chen, Sining and Sun, Yao and Zhu, Xiao Xiang},
  journal={International Journal of Applied Earth Observation and Geoinformation},
  volume={120},
  pages={103311},
  year={2023},
  publisher={Elsevier}
}

@article{xu2019pairwise,
  title={Pairwise coarse registration of point clouds in urban scenes using voxel-based 4-planes congruent sets},
  author={Xu, Yusheng and Boerner, Richard and Yao, Wei and Hoegner, Ludwig and Stilla, Uwe},
  journal={ISPRS journal of photogrammetry and remote sensing},
  volume={151},
  pages={106--123},
  year={2019},
  publisher={Elsevier}
}

@article{hoegner2018mobile,
  title={Mobile thermal mapping for matching of infrared images with 3D building models and 3D point clouds},
  author={Hoegner, L and Stilla, U},
  journal={Quantitative Infrared thermography journal},
  volume={15},
  number={2},
  pages={252--270},
  year={2018},
  publisher={Taylor \& Francis}
}

@article{xu2021voxel,
  title={Voxel-based representation of 3D point clouds: Methods, applications, and its potential use in the construction industry},
  author={Xu, Yusheng and Tong, Xiaohua and Stilla, Uwe},
  journal={Automation in Construction},
  volume={126},
  pages={103675},
  year={2021},
  publisher={Elsevier}
}

@article{pan2022enriching,
  title={Enriching geometric digital twins of buildings with small objects by fusing laser scanning and AI-based image recognition},
  author={Pan, Yuandong and Braun, Alexander and Brilakis, Ioannis and Borrmann, Andr{\'e}},
  journal={Automation in Construction},
  volume={140},
  pages={104375},
  year={2022},
  publisher={Elsevier}
}

@techreport{gilbert2020built,
  title={Built environment data standards and their integration: an analysis of IFC, CityGML and LandInfra},
  author={Gilbert, Thomas and R{\"o}nsdorf, Carsten and Plume, Jim and Simmons, Scott and Nisbet, Nick and Gruler, Hans-Christoph and Kolbe, Thomas H and van Berlo, L{\'e}on and Mercer, Aidan and others},
  year={2020},
  institution={Lehrstuhl f{\"u}r Geoinformatik}
}

@article{kolbe2021semantic,
  title={Semantic 3D city modeling and BIM},
  author={Kolbe, Thomas H and Donaubauer, Andreas},
  journal={Urban informatics},
  pages={609--636},
  year={2021},
  publisher={Springer}
}

@article{biljecki2021extending,
  title={Extending CityGML for IFC-sourced 3D city models},
  author={Biljecki, Filip and Lim, Joie and Crawford, James and Moraru, Diana and Tauscher, Helga and Konde, Amol and Adouane, Kamel and Lawrence, Simon and Janssen, Patrick and Stouffs, Rudi},
  journal={Automation in Construction},
  volume={121},
  pages={103440},
  year={2021},
  publisher={Elsevier}
}

@article{floros2018investigating,
  title={Investigating interoperability capabilities between IFC and CityGML LOD 4--retaining semantic information},
  author={Floros, GS and Ellul, Claire and Dimopoulou, Efi},
  journal={The International Archives of the Photogrammetry, Remote Sensing and Spatial Information Sciences},
  volume={42},
  pages={33--40},
  year={2018},
  publisher={Copernicus GmbH}
}

@article{boeters2015automatically,
  title={Automatically enhancing CityGML LOD2 models with a corresponding indoor geometry},
  author={Boeters, Roeland and Arroyo Ohori, Ken and Biljecki, Filip and Zlatanova, Sisi},
  journal={International journal of geographical information science},
  volume={29},
  number={12},
  pages={2248--2268},
  year={2015},
  publisher={Taylor \& Francis}
}

@article{pantoja2022generating,
  title={Generating LOD3 building models from structure-from-motion and semantic segmentation},
  author={Pantoja-Rosero, Bryan G and Achanta, Radhakrishna and Kozinski, Mateusz and Fua, Pascal and Perez-Cruz, Fernando and Beyer, Katrin},
  journal={Automation in Construction},
  volume={141},
  pages={104430},
  year={2022},
  publisher={Elsevier}
}

@article{donkers2016automatic,
  title={Automatic conversion of IFC datasets to geometrically and semantically correct CityGML LOD3 buildings},
  author={Donkers, Sjors and Ledoux, Hugo and Zhao, Junqiao and Stoter, Jantien},
  journal={Transactions in GIS},
  volume={20},
  number={4},
  pages={547--569},
  year={2016},
  publisher={Wiley Online Library}
}

@article{biljecki2019raise,
  title={Raise the roof: Towards generating LOD2 models without aerial surveys using machine learning},
  author={Biljecki, FILIP and Dehbi, YOUNESS},
  journal={ISPRS Annals of the Photogrammetry, Remote Sensing and Spatial Information Sciences},
  volume={4},
  pages={27--34},
  year={2019},
  publisher={Copernicus GmbH}
}

@article{dehbi2021optimal,
  title={Optimal scan planning with enforced network connectivity for the acquisition of three-dimensional indoor models},
  author={Dehbi, Youness and Leonhardt, Johannes and Oehrlein, Johannes and Haunert, Jan-Henrik},
  journal={ISPRS Journal of Photogrammetry and Remote Sensing},
  volume={180},
  pages={103--116},
  year={2021},
  publisher={Elsevier}
}

@article{wang2024framework,
  title={A framework for fully automated reconstruction of semantic building model at urban-scale using textured LoD2 data},
  author={Wang, Yuefeng and Jiao, Wei and Fan, Hongchao and Zhou, Guoqing},
  journal={ISPRS Journal of Photogrammetry and Remote Sensing},
  volume={216},
  pages={90--108},
  year={2024},
  publisher={Elsevier}
}

@article{hu2023location,
  title={Location reference recognition from texts: A survey and comparison},
  author={Hu, Xuke and Zhou, Zhiyong and Li, Hao and Hu, Yingjie and Gu, Fuqiang and Kersten, Jens and Fan, Hongchao and Klan, Friederike},
  journal={ACM Computing Surveys},
  volume={56},
  number={5},
  pages={1--37},
  year={2023},
  publisher={ACM New York, NY}
}

@article{gui2021automated,
  title={Automated LoD-2 model reconstruction from very-high-resolution satellite-derived digital surface model and orthophoto},
  author={Gui, Shengxi and Qin, Rongjun},
  journal={ISPRS Journal of Photogrammetry and Remote Sensing},
  volume={181},
  pages={1--19},
  year={2021},
  publisher={Elsevier}
}

@article{li2024review,
  title={A Review of Building Extraction from Remote Sensing Imagery: Geometrical Structures and Semantic Attributes},
  author={Li, Qingyu and Mou, Lichao and Sun, Yao and Hua, Yuansheng and Shi, Yilei and Zhu, Xiao Xiang},
  journal={IEEE Transactions on Geoscience and Remote Sensing},
  year={2024},
  publisher={IEEE}
}

@inproceedings{mooney2018coordinating,
  title={Coordinating improved communication between the academic and OpenStreetMap communities},
  author={Mooney, Peter and Schouppe, Joost and Ostermann, FO},
  booktitle={OpenStreetMap State of the Map 2018},
  year={2018}
}

@article{senaratne2017review,
  title={A review of volunteered geographic information quality assessment methods},
  author={Senaratne, Hansi and Mobasheri, Amin and Ali, Ahmed Loai and Capineri, Cristina and Haklay, Mordechai},
  journal={International Journal of Geographical Information Science},
  volume={31},
  number={1},
  pages={139--167},
  year={2017},
  publisher={Taylor \& Francis}
}

@article{corcoran2013analysing,
  title={Analysing the growth of OpenStreetMap networks},
  author={Corcoran, Padraig and Mooney, Peter and Bertolotto, Michela},
  journal={Spatial Statistics},
  volume={3},
  pages={21--32},
  year={2013},
  publisher={Elsevier}
}

@article{jackson2013assessing,
  title={Assessing completeness and spatial error of features in volunteered geographic information},
  author={Jackson, Steven P and Mullen, William and Agouris, Peggy and Crooks, Andrew and Croitoru, Arie and Stefanidis, Anthony},
  journal={ISPRS International Journal of Geo-Information},
  volume={2},
  number={2},
  pages={507--530},
  year={2013},
  publisher={MDPI}
}

@article{mooney2012annotation,
  title={The annotation process in OpenStreetMap},
  author={Mooney, Peter and Corcoran, Padraig},
  journal={Transactions in GIS},
  volume={16},
  number={4},
  pages={561--579},
  year={2012},
  publisher={Wiley Online Library}
}

@article{zielstra2012using,
  title={Using free and proprietary data to compare shortest-path lengths for effective pedestrian routing in street networks},
  author={Zielstra, Dennis and Hochmair, Hartwig H},
  journal={Transportation Research Record},
  volume={2299},
  number={1},
  pages={41--47},
  year={2012},
  publisher={SAGE Publications Sage CA: Los Angeles, CA}
}

@inproceedings{palen2015success,
  title={Success \& scale in a data-producing organization: The socio-technical evolution of OpenStreetMap in response to humanitarian events},
  author={Palen, Leysia and Soden, Robert and Anderson, T Jennings and Barrenechea, Mario},
  booktitle={Proceedings of the 33rd annual ACM conference on human factors in computing systems},
  pages={4113--4122},
  year={2015}
}

@inproceedings{poiani2016potential,
  title={Potential of collaborative mapping for disaster relief: A case study of OpenStreetMap in the Nepal earthquake 2015},
  author={Poiani, Thiago Henrique and Rocha, Roberto Dos Santos and Degrossi, L{\'\i}via Castro and De Albuquerque, Joao Porto},
  booktitle={2016 49th Hawaii International Conference on System Sciences (HICSS)},
  pages={188--197},
  year={2016},
  organization={IEEE}
}

@article{herfort2021evolution,
  title={The evolution of humanitarian mapping within the OpenStreetMap community},
  author={Herfort, Benjamin and Lautenbach, Sven and Porto de Albuquerque, Jo{\~a}o and Anderson, Jennings and Zipf, Alexander},
  journal={Scientific reports},
  volume={11},
  number={1},
  pages={3037},
  year={2021},
  publisher={Nature Publishing Group UK London}
}

@article{ballatore2013geographic,
  title={Geographic knowledge extraction and semantic similarity in OpenStreetMap},
  author={Ballatore, Andrea and Bertolotto, Michela and Wilson, David C},
  journal={Knowledge and Information Systems},
  volume={37},
  pages={61--81},
  year={2013},
  publisher={Springer}
}

@article{graser2014towards,
  title={Towards an open source analysis toolbox for street network comparison: Indicators, tools and results of a comparison of OSM and the official A ustrian reference graph},
  author={Graser, Anita and Straub, Markus and Dragaschnig, Melitta},
  journal={Transactions in GIS},
  volume={18},
  number={4},
  pages={510--526},
  year={2014},
  publisher={Wiley Online Library}
}

@article{boeing2017osmnx,
  title={OSMnx: New methods for acquiring, constructing, analyzing, and visualizing complex street networks},
  author={Boeing, Geoff},
  journal={Computers, environment and urban systems},
  volume={65},
  pages={126--139},
  year={2017},
  publisher={Elsevier}
}

@article{liu2016automated,
  title={Automated identification and characterization of parcels with OpenStreetMap and points of interest},
  author={Liu, Xingjian and Long, Ying},
  journal={Environment and Planning B: Planning and Design},
  volume={43},
  number={2},
  pages={341--360},
  year={2016},
  publisher={SAGE Publications Sage UK: London, England}
}

@article{schultz2017open,
  title={Open land cover from OpenStreetMap and remote sensing},
  author={Schultz, Michael and Voss, Janek and Auer, Michael and Carter, Sarah and Zipf, Alexander},
  journal={International journal of applied earth observation and geoinformation},
  volume={63},
  pages={206--213},
  year={2017},
  publisher={Elsevier}
}

@article{kaiser2017learning,
  title={Learning aerial image segmentation from online maps},
  author={Kaiser, Pascal and Wegner, Jan Dirk and Lucchi, Aur{\'e}lien and Jaggi, Martin and Hofmann, Thomas and Schindler, Konrad},
  journal={IEEE Transactions on Geoscience and Remote Sensing},
  volume={55},
  number={11},
  pages={6054--6068},
  year={2017},
  publisher={IEEE}
}

@article{johnson2016integrating,
  title={Integrating OpenStreetMap crowdsourced data and Landsat time-series imagery for rapid land use/land cover (LULC) mapping: Case study of the Laguna de Bay area of the Philippines},
  author={Johnson, Brian A and Iizuka, Kotaro},
  journal={Applied Geography},
  volume={67},
  pages={140--149},
  year={2016},
  publisher={Elsevier}
}

@article{Fonte2017,
  title={Generating Up-to-Date and Detailed Land Use and Land Cover Maps Using OpenStreetMap and GlobeLand30},
  author={Fonte, Cidália Costa and Minghini, Marco and Patriarca, Joaquim and Antoniou, Vyron and See, Linda and  Skopeliti, Andriani},
  journal={ISPRS International Journal of Geo-Information},
  volume={6},
  number={4},
  pages={125},
  year={2017}
}

@article{bakillah2014fine,
  title={Fine-resolution population mapping using OpenStreetMap points-of-interest},
  author={Bakillah, Mohamed and Liang, Steve and Mobasheri, Amin and Jokar Arsanjani, Jamal and Zipf, Alexander},
  journal={International Journal of Geographical Information Science},
  volume={28},
  number={9},
  pages={1940--1963},
  year={2014},
  publisher={Taylor \& Francis}
}

@article{goetz2013towards,
  title={Towards generating highly detailed 3D CityGML models from OpenStreetMap},
  author={Goetz, Marcus},
  journal={International Journal of Geographical Information Science},
  volume={27},
  number={5},
  pages={845--865},
  year={2013},
  publisher={Taylor \& Francis}
}

@article{jokar2013toward,
  title={Toward mapping land-use patterns from volunteered geographic information},
  author={Jokar Arsanjani, Jamal and Helbich, Marco and Bakillah, Mohamed and Hagenauer, Julian and Zipf, Alexander},
  journal={International Journal of Geographical Information Science},
  volume={27},
  number={12},
  pages={2264--2278},
  year={2013},
  publisher={Taylor \& Francis}
}

@article{hagenauer2012mining,
  title={Mining urban land-use patterns from volunteered geographic information by means of genetic algorithms and artificial neural networks},
  author={Hagenauer, Julian and Helbich, Marco},
  journal={International Journal of Geographical Information Science},
  volume={26},
  number={6},
  pages={963--982},
  year={2012},
  publisher={Taylor \& Francis}
}

@article{zielstra2011comparative,
  title={Comparative study of pedestrian accessibility to transit stations using free and proprietary network data},
  author={Zielstra, Dennis and Hochmair, Hartwig H},
  journal={Transportation Research Record},
  volume={2217},
  number={1},
  pages={145--152},
  year={2011},
  publisher={SAGE Publications Sage CA: Los Angeles, CA}
}

@article{over2010generating,
  title={Generating web-based 3D City Models from OpenStreetMap: The current situation in Germany},
  author={Over, Martin and Schilling, Arne and Neubauer, S and Zipf, Alexander},
  journal={Computers, Environment and urban systems},
  volume={34},
  number={6},
  pages={496--507},
  year={2010},
  publisher={Elsevier}
}

@article{anderson2019corporate,
  title={Corporate editors in the evolving landscape of OpenStreetMap},
  author={Anderson, Jennings and Sarkar, Dipto and Palen, Leysia},
  journal={ISPRS International Journal of Geo-Information},
  volume={8},
  number={5},
  pages={232},
  year={2019},
  publisher={MDPI}
}

@article{jokar2015emergence,
  title={The emergence and evolution of OpenStreetMap: a cellular automata approach},
  author={Jokar Arsanjani, Jamal and Helbich, Marco and Bakillah, Mohamed and Loos, Lukas},
  journal={International Journal of Digital Earth},
  volume={8},
  number={1},
  pages={76--90},
  year={2015},
  publisher={Taylor \& Francis}
}

@article{arsanjani2015exploration,
  title={An exploration of future patterns of the contributions to OpenStreetMap and development of a Contribution Index},
  author={Arsanjani, Jamal Jokar and Mooney, Peter and Helbich, Marco and Zipf, Alexander},
  journal={Transactions in GIS},
  volume={19},
  number={6},
  pages={896--914},
  year={2015},
  publisher={Wiley Online Library}
}

@article{mooney2014has,
  title={Has OpenStreetMap a role in Digital Earth applications?},
  author={Mooney, Peter and Corcoran, Padraig},
  journal={International Journal of Digital Earth},
  volume={7},
  number={7},
  pages={534--553},
  year={2014},
  publisher={Taylor \& Francis}
}

@article{budhathoki2013motivation,
  title={Motivation for open collaboration: Crowd and community models and the case of OpenStreetMap},
  author={Budhathoki, Nama R and Haythornthwaite, Caroline},
  journal={American Behavioral Scientist},
  volume={57},
  number={5},
  pages={548--575},
  year={2013},
  publisher={Sage Publications Sage CA: Los Angeles, CA}
}

@article{mooney2014analysis,
  title={Analysis of Interaction and Co-editing Patterns amongst OpenStreetMap Contributors},
  author={Mooney, Peter and Corcoran, Padraig},
  journal={Transactions in GIS},
  volume={18},
  number={5},
  pages={633--659},
  year={2014},
  publisher={Wiley Online Library}
}

@article{neis2012towards,
  title={Towards automatic vandalism detection in OpenStreetMap},
  author={Neis, Pascal and Goetz, Marcus and Zipf, Alexander},
  journal={ISPRS International Journal of Geo-Information},
  volume={1},
  number={3},
  pages={315--332},
  year={2012},
  publisher={MDPI}
}

@article{neis2012analyzing,
  title={Analyzing the contributor activity of a volunteered geographic information project—The case of OpenStreetMap},
  author={Neis, Pascal and Zipf, Alexander},
  journal={ISPRS International Journal of Geo-Information},
  volume={1},
  number={2},
  pages={146--165},
  year={2012},
  publisher={Molecular Diversity Preservation International}
}

@article{lin2011qualitative,
  title={A qualitative enquiry into OpenStreetMap making},
  author={Lin, Yu-Wei},
  journal={New Review of Hypermedia and Multimedia},
  volume={17},
  number={1},
  pages={53--71},
  year={2011},
  publisher={Taylor \& Francis}
}

@article{zhou2018exploring,
  title={Exploring the relationship between density and completeness of urban building data in OpenStreetMap for quality estimation},
  author={Zhou, Qi},
  journal={International Journal of Geographical Information Science},
  volume={32},
  number={2},
  pages={257--281},
  year={2018},
  publisher={Taylor \& Francis}
}

@article{degrossi2018taxonomy,
  title={A taxonomy of quality assessment methods for volunteered and crowdsourced geographic information},
  author={Degrossi, L{\'\i}via Castro and Porto de Albuquerque, Jo{\~a}o and Santos Rocha, Roberto dos and Zipf, Alexander},
  journal={Transactions in GIS},
  volume={22},
  number={2},
  pages={542--560},
  year={2018},
  publisher={Wiley Online Library}
}

@article{brovelli2018new,
  title={A new method for the assessment of spatial accuracy and completeness of OpenStreetMap building footprints},
  author={Brovelli, Maria Antonia and Zamboni, Giorgio},
  journal={ISPRS International Journal of Geo-Information},
  volume={7},
  number={8},
  pages={289},
  year={2018},
  publisher={MDPI}
}

@article{arsanjani2015assessment,
  title={An assessment of a collaborative mapping approach for exploring land use patterns for several European metropolises},
  author={Arsanjani, Jamal Jokar and Vaz, Eric},
  journal={International Journal of Applied Earth Observation and Geoinformation},
  volume={35},
  pages={329--337},
  year={2015},
  publisher={Elsevier}
}

@article{dorn2015quality,
  title={Quality evaluation of VGI using authoritative data—A comparison with land use data in Southern Germany},
  author={Dorn, Helen and T{\"o}rnros, Tobias and Zipf, Alexander},
  journal={ISPRS International Journal of Geo-Information},
  volume={4},
  number={3},
  pages={1657--1671},
  year={2015},
  publisher={MDPI}
}

@article{antoniou2015measures,
  title={Measures and indicators of VGI quality: An overview},
  author={Antoniou, Vyron and Skopeliti, Andriani},
  journal={ISPRS annals of the photogrammetry, remote sensing and spatial information sciences},
  volume={2},
  pages={345--351},
  year={2015},
  publisher={Copernicus GmbH}
}

@article{barrington2017world,
  title={The world’s user-generated road map is more than 80\% complete},
  author={Barrington-Leigh, Christopher and Millard-Ball, Adam},
  journal={PloS one},
  volume={12},
  number={8},
  pages={e0180698},
  year={2017},
  publisher={Public Library of Science San Francisco, CA USA}
}

@article{hochmair2015assessing,
  title={Assessing the Completeness of Bicycle Trail and Lane Features in O pen S treet M ap for the U nited S tates},
  author={Hochmair, Hartwig H and Zielstra, Dennis and Neis, Pascal},
  journal={Transactions in GIS},
  volume={19},
  number={1},
  pages={63--81},
  year={2015},
  publisher={Wiley Online Library}
}

@article{brovelli2017towards,
  title={Towards an automated comparison of OpenStreetMap with authoritative road datasets},
  author={Brovelli, Maria Antonia and Minghini, Marco and Molinari, Monia and Mooney, Peter},
  journal={Transactions in GIS},
  volume={21},
  number={2},
  pages={191--206},
  year={2017},
  publisher={Wiley Online Library}
}

@article{forghani2014quality,
  title={A quality study of the OpenStreetMap dataset for Tehran},
  author={Forghani, Mohammad and Delavar, Mahmoud Reza},
  journal={ISPRS International Journal of Geo-Information},
  volume={3},
  number={2},
  pages={750--763},
  year={2014},
  publisher={MDPI}
}

@article{barron2014comprehensive,
  title={A comprehensive framework for intrinsic OpenStreetMap quality analysis},
  author={Barron, Christopher and Neis, Pascal and Zipf, Alexander},
  journal={Transactions in GIS},
  volume={18},
  number={6},
  pages={877--895},
  year={2014},
  publisher={Wiley Online Library}
}

@incollection{kessler2013trust,
  title={Trust as a proxy measure for the quality of volunteered geographic information in the case of OpenStreetMap},
  author={Ke{\ss}ler, Carsten and De Groot, Ren{\'e} Theodore Anton},
  booktitle={Geographic information science at the heart of Europe},
  pages={21--37},
  year={2013},
  publisher={Springer}
}

@article{zielstra2013assessing,
  title={Assessing the effect of data imports on the completeness of openstreetmap--au nited s tates case study},
  author={Zielstra, Dennis and Hochmair, Hartwig H and Neis, Pascal},
  journal={Transactions in GIS},
  volume={17},
  number={3},
  pages={315--334},
  year={2013},
  publisher={Wiley Online Library}
}

@article{hecht2013measuring,
  title={Measuring completeness of building footprints in OpenStreetMap over space and time},
  author={Hecht, Robert and Kunze, Carola and Hahmann, Stefan},
  journal={ISPRS International Journal of Geo-Information},
  volume={2},
  number={4},
  pages={1066--1091},
  year={2013},
  publisher={MDPI}
}

@article{koukoletsos2012assessing,
  title={Assessing data completeness of VGI through an automated matching procedure for linear data},
  author={Koukoletsos, Thomas and Haklay, Mordechai and Ellul, Claire},
  journal={Transactions in GIS},
  volume={16},
  number={4},
  pages={477--498},
  year={2012},
  publisher={Wiley Online Library}
}

@article{haklay2010many,
  title={How many volunteers does it take to map an area well? The validity of Linus’ law to volunteered geographic information},
  author={Haklay, Mordechai and Basiouka, Sofia and Antoniou, Vyron and Ather, Aamer},
  journal={The cartographic journal},
  volume={47},
  number={4},
  pages={315--322},
  year={2010},
  publisher={Taylor \& Francis}
}

@article{girres2010quality,
  title={Quality assessment of the French OpenStreetMap dataset},
  author={Girres, Jean-Fran{\c{c}}ois and Touya, Guillaume},
  journal={Transactions in GIS},
  volume={14},
  number={4},
  pages={435--459},
  year={2010},
  publisher={Wiley Online Library}
}

@article{haklay2010good,
  title={How good is volunteered geographical information? A comparative study of OpenStreetMap and Ordnance Survey datasets},
  author={Haklay, Mordechai},
  journal={Environment and planning B: Planning and design},
  volume={37},
  number={4},
  pages={682--703},
  year={2010},
  publisher={SAGE Publications Sage UK: London, England}
}

@article{haklay2008openstreetmap,
  title={Openstreetmap: User-generated street maps},
  author={Haklay, Mordechai and Weber, Patrick},
  journal={IEEE Pervasive computing},
  volume={7},
  number={4},
  pages={12--18},
  year={2008},
  publisher={Ieee}
}

@article{zadeh2023improving,
  title={Improving the accuracy of earthquake risk estimates with OpenStreetMap building data},
  author={Zadeh, Tara Evaz and Oostwegel, Laurens JN and Lingner, Lars and Shinde, Simantini and Cotton, Fabrice and Schorlemmer, Danijel},
  journal={Proceedings of OSM Science 2023},
  pages={18--21},
  year={2023}
}

@article{oostwegel2023global,
  title={A global and dynamic completeness assessment of the OpenStreetMap buildings},
  author={Oostwegel, Laurens JN and Evaz Zadeh, T and Lingner, Lars and Schorlemmer, Danijel},
  journal={Proceedings of the OSM Science 2023},
  pages={6--9},
  year={2023}
}

@article{biljecki2023quality,
  title={Quality of crowdsourced geospatial building information: A global assessment of OpenStreetMap attributes},
  author={Biljecki, Filip and Chow, Yoong Shin and Lee, Kay},
  journal={Building and Environment},
  volume={237},
  pages={110295},
  year={2023},
  publisher={Elsevier}
}

@article{melanda2023openstreetmap,
  title={OpenStreetMap Data for Automated Labelling Machine Learning Examples: The Challenge of Road Type Imbalance},
  author={Melanda, EA and Herfort, Benjamin and Ulrich, Veit and Andorful, Francis and Zipf, Alexander},
  journal={Proceedings of the OSM Science 2023},
  pages={65--68},
  year={2023}
}

@article{andorful2023exploring,
  title={Exploring road and points of interest (POIs) associations in OpenStreetMap, a new paradigm for OSM road class prediction},
  author={Andorful, Francis and Lautenbach, Sven and Ludwig, Christina and Herfort, Benjamin and Nir, Fulman and Zipf, A},
  journal={Proceedings of the OSM Science 2023},
  pages={69--72},
  year={2023}
}

@article{grinberger2023openstreetmap,
  title={OpenStreetMap as an emerging scientific field: Reflections from OSM Science 2023},
  author={Grinberger, A Yair and Liu, Hao and Liu, Pengyuan and Yeboah, Godwin and Juhasz, Levente and Coetzee, Serena and Mooney, Peter and Sarretta, Alessandro and Anderson, Jennings and Minghini, Marco},
  journal={Proceedings of the OSM Science 2023},
  pages={1--5},
  year={2023},
  publisher={Zenodo}
}

@inproceedings{karlekar-bansal-2018-safecity,
    title = "{S}afe{C}ity: Understanding Diverse Forms of Sexual Harassment Personal Stories",
    author = "Karlekar, Sweta  and
      Bansal, Mohit",
    editor = "Riloff, Ellen  and
      Chiang, David  and
      Hockenmaier, Julia  and
      Tsujii, Jun{'}ichi",
    booktitle = "Proceedings of the 2018 Conference on Empirical Methods in Natural Language Processing",
    month = oct # "-" # nov,
    year = "2018",
    address = "Brussels, Belgium",
    publisher = "Association for Computational Linguistics",
    pages = "2805--2811",
}

@article{gardner2020quantifying,
  title={Quantifying gendered participation in OpenStreetMap: Responding to theories of female (under) representation in crowdsourced mapping},
  author={Gardner, Zoe and Mooney, Peter and De Sabbata, Stefano and Dowthwaite, Liz},
  journal={GeoJournal},
  volume={85},
  number={6},
  pages={1603--1620},
  year={2020},
  publisher={Springer}
}

@article{juhasz2020cartographic,
  title={Cartographic vandalism in the era of location-based games—The case of OpenStreetMap and Pok{\'e}mon GO},
  author={Juh{\'a}sz, Levente and Novack, Tessio and Hochmair, Hartwig H and Qiao, Sen},
  journal={ISPRS International Journal of Geo-Information},
  volume={9},
  number={4},
  pages={197},
  year={2020},
  publisher={MDPI}
}

@article{grinberger2019bridging,
  title={Bridging the map? Exploring interactions between the academic and mapping communities in OpenStreetMap},
  author={Grinberger, A Yair and Minghini, Marco and Juh{\'a}sz, Levente and Mooney, Peter and Yeboah, Godwin},
  journal={Proceedings of the Academic Track at the State of the Map 2019},
  pages={1--2},
  year={2019}
}

@article{van2010software,
  title={Software survey: VOSviewer, a computer program for bibliometric mapping},
  author={Van Eck, Nees and Waltman, Ludo},
  journal={scientometrics},
  volume={84},
  number={2},
  pages={523--538},
  year={2010},
  publisher={Akad{\'e}miai Kiad{\'o}, co-published with Springer Science+ Business Media BV~…}
}

@article{aria2017bibliometrix,
  title={bibliometrix: An R-tool for comprehensive science mapping analysis},
  author={Aria, Massimo and Cuccurullo, Corrado},
  journal={Journal of informetrics},
  volume={11},
  number={4},
  pages={959--975},
  year={2017},
  publisher={Elsevier}
}

@article{neis2014recent,
  title={Recent developments and future trends in volunteered geographic information research: The case of OpenStreetMap},
  author={Neis, Pascal and Zielstra, Dennis},
  journal={Future internet},
  volume={6},
  number={1},
  pages={76--106},
  year={2014},
  publisher={MDPI}
}

@article{mooney2017review,
  title={A review of OpenStreetMap data},
  author={Mooney, Peter and Minghini, Marco and others},
  journal={Mapping and the citizen sensor},
  pages={37--59},
  year={2017},
  publisher={Ubiquity Press}
}

@inproceedings{kaur2017systematic,
  title={Systematic literature review of data quality within openstreetmap},
  author={Kaur, Jasmeet and Singh, Jaiteg and Sehra, Sukhjit Singh and Rai, Hardeep Singh},
  booktitle={2017 International conference on next generation computing and information systems (ICNGCIS)},
  pages={177--182},
  year={2017},
  organization={IEEE}
}

@article{vargas2020openstreetmap,
  title={OpenStreetMap: Challenges and opportunities in machine learning and remote sensing},
  author={Vargas-Munoz, John E and Srivastava, Shivangi and Tuia, Devis and Falcao, Alexandre X},
  journal={IEEE Geoscience and Remote Sensing Magazine},
  volume={9},
  number={1},
  pages={184--199},
  year={2020},
  publisher={IEEE}
}

@misc{grinberger2022osm,
  title={OSM Science—The Academic Study of the OpenStreetMap Project, Data, Contributors, Community, and Applications},
  author={Grinberger, A Yair and Minghini, Marco and Juh{\'a}sz, Levente and Yeboah, Godwin and Mooney, Peter},
  journal={ISPRS International Journal of Geo-Information},
  volume={11},
  number={4},
  pages={230},
  year={2022},
  publisher={MDPI}
}

@article{grinberger2022bridges,
  title={Bridges and barriers: An exploration of engagements of the research community with the OpenStreetMap community},
  author={Grinberger, A Yair and Minghini, Marco and Yeboah, Godwin and Juh{\'a}sz, Levente and Mooney, Peter},
  journal={ISPRS International Journal of Geo-Information},
  volume={11},
  number={1},
  pages={54},
  year={2022},
  publisher={MDPI}
}

@article{garfield2004historiographic,
  title={Historiographic mapping of knowledge domains literature},
  author={Garfield, Eugene},
  journal={Journal of information science},
  volume={30},
  number={2},
  pages={119--145},
  year={2004},
  publisher={Sage Publications Sage CA: Thousand Oaks, CA}
}

@article{sun2023qqb,
  author={Yao Sun and Wang, Yi and Eineder, Michael},
  journal={IEEE Geoscience and Remote Sensing Letters}, 
  title={QuickQuakeBuildings: Post-Earthquake SAR-Optical Dataset for Quick Damaged-Building Detection}, 
  year={2024},
  volume={21},
  pages={1-5},
  doi={10.1109/LGRS.2024.3406966}}

@Article{sun2023flickrstr,
AUTHOR = {Sun, Y. and Kruspe, A. and Meng, L. and Tian, Y. and Hoffmann, E. J. and Auer, S. and Zhu, X. X.},
TITLE = {TOWARDS LARGE-SCALE BUILDING ATTRIBUTE MAPPING USING CROWDSOURCED IMAGES: SCENE TEXT RECOGNITION ON FLICKR AND PROBLEMS TO BE SOLVED},
JOURNAL = {The International Archives of the Photogrammetry, Remote Sensing and Spatial Information Sciences},
VOLUME = {XLVIII-1/W2-2023},
YEAR = {2023},
PAGES = {225--232}
}

@article{li2023beyond,
  title={Beyond Two Dimensions: Large-Scale Building Height Mapping in OpenStreetMap via Synthetic Aperture Radar and Street-View Imagery},
  author={Li, Hao and Sun, Yao},
  journal={Proceedings of the OSM Science 2023},
  pages={38--41},
  year={2023}
}

@ARTICLE{sun2020cgnet,
  author={Sun, Yao and Hua, Yuansheng and Mou, Lichao and Zhu, Xiao Xiang},
  journal={IEEE Transactions on Geoscience and Remote Sensing}, 
  title={{CG-Net}: {Conditional} {GIS}-Aware Network for Individual Building Segmentation in {VHR SAR} Images}, 
  year={2021},
  volume={},
  number={},
  pages={1-15}
  }

@article{sun2021bbox,
title = {Large-scale building height retrieval from single SAR imagery based on bounding box regression networks},
journal = {ISPRS Journal of Photogrammetry and Remote Sensing},
volume = {184},
pages = {79-95},
year = {2022},
author = {Yao Sun and Lichao Mou and Yuanyuan Wang and Sina Montazeri and Xiao Xiang Zhu},
}

@inproceedings{li2020instance,
  title={Instance Segmentation of Buildings Using Keypoints}, author={Li, Qingyu and Mou, Lichao and Hua, Yuansheng and Sun, Yao and Jin, Pu and Shi, Yilei and Zhu, Xiao Xiang},
  booktitle={IEEE International Geoscience and Remote Sensing Symposium (IGARSS)}, 
  year={2020},
}

@inproceedings{chen2021maskH, 
    title = {{Mask-height R-CNN: An end-to-end network for 3D building reconstruction from monocular remote sensing imagery}},
    author = {Chen, Sining and Mou, Lichao and Li, Qingyu and Sun, Yao and Zhu, Xiao Xiang}, 
    booktitle={IEEE International Geoscience and Remote Sensing Symposium (IGARSS)},
    year={2021}
  }

@article{sun2020auto,
    title = {Automatic registration of a single 
    {SAR} image and {GIS} building footprints in a large-scale urban area},
    author = {Sun, Yao and Montazeri, Sina and Wang, Yuanyuan and Zhu, Xiao Xiang},
    journal = {ISPRS Journal of Photogrammetry and Remote Sensing},
    volume = {170},
    pages = {1-14},
    year = {2020},
}

@article{fan2014quality,
  title={Quality assessment for building footprints data on OpenStreetMap},
  author={Fan, Hongchao and Zipf, Alexander and Fu, Qing and Neis, Pascal},
  journal={International Journal of Geographical Information Science},
  volume={28},
  number={4},
  pages={700--719},
  year={2014},
  publisher={Taylor \& Francis}
}

@inproceedings{sun2017Building,
  title = {Building height estimation in single {SAR} image using {OSM} building footprints},
  booktitle = {Joint Urban Remote Sensing Event (JURSE)},
  author = {Sun, Yao and Shahzad, Muhammad and Zhu, Xiao Xiang},  
  year = {2017}
}

@misc{OpenStreetMapWiki2024,
  author = {OpenStreetMap},
  title = {History of {OpenStreetMap} - {OpenStreetMap} {Wiki}},
  year = 2024,
  url  = {https://wiki.openstreetmap.org/wiki/History_of_OpenStreetMap},
  lastchecked = {2024-08-30}
}

@article{hagen2019,
  author={Hagen, Erica},
  title={Sustainability in OpenStreetMap Building a More Stable Ecosystem in OSM for Development and Humanitarianism},
  year={2019},
  publisher={World Bank Group}
}

@incollection{schott2023analyzing,
  title={Analyzing and improving the quality and fitness for purpose of OpenStreetMap as labels in remote sensing applications},
  author={Schott, Moritz and Zell, Adina and Lautenbach, Sven and Sumbul, Gencer and Schultz, Michael and Zipf, Alexander and Demir, Beg{\"u}m},
  booktitle={Volunteered geographic information: Interpretation, visualization and social context},
  pages={21--42},
  year={2023},
  publisher={Springer Nature Switzerland Cham}
}

\end{document}